\documentclass[a4paper,12pt,twoside]{article}
\usepackage{graphicx}\usepackage{a4wide}
\usepackage{amssymb,amsmath,amsthm,mathrsfs}
\usepackage{siunitx}
\usepackage{subfigure,color,colordvi,graphicx,xcolor}
\usepackage[only,llbracket,rrbracket,interleave]{stmaryrd}
\usepackage[sort,compress]{cite}
\usepackage{enumitem}

\makeatletter
\newcommand*\wt[2][0.2ex]{%
        \begingroup
        \mathchoice{\wt@helper{#1}{#2}{\displaystyle}{\textfont}}
                   {\wt@helper{#1}{#2}{\textstyle}{\textfont}}
                   {\wt@helper{#1}{#2}{\scriptstyle}{\scriptfont}}
                   {\wt@helper{#1}{#2}{\scriptscriptstyle}{\scriptscriptfont}}%
        \endgroup
        #2%
}
\newcommand*\wt@helper[4]{%
        \def\currentfont{\the#41}%
        \def\currentskewchar{\char\the\skewchar\currentfont}%
        \setbox\tw@\hbox{\currentfont#2\currentskewchar}%
        \dimen@ii\wd\tw@
        \setbox\tw@\hbox{\currentfont#2{}\currentskewchar}%
        \advance\dimen@ii-\wd\tw@
        \rlap{\raisebox{-#1}{$\m@th#3\kern\dimen@ii\widetilde{\phantom{#2}}$}}%
}
\makeatother

\newcommand{\bm}[1]{\text{\boldmath $#1$\unboldmath}}
\newcommand{\mat}[1]{\mathbf{#1}}
\newcommand{\grad}{\bm{\nabla}}
\newcommand{\RR}{\mathbb{R}}

\newcommand{\eltwo}{\ensuremath{\mathcal{L}_2}}

\newcommand{\bu}{\bm{u}}

\newcommand{\bn}{\bm{n}}
\newcommand{\bx}{\bm{x}}

\newcommand{\bmu}{\bm{\mu}}

\newcommand{\btau}{\bm{\tau}}

\newcommand{\Insd}{\mat{I}_{d}}

\newcommand{\bU}{\bm{U}}
\newcommand{\bUin}{\bm{U}_{\!\! \text{in}}}
\newcommand{\nuT}{\widetilde{\nu}}

\newcommand{\GammaI}{\Gamma_{\!\! \text{in}}}
\newcommand{\GammaO}{\Gamma_{\!\! \text{out}}}
\newcommand{\GammaW}{\Gamma_{\!\! \text{w}}}

\newcommand{\upgd}  {\bU_{_{\!\! \texttt{PGD}}}}
\newcommand{\ppgd}  {P_{_{\!\! \texttt{PGD}}}}
\newcommand{\nuTpgd}  {\nuT_{_{\!\! \texttt{PGD}}}}
\newcommand{\turbNUpgd}  {\nu_{_{\!\! t,\texttt{PGD}}}}
\newcommand{\sigmaU}  {\sigma_U}
\newcommand{\sigmaP}  {\sigma_P}
\newcommand{\sigmaNU}  {\sigma_{\nu}}
\newcommand{\sigmaT}  {\sigma_t}
\newcommand{\fu}{\bm{f}_{\!\! U}}
\newcommand{\fp}{f_{\!\! P}}
\newcommand{\fnu}{f_{\! \nu}}
\newcommand{\ft}{f_{\! t}}
\newcommand{\uBpgd}  {\overline{\bU}_{_{\!\! \texttt{PGD}}}}
\newcommand{\pBpgd}  {\overline{P}_{_{\!\! \texttt{PGD}}}}
\newcommand{\nuBpgd}  {\overline{\nu}_{_{\!\! \texttt{PGD}}}}
\newcommand{\uref}  {\bU_{_{\!\! \texttt{REF}}}}
\newcommand{\pref}  {P_{_{\!\! \texttt{REF}}}}

\newcommand{\de}{\delta}
\newcommand{\I}{\mathcal{I}}
\newcommand{\bI}{\bm{\mathcal{I}}}
\newcommand{\De}{\varDelta}
\newcommand{\Sx}[1]{\mathscr{S}_{_{\!\! \texttt{PGD}}}^{x} \! ( #1 )}
\newcommand{\Smu}[1]{\mathscr{S}_{_{\!\! \texttt{PGD}}}^{\mu} \! ( #1 )}
\newcommand{\A}{\mathcal{A}}
\newcommand{\V}{\mathcal{V}}
\newcommand{\Vmu}{\mathcal{V}_{\mu}}
\newcommand{\Amu}{\mathscr{A}_{\mu}}
\newcommand{\Lmu}{\mathscr{L}_{\mu}}
\newcommand{\vpgd}  {v_{_{\! \texttt{PGD}}}}
\newcommand{\fv}{f_{\! v}}
\newcommand{\sigmaV}  {\sigma_v}

\newenvironment{keywords}{\begin{quote}\emph{\textbf{Keywords:}}}{\end{quote}}

\theoremstyle{definition}

\newtheorem{remark}{Remark}

\RequirePackage{algorithm}[0.1] %Following SIAM cls
\usepackage{algorithmic} %Algorithm style, could alternatively use algpseudocode 

\graphicspath{{figsPDF/}{figsPNG/}{figs/}}

\usepackage{hyphenat}
\begin{document}
%==========================================================================
\title{Parametric solutions of turbulent incompressible flows in OpenFOAM via the proper generalised decomposition}

\author{
\renewcommand{\thefootnote}{\arabic{footnote}}
			  Vasileios Tsiolakis\footnotemark[1]\textsuperscript{ \ ,}\footnotemark[2]\textsuperscript{ \ ,}\footnotemark[3] \ ,
			  Matteo Giacomini\footnotemark[2]\textsuperscript{ \ ,}\footnotemark[4]\textsuperscript{ \ ,}*  ,
			  Ruben Sevilla\footnotemark[3], \\
\renewcommand{\thefootnote}{\arabic{footnote}}
             Carsten Othmer\footnotemark[1] \ and
             Antonio Huerta\footnotemark[2]\textsuperscript{ \ ,}\footnotemark[4]
}

\date{\today}
%________________________________________________________________________
\maketitle

\renewcommand{\thefootnote}{\arabic{footnote}}

\footnotetext[1]{Volkswagen AG, Brieffach 011/1777, D-38436, Wolfsburg, Germany}
\footnotetext[2]{Laboratori de C\`alcul Num\`eric (LaC\`aN), ETS de Ingenieros de Caminos, Canales y Puertos, Universitat Polit\`ecnica de Catalunya, Barcelona, Spain}
\footnotetext[3]{Zienkiewicz Centre for Computational Engineering, College of Engineering, Swansea University, Wales, UK}
\footnotetext[4]{Centre Internacional de M\`etodes Num\`erics en Enginyeria (CIMNE), Barcelona, Spain.
\vspace{5pt}\\
* Corresponding author: Matteo Giacomini. \textit{E-mail:} \texttt{matteo.giacomini@upc.edu}
}

%________________________________________________________________________
\begin{abstract}
An \emph{a priori} reduced order method based on the proper generalised decomposition (PGD) is proposed to compute parametric solutions involving turbulent incompressible flows of interest in an industrial context, using OpenFOAM.
The PGD framework is applied for the first time to the incompressible Navier-Stokes equations in the turbulent regime, to compute a generalised solution for velocity, pressure and turbulent viscosity, explicitly depending on the design parameters of the problem.
In order to simulate flows of industrial interest, a minimally intrusive implementation based on OpenFOAM SIMPLE algorithm applied to the Reynolds-averaged Navier-Stokes equations with the Spalart-Allmaras turbulence model is devised.
The resulting PGD strategy is applied to parametric flow control problems and achieves both qualitative and quantitative agreement with the full order OpenFOAM solution for convection-dominated fully-developed turbulent incompressible flows, with Reynolds number up to one million.
\end{abstract}

%________________________________________________________________________

%==========================================================================
\section{Introduction}
\label{sc:Intro}
%==========================================================================

Parametric studies involving flows of industrial interest require robust computational fluid dynamics (CFD) solvers and efficient strategies to simulate multiple queries of the same problem.

Finite volume (FV) methods represent the most common approach in industry to perform flow simulations~\cite{LeVeque:02,Toro:09,Sonar-MS:07,Ohlberger-BHO:17,Eymard-EGH:00,RS-SGH:18,RS-VGSH:20,MG-RS-20} and different strategies have been proposed to simulate flows in the turbulent regime~\cite{Deardorff-70,Hughes-HMJ-00,Spalart-TSSS-00}. 
A widespread approach is represented by the Reynolds-averaged Navier-Stokes (RANS) equations~\cite{Reynolds-1895} coupled with the one-equation Spalart-Allmaras (SA) turbulence model~\cite{SpalartAllmaras-92}.
This work focuses on such strategy and relies on its cell-centred FV implementation available in OpenFOAM~\cite{OpenFOAM} and validated by the industry.

When the simulation requires testing a large number of different configurations - e.g. for shape optimisation, uncertainty quantification, inverse and control problems - numerical strategies to reduce the cost of the overall computation are critical. Reduced order models (ROM)~\cite{AH-CHRW:17,Gunzburger-PWG-18} construct an approximation of the solution in a lower dimensional space, for which an appropriate basis needs to be devised.

It is known that numerical difficulties arise when reduced basis (RB) and proper orthogonal decomposition (POD) techniques are applied to convection-dominated problems~\cite{Iliescu-IW-13,Pacciarini-PR-14,Iliescu-GIJW-15}.
This is especially critical in the context of flow simulations when the Reynolds number is increased and turbulent phenomena need to be accounted for. 
More precisely, the most relevant POD modes are associated with the highest energy scales of the problem under analysis, whereas small scales, which play a critical role in the dissipation of turbulent kinetic energy, are poorly represented by POD-ROM~\cite{Aubry-AHLS-88}.
To remedy this issue, closure models stemming from traditional description of turbulence have been extended to ROMs, leading to Galerkin projection-based POD with dynamic subgrid-scale~\cite{Iliescu-WABI-12}, variational multiscale~\cite{Iliescu-IW-14,Rozza-SBZR-19}, $k-\omega$ SST~\cite{Rozza-LCLR-16} models and to a certified Smagorinsky RB strategy~\cite{Ballarin-RAMBR-17}.
Moreover, strategies to improve efficiency and accuracy of POD-ROM in the context of realistic and turbulent flows have been proposed by coupling the projection-based framework with residual minimisation~\cite{Farhat-CFCA-13}, nonlinear least-squares optimisation~\cite{Zimmermann-ZG-12}, interpolation based on radial basis functions~\cite{Rozza-GSSRB} and a constrained greedy approach~\cite{Taddei-FMPT-18}.
In the context of machine learning-based reduced order models~\cite{Iliescu-XMRI-18,Hesthaven-GH-19,Hesthaven-WHR-19}, a strategy coupling a traditional projection-based POD for velocity and pressure with a data-driven technique for the eddy viscosity has been recently proposed in~\cite{Rozza-HSMR-20}.

All above contributions involve the development of \emph{a posteriori} ROMs, namely RB and POD, in which the basis of the low-dimensional approximation space is computed starting from a set of snapshots. On the contrary, PGD~\cite{Chinesta-Keunings-Leygue,Chinesta-CLBACGAAH:13} constructs a reduced basis of separable functions explicitly depending on space and on user-defined parameters, with no \emph{a priori} knowledge of the solution of the problem.
The resulting PGD \emph{computational vademecum} provides a generalised solution which can thus be efficiently evaluated in the online phase via interpolation in the parametric space, that is, no extra problem needs to be solved in the low-dimensional reduced space as in POD.
In the context of flow problems, PGD was originally utilised to develop efficient solvers for the incompressible Navier-Stokes equations by separating spatial directions~\cite{Allery-DAA-10} and space and time~\cite{Allery-DAA-11,Allery-LA-14}.
In addition, problems involving parametrised geometries have been solved using PGD~\cite{AH-AHCCL:14,SZ-ZDMH:15}, with special emphasis on incompressible flows in geometrically parametrised domains~\cite{PD-DZH:17,RS-SZH:20,RS-SBGH:20,MG-GBSH-20}.
To foster the application of \emph{a priori} model order reduction techniques to problems of industrial interest, a non-intrusive PGD implementation in the CFD software OpenFOAM has been recently proposed in~\cite{Tsiolakis-TGSOH-20} to solve parametrised incompressible Navier-Stokes equations in the laminar regime.

Following the work on PGD for convection phenomena~\cite{2010-IJMF-GDCCDH,Gonzalez-GCCDH-13} and for viscous incompressible Navier-Stokes flows~\cite{Tsiolakis-TGSOH-20}, the present contribution proposes the first \emph{a priori} ROM for turbulent incompressible flows.
Similarly to the previously cited contributions on \emph{a posteriori} ROMs~\cite{Iliescu-WABI-12,Iliescu-IW-14,Rozza-SBZR-19,Rozza-LCLR-16,Ballarin-RAMBR-17,Farhat-CFCA-13,Zimmermann-ZG-12,Rozza-GSSRB,Taddei-FMPT-18,Iliescu-XMRI-18,Hesthaven-GH-19,Hesthaven-WHR-19,Rozza-HSMR-20}, \emph{a priori} ROMs can also be devised using a variety of techniques to model turbulence. 
The proposed PGD strategy considers the one-equation SA turbulence model and it constructs a separated representation of velocity, pressure and eddy viscosity to solve the RANS-SA equations in OpenFOAM. More precisely, the PGD-ROM methodology mimics the structure of the \texttt{simpleFoam} algorithm with SA turbulence model, resulting in a minimally intrusive approach within OpenFOAM. The resulting strategy thus provides a generalised expression of the velocity, pressure and eddy viscosity fields, explicitly depending on user-defined parameters, for convection-dominated incompressible flows.
Alternative PGD-ROM strategies may thus be obtained by substituting the proposed separated form of the SA equation with an appropriate PGD solver for the selected turbulence model.

The remainder of this paper is organised as follows. Section~\ref{sc:RANS-SA} recalls the full order RANS-SA equations and the corresponding cell-centred FV approximation utilised by OpenFOAM. The rationale of the PGD-ROM for the turbulent incompressible Navier-Stokes equations is introduced in section~\ref{sc:Par-RANS-SA}, where the details of the algorithms to devise the separated velocity-pressure approximation of the flow equations (\texttt{PGD-NS}) and the separated form of the eddy (\texttt{PGD-SA}) and turbulent (\texttt{PGD-$\nu_t$}) viscosities via the SA equation are presented. Numerical experiments involving flow control in external aerodynamics, in two and three dimensions, with Reynolds number ranging up to $1,000,000$ are reported in section~\ref{sc:simulations}. 
Finally, section~\ref{sc:Conclusion} summarises the contributions of this work and two appendices report additional technical details on the employed PGD algorithm and on the expressions of the coefficients appearing in the spatial and parametric iterations of the alternating direction scheme for the PGD solvers of the RANS and the SA equations.

%==========================================================================
\section{The Reynolds-averaged Navier-Stokes equations and the Spalart-Allmaras turbulence model}
\label{sc:RANS-SA}
%==========================================================================

To simulate turbulent incompressible flows using the RANS equations, the velocity-pressure pair $(\bu,p)$ is decomposed into a mean flow component $(\bU,P)$ and a perturbation $(\bu',p')$, that is $\bu {=} \bU {+} \bu'$ and $p {=} P {+} p'$.
Given an open bounded computational domain $\Omega \subset \mathbb{R}^{d}$ in $d$ spatial dimensions, the boundary $\partial\Omega$ is partitioned such that $\partial\Omega {=} \GammaI \cup \GammaW \cup \GammaO$, where the three disjoint portions $\GammaI$, $\GammaW$ and $\GammaO$ denote inlet surfaces, material walls and outlet surfaces, respectively.
%Moreover, $\Tend>0$ denotes the final time of interest. 
%
The steady-state RANS equations for the mean flow variables $(\bU,P)$ are given by
%
%\begin{equation} \label{eq:RANS}
%\left\{\begin{aligned}
%\begin{aligned}[b]
%\frac{\partial \bU}{\partial t} +& \grad {\cdot} (\bU {\otimes} \bU) \\
%&- \grad {\cdot} ((\nu {+} \nu_t)  \grad \bU) + \grad P 
%\end{aligned}
%&= \bm{0}       &&\text{in $\Omega \times (0,\Tend]$,}\\
%\grad {\cdot} \bU &= 0  &&\text{in $\Omega \times (0,\Tend]$,}\\
%\bU &= \bU^0  &&\text{in $\Omega \times \{0\}$,}\\
%\bU &= \bm{0}  &&\text{on $\GammaW \times (0,\Tend]$,}\\
%\bU &= \bUin  &&\text{on $\GammaI \times (0,\Tend]$,}\\
%(\nu \grad \bU {-} p \Insd ) \bn &= \bm{0}  &&\text{on $\GammaO \times (0,\Tend]$,}
%\end{aligned}\right.
%\end{equation}
%
\begin{equation} \label{eq:RANS}
\left\{\begin{aligned}
\grad {\cdot} (\bU {\otimes} \bU) - \grad {\cdot} ((\nu {+} \nu_t)  \grad \bU) + \grad P &= \bm{0}       &&\text{in $\Omega$,}\\
\grad {\cdot} \bU &= 0  &&\text{in $\Omega$,}\\
\bU &= \bUin  &&\text{on $\GammaI$,}\\
\bU &= \bm{0}  &&\text{on $\GammaW$,}\\
(\nu \grad \bU {-} p \Insd ) \bn &= \bm{0}  &&\text{on $\GammaO$,}
\end{aligned}\right.
\end{equation}
where $\Insd$ denotes the $d {\times} d$ identity matrix, $\nu$ represents the physical viscosity of the fluid and $\nu_t$ is the turbulent viscosity introduced in the momentum equation to model the perturbations to the mean flow due to turbulence. The boundary conditions for the flow equations~\eqref{eq:RANS} impose the velocity profile $\bUin$ on the inlet surface $\GammaI$, no-slip Dirichlet data on fixed material walls $\GammaW$ and homogeneous Neumann data on the outlet surface $\GammaO$.

In order to describe the turbulent viscosity $\nu_t$, the one-equation SA turbulence model~\cite{SpalartAllmaras-92} introduces the relation 
\begin{equation}\label{eq:nuT}
\nu_t = \nuT f_{v1} ,
\end{equation}
where $\nuT$ is the eddy viscosity and $f_{v1}$ is an appropriately defined spatial function, see e.g.~\cite{Mellor-Herring-68,Cebeci-Smith-70,SpalartAllmaras-92}, reported in equation~\eqref{eq:functionsSA}.  
Under the assumption of fully turbulent flows, the trip term controlling the transition between laminar and turbulent regimes in the SA model is neglected and the eddy viscosity $\nuT$ is obtained as the solution of
%
%\begin{equation}\label{eq:SA}
%\left\{\begin{aligned}
%\begin{aligned}[b]
%\frac{\partial \nuT}{\partial t} + \grad {\cdot} (\bU \nuT) - \frac{1}{\sigma} \grad {\cdot} & ( (\nu {+} \nuT) \grad \nuT ) \\
%&- \frac{c_{b2}}{\sigma} \grad \nuT {\cdot} \grad \nuT
%\end{aligned}
% &= s       &&\text{in $\Omega \times (0,\Tend]$,}\\
%\nuT &= \nuT^0  &&\text{in $\Omega \times \{0\}$,}\\
%\nuT &= \nuT_D  &&\text{on $\GammaI \cup \GammaW \times (0,\Tend]$,}\\
%\grad \nuT {\cdot} \bn &= 0  &&\text{on $\GammaO \times (0,\Tend]$,}
%\end{aligned}\right.
%\end{equation}
%
\begin{equation}\label{eq:SA}
\left\{\begin{aligned}
\grad {\cdot} (\bU \nuT) - \frac{1}{\sigma} \grad {\cdot} ( (\nu {+} \nuT) \grad \nuT ) -& \frac{c_{b2}}{\sigma} \grad \nuT {\cdot} \grad \nuT  &&\\
&= c_{b1} \widetilde{S} \nuT - c_{w1} \frac{f_w}{\widetilde{d}^2} \nuT^2 &&\text{in $\Omega$,}\\
\nuT &= \nuT_D  &&\text{on $\GammaI \cup \GammaW$,}\\
\grad \nuT {\cdot} \bn &= 0  &&\text{on $\GammaO$,}
\end{aligned}\right.
\end{equation}
where $\nuT_D$ is the eddy viscosity Dirichlet datum, $\widetilde{d}$ represents the distance of a given point in the domain from the closest physical wall and $c_{b1}$, $c_{b2}$, $c_{w1}$ and $\sigma$ are four scalar constants. Moreover, $\widetilde{S}$ and $f_w$ are the spatial functions associated with the production and the destruction of eddy viscosity, respectively, whereas the operators on the left-hand side of equation~\eqref{eq:SA} model convection, diffusion and cross-diffusion phenomena~\cite{SpalartAllmaras-92}.

The SA turbulence model, derived by means of dimensional analysis and empirical observations~\cite{SpalartAllmaras-92}, is thus closed by introducing the definition of the quantities
\begin{equation}\label{eq:functionsSA}
\hspace{-7pt}
\begin{alignedat}{5}
\bm{\omega} &:= \frac{\grad \bU - \grad \bU^T}{2} , \; && \widetilde{S} &&:= \left[ 2 \langle \bm{\omega} , \bm{\omega} \rangle_F \right]^{1/2} + \frac{\nuT}{\kappa^2 \widetilde{d}^2} f_{v2} , \; && \chi &&:= \frac{\nuT}{\nu} , \\
f_w &:= g \left[ \frac{1 + c_{w3}^6}{g^6 + c_{w3}^6} \right]^{1/6} ,  \; && f_{v2} &&:= 1 - \frac{\chi}{1+ \chi f_{v1}} , \; && f_{v1} &&:= \frac{\chi^3}{\chi^3 + c_{v1}^3} , \\
c_{w1} &:= \frac{c_{b1}}{\kappa^2} + \frac{1 + c_{b2}}{\sigma} , \; && g &&:= r + c_{w2}(r^6 -r) , \; && r &&:= \frac{\nuT}{\widetilde{S} \kappa^2 \widetilde{d}^2} ,
\end{alignedat}
\end{equation}
where $\langle \cdot , \cdot \rangle_F$ denotes the Frobenius inner product and the scalar constants $\sigma {=} 2/3$, $\kappa {=} 0.41$, $c_{b1} {=} 0.1355$, $c_{b2} {=} 0.622$, $c_{v1} {=} 7.1$, $c_{w2} {=} 0.3$ and $c_{w3} {=} 2$ are selected~\cite{SpalartAllmaras-92}.
%
%Following the work in~\cite{Tsiolakis-TGSOH-20}, this paper focuses on the solution of parametric problems involving steady state turbulent flows. Henceforth, the time dependence in equations~\eqref{eq:RANS} and~\eqref{eq:SA} is thus omitted.

%==========================================================================
\subsection{A finite volume formulation of the RANS-SA equations}
\label{sc:CCFV-NS-SA}
%==========================================================================

In order to discretise the turbulent Navier-Stokes equations, OpenFOAM cell-centred finite volume rationale is considered~\cite{OpenFOAM}. The computational domain is subdivided in $N$ cells $V_i, \, i{=}1,\ldots,N$ such that $V_i {\cap} V_j {=} \emptyset, \text{for $i {\neq} j$}$ and $\Omega {=} \bigcup_{i=1}^{N} V_i$.
In each cell $V_i$, the integral form of equation~\eqref{eq:RANS} is defined as 
\begin{equation} \label{eq:weak-RANS}
\!\left\{\begin{aligned}
\int_{V_i}{\!\! \grad {\cdot} (\bU {\otimes} \bU) \, dV} 
- \int_{V_i}{\!\!  \grad {\cdot} ((\nu {+} \nu_t)  \grad \bU) \, dV} 
+ \int_{V_i}{\!\! \grad P \, dV} 
&= \bm{0} , \\
\int_{V_i}{\!\! \grad {\cdot} \bU \, dV} &= 0 ,
\end{aligned}\right.
\end{equation}
where $(\bU,P)$ are cell-by-cell constant approximations of the velocity and pressure fields, respectively, and $\bU {=} \bUin \, \text{on} \, \GammaI$ and $\bU {=} \bm{0} \, \text{on} \, \GammaW$.

In a similar fashion, the cell-centred finite volume approximation of the SA equation~\eqref{eq:SA} is: compute $\nuT$ constant in each cell such that $\nuT = \nuT_D \, \text{on} \, \GammaI \cup \GammaW$ and it holds
\begin{equation} \label{eq:weak-SA}
\begin{aligned}
\int_{V_i}\!\! \grad {\cdot} (\bU \nuT) & \, dV
- \frac{1}{\sigma} \int_{V_i}{\!\!  \grad {\cdot} \left( (\nu {+} \nuT) \grad \nuT \right) \, dV} \\
&- \frac{c_{b2}}{\sigma} \int_{V_i}{\!\!  \grad \nuT {\cdot} \grad \nuT \, dV} 
- c_{b1} \int_{V_i}{\!\! \widetilde{S} \nuT \, dV} 
+ c_{w1}  \int_{V_i}{\!\! \frac{f_w}{\widetilde{d}^2} \nuT^2 \, dV}  = 0.
\end{aligned}
\end{equation}

%==========================================================================
\subsection{A turbulent Navier-Stokes solver in OpenFOAM}
\label{sc:OpenFOAM-NS-SA}
%==========================================================================

OpenFOAM strategy to solve the RANS equation with SA turbulence model relies on a staggered approach. 
First, the flow equations~\eqref{eq:weak-RANS} are solved using a seed value of $\nu_t$. More precisely, the integrals over each cell in~\eqref{eq:weak-RANS} are approximated by means of the corresponding fluxes across the boundaries of the cell~\cite{Ohlberger-BHO:17,Eymard-EGH:00}. In addition, the semi-implicit method for pressure linked equations (SIMPLE) algorithm~\cite{Patankar-PS:72}, that is, a fractional-step Chorin-Temam projection method~\cite[Sect. 6.7]{Donea-Huerta}, is utilised to handle incompressibility. Also, a relaxation approach is employed for the nonlinear convection term.
Second, the velocity field $\bU$ obtained using \texttt{simpleFoam} is employed to compute the quantities in~\eqref{eq:functionsSA} and to solve the SA equation~\eqref{eq:weak-SA}. It is worth noting that equation~\eqref{eq:weak-SA} is highly nonlinear and a relaxation strategy is also required in OpenFOAM to improve the convergence of the numerical algorithm~\cite{OpenFOAM}.
Finally, the updated value of the turbulent viscosity $\nu_t$ is determined according to equation~\eqref{eq:nuT} and the \texttt{simpleFoam} routine is utilised to recompute the turbulent velocity and pressure fields.

%==========================================================================
\section{Proper generalised decomposition for parametric turbulent flow problems}
\label{sc:Par-RANS-SA}
%==========================================================================

In the context of parametric studies, viscosity coefficient, reference velocity or boundary conditions of the problems may depend on a set of $M$ user-defined parameters $\bmu {=} (\mu_1,\ldots,\mu_M)^T$.
The solution of the RANS-SA equations is thus denoted by the velocity-pressure pair $(\bU(\bx,\bmu),P(\bx,\bmu))$ and the eddy viscosity $\nuT(\bx,\bmu)$, which are now functions of the spatial, $\bx \in \Omega \subset \RR^d$, and parametric, $\bmu\in\bI\subset\mathbb{R}^{M}$, variables. 
More precisely, $\bU(\bx,\bmu),P(\bx,\bmu)$ and $\nuT(\bx,\bmu)$ fulfil the high-dimensional RANS-SA equations obtained by integrating~\eqref{eq:weak-RANS}-\eqref{eq:weak-SA} in the parametric space $\bI$.

The PGD-ROM strategy described in this section relies on the construction of a separated approximation $(\upgd^n, \ppgd^n)$ of the velocity and pressure fields and a separated representation of any additional variable introduced by the employed turbulence model. 
The computation of the former is performed via a PGD solver for the incompressible Navier-Stokes equations, see section~\ref{sc:PGD-RANS}, whereas for the latter a separated formulation of the involved turbulence equations is required.

Considering the one-equation SA turbulence model introduced in section~\ref{sc:RANS-SA}, a separated representation $\nuTpgd^m$ of the eddy viscosity is obtained from a PGD solver of the SA equation as described in section~\ref{sc:PGD-SA}. In addition, a separated representation $\turbNUpgd^q$ of the turbulent viscosity is devised in section~\ref{sc:PGD-nuT} exploiting its relation with the eddy viscosity.
It is worth noticing that this framework is general and could be adapted to other descriptions of turbulence by devising separated representations of the involved variables via appropriately defined PGD solvers of the equations in the turbulence model.

The global set of variables involved in the PGD approximation of the RANS-SA equations is thus $(\upgd^n, \ppgd^n)$, $\nuTpgd^m$ and $\turbNUpgd^q$. As classical in PGD~\cite{Chinesta-Keunings-Leygue}, each variable is constructed as a sum of separable modes, each being the product of functions that depend on either the spatial or one of the parametric variables $\mu_j, \ j=1,\ldots,M$. For the sake of simplicity, only space, $\bx$, and parameters, $\bmu$, are henceforth separated.
It is worth noticing that the final number of modes needed for the PGD approximations, denoted by the super-indexes $n$ for the velocity and the pressure, $m$ for the eddy viscosity and $q$ for the turbulent viscosity, is not known \emph{a priori} and, in general, it is different for each of the involved variables.
More precisely, the number of terms in the PGD expansion is automatically determined by the algorithm which stops the enrichment procedure when a user-defined stopping criterion is fulfilled~\cite{Chinesta-Keunings-Leygue}. Classical definitions of this stopping criterion include the relative amplitude of the last computed mode with respect to the first one or to the sum of all previously computed terms~\cite{Tsiolakis-TGSOH-20}.

%==========================================================================
\subsection{Separated representation of the flow and the turbulent variables}
\label{sc:PGD}
%==========================================================================

First, the rank-$n$ separated representation $(\upgd^n, \ppgd^n)$ of the flow variables is introduced. Following~\cite{Tsiolakis-TGSOH-20}, the computation of each PGD mode is split into a \emph{prediction} and a \emph{correction} step. More precisely, the PGD approximation for the flow variables is defined as 
\begin{subequations}\label{eq:sep-increment}
\begin{equation}\label{eq:sep-increment-up}
\left\{\begin{aligned}
 \upgd^n(\bx,\bmu)   &= 
   \upgd^{n-1}(\bx,\bmu) + \sigmaU^n \left[ \fu^n(\bx)\phi^n(\bmu) + \De( \fu^n(\bx)\phi^n(\bmu) ) \right] , \\
 \ppgd^n(\bx,\bmu)   &= 
  \ppgd^{n-1}(\bx,\bmu) +\sigmaP^n \left[ \fp^n(\bx)\phi^n(\bmu) + \De( \fp^n(\bx)\phi^n(\bmu) ) \right] ,
 \end{aligned}\right.
\end{equation}
where $\upgd^{n-1}$ and $\ppgd^{n-1}$ feature the contributions of the previous $n{-}1$ PGD modes, $\sigmaU^n\fu^n\phi^n$ and $\sigmaP^n\fp^n\phi^n$ represent the predictions of the $n$-th mode and $\sigmaU^n \De( \fu^n \phi^n )$ and $\sigmaP^n \De( \fp^n \phi^n )$ are the corresponding correction terms. The coefficients $\sigmaU^n$ and $\sigmaP^n$ denote the amplitudes of the $n$-th velocity and pressure mode, respectively. 
\\
Similarly, the rank-$m$ PGD separated form of the eddy viscosity is given by
\begin{equation}\label{eq:sep-increment-nu}
 \nuTpgd^m(\bx,\bmu)  = 
  \nuTpgd^{m-1}(\bx,\bmu) +\sigmaNU^m \left[ \fnu^m(\bx)\psi^m(\bmu) + \De( \fnu^m(\bx)\psi^m(\bmu) ) \right] ,
\end{equation}
with $\nuTpgd^{m-1}$ containing the previous $m{-}1$ terms in the PGD approximation, $\sigmaNU^m\fnu^m\psi^m$ and $\sigmaNU^m \De( \fnu^m \psi^m )$ being the prediction and the correction of the $m$-th mode, respectively, and $\sigmaNU^m$ its amplitude.

Both the PGD approximations of the flow variables~\eqref{eq:sep-increment-up} and of the eddy viscosity~\eqref{eq:sep-increment-nu} are devised solving appropriate separated forms of the corresponding equations, as presented in the following sections. More precisely, $\upgd^n$ and $\ppgd^n$ are obtained from a separated form of the Navier-Stokes equations~\eqref{eq:weak-RANS}, see section~\ref{sc:PGD-RANS}, whereas $\nuTpgd^m$ is determined via a PGD projection of the SA equation~\eqref{eq:weak-SA}, as reported in section~\ref{sc:PGD-SA}.

Finally, a separated representation of the turbulent viscosity is obtained from the relation~\eqref{eq:nuT}, leading to the rank-$q$ PGD expansion
\begin{equation}\label{eq:sep-turbNu}
 \turbNUpgd^q(\bx,\bmu)  =  \turbNUpgd^{q-1}(\bx,\bmu) +\sigmaT^q\ft^q(\bx)\xi^q(\bmu) ,
\end{equation}
\end{subequations}
where $\turbNUpgd^{q-1}$ represents the PGD approximation obtained from the previous $q {-} 1$ terms, $\ft^q$ and $\xi^q$ denote the $q$-th normalised spatial and parametric modes, respectively, and $\sigmaT^q$ its corresponding amplitude. 
It is worth noticing that, contrary to the flow variables and the eddy viscosity, the computation of the turbulent viscosity $\turbNUpgd^q$ does not involve the solution of a differential equation and only requires elementary arithmetic operations, whence the predictor-corrector approach is substituted by the classical PGD separation in equation~\eqref{eq:sep-turbNu}.

\begin{remark}\label{rmrk:equalParamFun}
Following~\cite{PD-DZH:17}, the same scalar parametric function $\phi(\bmu)$ is selected for both velocity and pressure. On the contrary, different scalar functions $\psi(\bmu)$ and $\xi(\bmu)$ are considered for the separated approximations of the eddy and turbulent viscosity, respectively.
\end{remark}

The corrections of the PGD modes introduced in equation~\eqref{eq:sep-increment-up} and~\eqref{eq:sep-increment-nu} for the computation of the current mode feature variations $\De$ in the spatial and in the parametric functions, namely
%
%\begin{equation}\label{eq:PGD-corr}
%\left\{\begin{aligned}
%\De( \fu^n(\bx)\phi^n(\bmu) ) &:= \De\fu(\bx)\phi^n(\bmu)+\fu^n(\bx)\De\phi(\bmu)+\De\fu(\bx)\De\phi(\bmu) , \\
%\De( \fp^n(\bx)\phi^n(\bmu) ) &:= \De\fp(\bx)\phi^n(\bmu)+\fp^n(\bx)\De\phi(\bmu)+\De\fp(\bx)\De\phi(\bmu) , \\
%\De( \fnu^m(\bx)\psi^m(\bmu) ) &:= \De\fnu(\bx)\psi^m(\bmu)+\fnu^m(\bx)\De\psi(\bmu)+\De\fnu(\bx)\De\psi(\bmu) .
%\end{aligned}\right.
%\end{equation}
%
\begin{equation}\label{eq:PGD-corr}
\left\{\begin{aligned}
\De( \fu^n(\bx)\phi^n(\bmu) ) &:= \De\fu(\bx)\phi^n(\bmu)+\fu^n(\bx)\De\phi(\bmu)  , \\
\De( \fp^n(\bx)\phi^n(\bmu) ) &:= \De\fp(\bx)\phi^n(\bmu)+\fp^n(\bx)\De\phi(\bmu)  , \\
\De( \fnu^m(\bx)\psi^m(\bmu) ) &:= \De\fnu(\bx)\psi^m(\bmu)+\fnu^m(\bx)\De\psi(\bmu)  .
\end{aligned}\right.
\end{equation}
Of course, higher order contributions (e.g., $\De\fu(\bx)\De\phi(\bmu)$ for the velocity) could also be considered in the definition of the corrections~\eqref{eq:PGD-corr} of the PGD modes. Nonetheless, the importance of these extra terms is negligigle with respect to both the predictions $\fu^n$, $\fp^n$, $\phi^n$, $\fnu^m$ and $\psi^m(\bmu)$ of the modes and the first-order corrections introduced in equation~\eqref{eq:PGD-corr} and they are thus neglected henceforth. Finally, for the sake of readability, the following compact expressions are introduced to define the corrections of the velocity, pressure and eddy viscosity to be computed by the PGD algorithm
\begin{equation}\label{eq:PGD-corr-computed}
\left\{\begin{aligned}
\de\upgd^n &:= \De\fu \phi^n +\fu^n \De\phi , \\
\de\ppgd^n &:= \De\fp \phi^n +\fp^n \De\phi  , \\
\de\nuTpgd^m &:= \De\fnu \psi^m +\fnu^m \De\psi ,
\end{aligned}\right.
\end{equation}
where the dependence of the spatial and parametric modes on $\bx$ and $\bmu$ has been omitted.
A comparison of the classical PGD algorithm, see e.g.~\cite{Chinesta-Keunings-Leygue}, and the \emph{predictor-corrector} strategy employed in this work is reported in~\ref{sc:appPredCorr}.

\begin{remark}\label{rmrk:amplitude}
Upon convergence of the alternating direction algorithm, the above mentioned predictions and corrections are combined  into  the $n$-th spatial and parametric modes denoted by $\fu^n$ and $\phi^n$, with $\| \fu^n \| {=}\| \phi^n \| {=} 1$. In this context, the  coefficient $\sigmaU^n$ encapsulates the information on the amplitude of the mode.
From a practical viewpoint, the solution of the alternating direction method is the pair $(\tilde{\fu}^{\!\! n},\phi^n)$, $\tilde{\fu}^{\!\! n}$ being the spatial mode before the normalisation procedure. The corresponding normalised spatial mode is thus given by $\fu^n {=} \tilde{\fu}^{\!\! n}/\sigmaU^n$, with $\sigmaU^n {:=} \| \tilde{\fu}^{\!\! n} \|$.
The details of  the normalisation procedure for the classical  and the \emph{predictor-corrector} PGD strategies are presented in~\ref{sc:appPredCorr}.
For the simulations in section~\ref{sc:simulations}, the normalisation procedure has been performed using the $\eltwo$ norm.
\end{remark}

%==========================================================================
\subsection{A minimally intrusive PGD implementation of a parametric solver for turbulent incompressible Navier-Stokes flows in OpenFOAM}
\label{sc:PGD-RANS-SA}
%==========================================================================

The proposed minimally intrusive parametric solver for turbulent incompressible Navier-Stokes flows in OpenFOAM constructs the separated expressions $\upgd^n$, $\ppgd^n$, $\nuTpgd^m$ and $\turbNUpgd^q$ exploiting the numerical discretisation techniques natively implemented in OpenFOAM and validated by the CFD community. For this purpose, the PGD algorithm mimics the segregated structure of \texttt{simpleFoam} with SA turbulence model which involves the following steps~\cite{OpenFOAM}:
\begin{enumerate}[label=(\Alph*)]
\item\label{step1} Compute velocity and pressure using a fractional step projection approach to solve~\eqref{eq:weak-RANS}, given the current approximation of the turbulent viscosity (RANS solver via \texttt{simpleFoam}).
\item Use the value of the computed velocity to solve~\eqref{eq:weak-SA} and determine the eddy viscosity (SA solver).
\item Update the turbulent viscosity according to~\eqref{eq:nuT} and go to step~\ref{step1} to restart the computation.
\end{enumerate}

\begin{algorithm}
\caption{An OpenFOAM implementation of turbulent \texttt{pgdFoam}}\label{alg:PGD-RANS-SA-OF}
\begin{algorithmic}[1]
\REQUIRE{Stopping criteria $\eta_\diamond^\star$ ($\diamond=U,P$) and $\eta_\nu^\star$ for the PGD enrichment of the flow equations and the turbulence model. Initial accuracy level $\gamma$ for the PGD enrichment of the turbulent viscosity.}
\STATE{Compute boundary condition modes: the spatial mode is solution of~\eqref{eq:weak-RANS} using \texttt{simpleFoam} with SA turbulence model and the parametric modes are set equal to $1$.}
\STATE{Set $n \gets 1$, $i \gets 0$ and initialise the amplitudes $\sigma_\diamond^1 \gets 1$.}
%___Enrichment for velocity and pressure
\WHILE{$\sigma_\diamond^n > \eta_\diamond^\star\,\sigma_\diamond^1$}
\STATE{Set the enrichment threshold $\eta_t^i$ for the turbulent viscosity according to equation~\eqref{eq:criterionTurbUpdate}.}
\STATE{Call \texttt{PGD-NS} to compute the spatial, $(\fu^n,\fp^n)$, and parametric, $\phi^n$, modes of velocity and pressure and the amplitudes $\sigmaU^n$ and $\sigmaP^n$.}
%___PGD turbulent viscosity
\IF{$\sigma_\diamond^n < \eta_t^i$}
\STATE{Set $m \gets 1$ and initialise the amplitude $\sigma_\nu^1 \gets 1$.}
%___Enrichment for eddy viscosity
\WHILE{$\sigma_\nu^m > \eta_\nu^\star\,\sigma_\nu^1$}
\STATE{Call \texttt{PGD-SA} to compute the spatial, $\fnu^m$, and parametric, $\psi^m$, modes of the eddy viscosity and the amplitude $\sigmaNU^m$.}
\STATE{Update the mode counter $m \gets m+1$.}
\ENDWHILE
%
%___Update for turbulent viscosity
\STATE{Call \texttt{PGD-$\nu_t$} to compute the spatial, $\ft^q$, and parametric, $\xi^q$, modes of the turbulent viscosity and the amplitude $\sigmaT^q$.}
\STATE{Increment viscosity update counter $i \gets i+1$.}
\STATE{Reinitialise the mode counter $n \gets 1$. Reset $\upgd^n$ and $\ppgd^n$.}
\ELSE
\STATE{Update the mode counter $n \gets n+1$.}
\ENDIF
\ENDWHILE
\end{algorithmic}
\end{algorithm}

Following this rationale, the corresponding parametric solver is described in algorithm~\ref{alg:PGD-RANS-SA-OF}.
To this end, three routines are introduced to compute $\upgd^n$, $\ppgd^n$, $\nuTpgd^m$ and $\turbNUpgd^q$:
\begin{itemize}
\item \texttt{PGD-NS} solves a separated form of the incompressible Navier-Stokes equations to determine the velocity-pressure pair $(\upgd^n,\ppgd^n)$, given the current PGD expression $\turbNUpgd^q$ of the turbulent viscosity (Section~\ref{sc:PGD-RANS});
\item \texttt{PGD-SA} solves a separated form of the SA turbulence model to determine the eddy viscosity $\nuTpgd^m$ (Section~\ref{sc:PGD-SA});
\item \texttt{PGD-$\nu_t$} computes the updated approximation $\turbNUpgd^q$ of the turbulent viscosity by means of a separated expression of its relation with the eddy viscosity (Section~\ref{sc:PGD-nuT}) and go to \texttt{PGD-NS} to restart the computation.
\end{itemize}
The minimally intrusive PGD solver in OpenFOAM is thus obtained by integrating the \texttt{PGD-NS}, \texttt{PGD-SA} and \texttt{PGD-$\nu_t$} routines in the following computational framework.
More precisely, in step (A), the PGD algorithm solves the parametrised flow equations (Algorithm 1 - Line 5) via the non-intrusive \texttt{PGD-NS} strategy, given the current PGD approximation of the turbulent viscosity. The PGD enrichment procedure for the RANS equations continues by alternately solving the spatial and parametric problems until a user-prescribed threshold is achieved by the amplitude of the computed velocity and pressure modes. 
Once a \emph{sufficiently accurate} PGD approximation of velocity and pressure is obtained, step (B) computes a separated representation of the eddy viscosity by means of the minimally intrusive \texttt{PGD-SA} block (Algorithm 1 - Line 9). In step (C), the separated representation of the turbulent viscosity is recomputed by means of the \texttt{PGD-$\nu_t$} routine (Algorithm 1 - Line 12).
Finally, the PGD approximation of the flow equations is reset (Algorithm 1 - Line 14) and step (A) restarts the PGD computation of the flow variables using the newly computed separated expression of the turbulent viscosity.

\begin{remark}\label{rmrk:turbTolerance}
The overall cost of the PGD solver for the parametric turbulent Navier-Stokes equations depends on the number of updates performed to correct the PGD expression of the turbulent viscosity (Algorithm 1 - Lines from 6 to 15). 
Indeed, the computation of the velocity and pressure modes is restarted from scratch after each update of the turbulent viscosity (Algorithm 1 - Line 14). In this context, computing a large number of modes for velocity and pressure using early approximations of the turbulent viscosity might significantly increase the computational effort of the algorithm with limited accuracy gain. To remedy this issue, effective numerical strategies are devised by introducing an appropriately defined criterion that limits the number of velocity and pressure modes determined in the early stages of the algorithm while allowing a larger number of modes to be computed when the precision of the approximation of the turbulent viscosity increases.
More precisely, the PGD enrichment of velocity and pressure for a given expression of the turbulent viscosity stops when the amplitude of the computed modes drops below a user-defined tolerance $\eta_t^i$ (Algorithm 1 - Line 6). In addition, this threshold is gradually reduced after each update of the turbulent viscosity (Algorithm 1 - Line 4) allowing to increase the accuracy of the velocity and pressure modes computed by means of the \texttt{PGD-NS} routine, when the \texttt{PGD-$\nu_t$} algorithm provides an improved PGD representation of the turbulent viscosity.
In the simulations in section~\ref{sc:simulations}, the threshold to control the number of velocity and pressure modes computed by the \texttt{PGD-NS} algorithm at the $i$-th iteration is given by
\begin{equation}\label{eq:criterionTurbUpdate}
\eta_t^i = 10^{-(i+\gamma)} ,
\end{equation}
that is, starting from an initial accuracy of $10^{-\gamma}$, an exponentially decreasing tolerance is defined after each update of the turbulent viscosity.
An alternative approach to control the accuracy of the separated representation of eddy and turbulent viscosities may be devised modifying line 6 of algorithm 1 and fixing \emph{a priori} the number of modes in the PGD approximation of the velocity field required to run \texttt{PGD-SA} and $\texttt{PGD-$\nu_t$}$ routines.
\end{remark}

In the following sections, the structure of the \texttt{PGD-NS}, \texttt{PGD-SA} and \texttt{PGD-$\nu_t$} routines for the computation of  $(\upgd^n,\ppgd^n)$, $\nuTpgd^m$ and $\turbNUpgd^q$, respectively, will be detailed.

%==========================================================================
\subsection{Proper generalised decomposition of the flow equations}
\label{sc:PGD-RANS}
%==========================================================================

In this section, the spatial and parametric steps of the non-intrusive PGD algorithm applied to the turbulent Navier-Stokes equations~\eqref{eq:weak-RANS} are presented.
The integral form of the steady parametrised Navier-Stokes equations in the high-dimensional space $\Omega \times \bI$ is given by
\begin{equation} \label{eq:weak-NS-PGD}
\left\{\begin{aligned}
\begin{aligned}[b]
\int_{\bI}\int_{V_i}\!\! & \grad {\cdot} (\bU {\otimes} \bU) \, dV \, d\bI \\
&- \int_{\bI}\int_{V_i} \!\! \grad {\cdot} ((\nu {+} \nu_t)  \grad \bU) \, dV \, d\bI 
+\int_{\bI}\int_{V_i}{\!\! \grad P  \, dV \, d\bI} 
\end{aligned}
&= \bm{0} , \\
\int_{\bI}\int_{V_i}{\!\! \grad {\cdot} \bU \, dV \, d\bI} &= 0 ,
\end{aligned}\right.
\end{equation}
for each cell $V_i, \, i{=}1,\ldots,N$ of the spatial computational domain.

The PGD solver for the incompressible Navier-Stokes equations is obtained by replacing the separated approximations~\eqref{eq:sep-increment-up} into equation~\eqref{eq:weak-NS-PGD}. The first mode $(\upgd^0, \ppgd^0)$ is selected to verify the boundary condition on the inlet surface and, more generally, all inhomogeneous Dirichlet boundary data. It is worth recalling that, as classic in PGD approximations~\cite{Chinesta-Keunings-Leygue}, the first mode only prescribes the Dirichlet boundary conditions and it does not fulfill the equation under analysis. Then, the following terms of the PGD expansion of velocity and pressure are computed following the greedy strategy described in~\cite{Tsiolakis-TGSOH-20}, henceforth named \texttt{PGD-NS}. To compute the $n$-th modes $\sigmaU^n \left[ \fu^n\phi^n + \de\upgd^n  \right]$ and $\sigmaP^n \left[ \fp^n \phi^n + \de\ppgd^n \right]$, the high-dimensional problem is thus alternatively restricted to the tangent manifold associated with either the spatial or the parametric coordinate. More precisely, a first iteration of the alternating direction scheme is performed to determine the predictions $(\sigmaU^n\fu^n\phi^n,\sigmaP^n\fp^n\phi^n)$ using the classical PGD algorithm~\cite{PGD-CCH:14,Chinesta-Keunings-Leygue}, whereas the following iterations are devoted to compute the separated expressions of the corrections $(\sigmaU^n\de\upgd^n,\sigmaP^n\de\ppgd^n)$~\cite{Tsiolakis-TGSOH-20}.

\begin{remark}
The segregated algorithm employed by OpenFOAM to simulate turbulent flows requires an initial guess of the turbulent viscosity $\nu_t$ to solve the flow equations. The turbulent viscosity is later updated after the SA equation is solved. In a similar fashion, the proposed PGD solver utilises the last computed separated approximation $\turbNUpgd^q$ of the turbulent viscosity to solve the flow equations, before updating it with the result of the \texttt{PGD-SA} and \texttt{PGD-$\nu_t$} steps. Thus, the turbulent viscosity $\nu_t$ in equation~\eqref{eq:weak-NS-PGD} is henceforth replaced by its separated approximation $\turbNUpgd^q$.
\end{remark}

Let $\uBpgd^n {:=} \upgd^{n-1} {+} \sigmaU^n\fu^n\phi^n$ and $\pBpgd^n {:=} \ppgd^{n-1} {+} \sigmaP^n\fp^n\phi^n$ denote the estimated rank-$n$ PGD approximations for velocity and pressure obtained as the sum of the converged $n{-}1$ terms and the computed predictions of the $n$-th modes. The unknown increments $(\sigmaU^n\de\upgd^n,\sigmaP^n\de\ppgd^n)$ are determined from equation~\eqref{eq:weak-NS-PGD} solving
\begin{equation} \label{eq:PGD-NS-increment}
\left\{\begin{aligned}
\int_{\bI}\int_{V_i}{\!\! \grad {\cdot} (\sigmaU^n\de\upgd^n {\otimes} \sigmaU^n\de\upgd^n) \, dV \, d\bI} 
\hspace{200pt} & \\
+ \int_{\bI}\int_{V_i}{\!\! \grad {\cdot} (\sigmaU^n\de\upgd^n {\otimes} \upgd^n) \, dV \, d\bI} 
+ \int_{\bI}\int_{V_i}{\!\! \grad {\cdot} (\upgd^n {\otimes} \sigmaU^n\de\upgd^n) \, dV \, d\bI} 
\hspace{35pt} & \\
- \int_{\bI}\int_{V_i}{ \!\! \grad {\cdot} ( (\nu {+} \turbNUpgd^q) \grad (\sigmaU^n\de\upgd^n))  \, dV \, d\bI} 
+ \int_{\bI}\int_{V_i}{\!\! \grad (\sigmaP^n\de\ppgd^n)  \, dV \, d\bI} = \mathcal{R}_u , 
 & \\
\int_{\bI}\int_{V_i}{\!\! \grad {\cdot} (\sigmaU^n\de\upgd^n) \, dV \, d\bI} = \mathcal{R}_p , 
&
%\hspace{130pt} &
\end{aligned}\right.
\end{equation}
where the unknowns $(\sigmaU^n\de\upgd^n,\sigmaP^n\de\ppgd^n)$ have been gathered on the left-hand side, whereas the right-hand side features the residuals
\begin{equation}\label{eq:resNS}
\begin{aligned}
\mathcal{R}_u &:=
\begin{aligned}[t]
&- \int_{\bI}\int_{V_i}{\!\! \grad {\cdot} (\uBpgd^n {\otimes} \uBpgd^n) \, dV \, d\bI} \\
&+ \int_{\bI} \int_{V_i}{\!\! \grad {\cdot} ( (\nu {+} \turbNUpgd^q) \grad \uBpgd^n)  \, dV \, d\bI} 
- \int_{\bI}\int_{V_i}{\!\! \grad \pBpgd^n  \, dV \, d\bI} ,
\end{aligned}
\\
\mathcal{R}_p &:= - \int_{\bI}\int_{V_i}{\!\! \grad {\cdot} \uBpgd^n \, dV \, d\bI} .
\end{aligned}
\end{equation}

\begin{remark}\label{rmrk:relaxNS}
It is straightforward to observe that the equations~\eqref{eq:PGD-NS-increment} mimic the structure of the full order Navier-Stokes equations~\eqref{eq:weak-NS-PGD}. In order to devise a PGD algorithm non-intrusive with respect to the OpenFOAM spatial solver \texttt{simpleFOAM}, in~\cite{Tsiolakis-TGSOH-20} the authors proposed to relax the second and third terms in~\eqref{eq:PGD-NS-increment} during the PGD spatial step. More precisely, these terms were evaluated using the last known increment computed during the previous SIMPLE iteration. Hence, they were treated explicitly, as additional terms on the right-hand side of the momentum equation. This approach is also followed in this work, as detailed in section~\ref{sc:PGD-RANS-spatial}.
\end{remark}

An efficient implementation of this strategy can be devised by exploiting the affine decomposition of the forms in~\eqref{eq:PGD-NS-increment} and~\eqref{eq:resNS}, see~\cite{Patera-Rozza:07,Rozza:14}, and the separated structure of the unknowns~\eqref{eq:sep-increment-up}. 
Nonetheless, it is worth noticing that when the number of modes in the separated representation grows, the overall cost of the PGD strategy may significantly increase. In this context, several variations of the above approach have been proposed in the literature, e.g. by performing a PGD compression step~\cite{Modesto-MZH:15,Diez-DZGH-19} to eliminate the redundant information in the previously computed modes or by introducing an update step~\cite{Nouy:07,Nouy-NL-09} or an Arnoldi-type iteration~\cite{Nouy:08,LT-TMN:14} to exploit the intermediate solutions in the alternating direction algorithm.

In the following sections, the restriction of the high-dimensional problem~\eqref{eq:PGD-NS-increment} to the tangent manifold associated with the spatial and the parametric coordinates is described. More precisely, at each iteration of the alternating direction algorithm, the spatial step is solved using \texttt{simpleFoam}, whereas the parametric step leads to an algebraic problem solved using a collocation approach.

%==========================================================================
\subsubsection{\texttt{PGD-NS}: the spatial iteration}
\label{sc:PGD-RANS-spatial}
%==========================================================================

In order to construct a PGD approximation of the flow equations~\eqref{eq:PGD-NS-increment}, a separated representation of the data is required~\cite{SZ-ZDMH:15}. More precisely, the form $\nu(\bx,\bmu) {=} D(\bx)\zeta(\bmu)$ is assumed for the physical viscosity, whereas the PGD approximation $\turbNUpgd^q(\bx,\bmu) {=} \sum_{j=1}^q \sigmaT^j\ft^j(\bx) \xi^j(\bmu)$ is considered for the turbulent viscosity.

The high-dimensional problem~\eqref{eq:PGD-NS-increment} is thus restricted to the spatial direction multiplying it by $\phi^n$. In addition, setting the value of the parametric function $\phi^n$, the increments $(\sigmaU^n\de\upgd^n,\sigmaP^n\de\ppgd^n)$ reduce to $\phi^n (\sigmaU^n \De\fu, \sigmaP^n \De\fp)$, see equation~\eqref{eq:PGD-corr-computed}. The spatial increments $(\sigmaU^n \De\fu, \sigmaP^n \De\fp)$ are thus computed as the FV solution of the flow equations
\begin{equation} \label{eq:PGD-NS-spatial}
\left\{\begin{aligned}
\alpha_0 \!\! \int_{V_i}\!\! \grad {\cdot} (\sigmaU^n \De\fu {\otimes} \sigmaU^n \De & \fu) dV \\
- \!\! \int_{V_i}\!\! \grad {\cdot} \Big(\!\! \Big(\alpha_1 D + \sum_{j=1}^q \alpha_7^j & \sigmaT^j\ft^j \Big) \grad (\sigmaU^n \De\fu) \! \Big)  dV
 + \alpha_2 \!\! \int_{V_i}{\grad (\sigmaP^n \De\fp)  dV} \\
&= R_U^n - \!\! \int_{V_i}\!\! \grad {\cdot} \Big( \sum_{j=1}^n \! \alpha_3^j \sigmaU^j \fu^j {\otimes} \sigmaU^{k-1} \De\fu^{k-1} \Big)  dV   \\
& \hspace{35pt} - \!\! \int_{V_i}\!\! \grad {\cdot} \Big( \sigmaU^{k-1} \De\fu^{k-1} {\otimes}\!\sum_{j=1}^n \!\alpha_3^j \sigmaU^j \fu^j \Big)   dV , \\
\alpha_2 \!\! \int_{V_i}{\grad {\cdot} (\sigmaU^n \De\fu)  dV} &= R_P^n ,
\end{aligned}\right.
\end{equation}
where $R_U^n$ and $R_P^n$ denote the spatial residuals of the momentum and continuity equations, respectively, and the coefficients $\alpha_k, \, k{=}0,\ldots,3$ and $\alpha_7$, reported in~\ref{sc:appCoeff}, only depend on user-defined data and parametric functions and can thus be efficiently precomputed. 

Following remark~\ref{rmrk:relaxNS}, an implementation of the PGD spatial solver for the flow equations non-intrusive with respect to the OpenFOAM SIMPLE algorithm is obtained by relaxing the two linear contributions arising from the nonlinear convection term~\cite{Tsiolakis-TGSOH-20}. More precisely, the last two integrals on the right-hand side of the momentum equation in~\eqref{eq:PGD-NS-spatial} are evaluated using the last increment $\sigmaU^{k-1}\De\fu^{k-1}$ computed in the SIMPLE iterations. It is straightforward to observe that the resulting structure of the left-hand side of equations~\eqref{eq:PGD-NS-spatial} mimics the traditional Navier-Stokes equations~\eqref{eq:weak-RANS}, whence the PGD spatial iteration is solved using the \texttt{simpleFoam} algorithm, natively implemented in OpenFOAM~\cite{Tsiolakis-TGSOH-20}.

The spatial residuals $R_U^n$ and $R_P^n$ on the right-hand side of equations~\eqref{eq:PGD-NS-spatial} are determined starting from the previous PGD terms $(\upgd^{n-1},\ppgd^{n-1})$ and from the predictions $(\sigmaU^n\fu^n\phi^n,\sigmaP^n\fp^n\phi^n)$ of the $n$-th mode currently computed, namely
\begin{subequations}\label{eq:NSspatialRes-general}
\begin{align}
&
\begin{aligned}
R_U^n := 
&- \int_{\bI} \phi^n \int_{V_i}{\!\! \grad {\cdot} \left( [\upgd^{n-1} + \sigmaU^n\fu^n \phi^n ] {\otimes} [\upgd^{n-1} + \sigmaU^n\fu^n \phi^n ] \right) dV \, d\bI} \\
&+ \int_{\bI} \phi^n \int_{V_i}{\!\! \grad {\cdot} \left( (\nu + \turbNUpgd^q) \grad (\upgd^{n-1} + \sigmaU^n\fu^n \phi^n) \right)  dV \, d\bI} \\
&- \int_{\bI} \phi^n \int_{V_i}{\!\! \grad (\ppgd^{n-1} + \sigmaP^n\fp^n \phi^n)  \, dV \, d\bI} ,
\end{aligned}
\label{eq:NSspatialResU-general} \\
&
R_P^n := 
- \int_{\bI} \phi^n \int_{V_i}{\!\! \grad {\cdot} \left(\upgd^{n-1} +  \sigmaU^n\fu^n \phi^n \right) dV \, d\bI} .
\label{eq:NSspatialResP-general}
\end{align}
\end{subequations}

It is worth recalling that the factor $\phi^n$ in the expressions of $R_U^n$ and $R_P^n$ follows from the restriction of the residuals~\eqref{eq:resNS} defined in the high-dimensional space $\Omega {\times} \bI$ to the tangent manifold associated with the spatial direction~\cite{Tsiolakis-TGSOH-20}.
In order to perform an efficient computation of such residuals, the separated expressions of $(\upgd^{n-1},\ppgd^{n-1})$ as a product of spatial and parametric functions are exploited, leading to
\begin{subequations}\label{eq:NSspatialRes}
\begin{align}
&
\begin{aligned}
R_U^n
= &- \sum_{j=1}^n \sum_{\ell=1}^n \alpha_4^{j\ell} \!\! \int_{V_i}{\grad {\cdot} (\sigmaU^j\fu^j {\otimes} \sigmaU^\ell\fu^\ell) \, dV} \\
&+\sum_{\ell=1}^n \int_{V_i}{\!\! \grad {\cdot} \Big(\!\! \Big(\alpha_5^\ell D + \sum_{j=1}^q\alpha_8^{j\ell} \sigmaT^j\ft^j\Big) \grad (\sigmaU^\ell\fu^\ell ) \! \Big) \, dV} \\
&-   \sum_{\ell=1}^n \alpha_6^\ell \int_{V_i}{\!\! \grad (\sigmaP^\ell\fp^\ell )  \, dV} ,
\end{aligned}
\label{eq:NSspatialResU} \\
&
R_P^n 
= - \sum_{\ell=1}^n \alpha_6^\ell \int_{V_i}{\!\! \grad {\cdot} (\sigmaU^\ell \fu^\ell ) \, dV} ,
\label{eq:NSspatialResP}
\end{align}
\end{subequations}
where the coefficients $\alpha_k, \, k{=}4,\ldots,6$ and $\alpha_8$ encapsulate the information of the previously computed parametric modes and are defined in~\ref{sc:appCoeff}.

%==========================================================================
\subsubsection{\texttt{PGD-NS}: the parametric iteration}
%==========================================================================
  
In the parametric iteration of the PGD solver for the flow equations, the value of the spatial functions  $(\sigmaU^n\fu^n,\sigmaP^n\fp^n) {\gets} (\sigmaU^n [\fu^n {+} \De\fu], \sigmaP^n [\fp^n {+} \De\fp])$ is updated and fixed. It follows that the increments associated with the momentum and continuity equations in the parametric step are $\sigmaU^n\fu^n \De\phi$ and $\sigmaP^n\fp^n \De\phi$, respectively. 

Following remark~\ref{rmrk:equalParamFun}, a unique parametric increment $\De\phi$ is defined for both the velocity and the pressure mode. To compute $\Delta\phi$, an algebraic problem is obtained via the restriction of the high-dimensional equations~\eqref{eq:PGD-NS-increment} to the parametric direction $\bI$ multiplying the momentum and continuity equations by the spatial functions $\sigmaU^n\fu^n$ and $\sigmaP^n\fp^n$, respectively. The resulting parametric problem, solved by means of a collocation approach, is 
\begin{subequations}\label{eq:PGD-NS-param}
\begin{equation}\label{eq:NSparamMatrix}
a_0 (\De\phi)^2 + \left( - a_1 \zeta - \sum_{j=1}^q a_8^j \xi^j + a_2 + \sum_{j=1}^n a_3^j \phi^j \right) \De\phi 
= r_U^n + r_P^n ,
\end{equation}
where the residuals $r_U^n$ and $ r_P^n$ of the momentum and continuity equations in the parametric space are given by 
\begin{align}\label{eq:NSparamRes}
r_U^n & := \sum_{\ell=1}^n \left( - \sum_{j=1}^n a_4^{j\ell} \phi^j + a_5^\ell \zeta + \sum_{j=1}^q a_9^{j\ell} \xi^j - a_6^\ell \right) \phi^\ell , \\
r_P & := - \sum_{\ell=1}^n a_7^\ell \phi^\ell ,
\end{align}
\end{subequations}
being $a_k, \ k=0,\ldots,9$ a set of coefficients which depend on user-defined data and on previously computed spatial modes as reported in~\ref{sc:appCoeff}.

\begin{remark}\label{rmrk:diffLaminar}
Contrary to the parametric problem in~\cite{Tsiolakis-TGSOH-20}, the second-order term $(\De\phi)^2$ is maintained in equation~\eqref{eq:NSparamMatrix}. Although this term was negligible in laminar simulations, it has been verified numerically that its presence improves the stability of the solution of the parametric step of the Navier-Stokes equations in the turbulent regime.
\end{remark}

The alternating direction algorithm in the \texttt{PGD-NS} routine stops when the relevance of the computed increment is negligible with respect to the amplitude of the corresponding mode. Similarly, the global enrichment procedure stops when the contribution of the current mode in the PGD expansion, measured by means of its relative amplitude, is negligible.

\begin{remark}
As shown in~\cite{Tsiolakis-TGSOH-20}, the \emph{a priori} PGD algorithm devised starting from the separated formulation of problem~\eqref{eq:PGD-NS-increment} provides a stable approximation for both velocity and pressure, without the need for tailored pressure corrections required by \emph{a posteriori} ROMs for incompressible flows~\cite{Ballarin-BMQR:15,Stabile-SR:18}.
\end{remark}

%==========================================================================
\subsection{Proper generalised decomposition of the turbulence model}
\label{sc:PGD-SA}
%==========================================================================

The construction of a separated expression $\turbNUpgd^q$ of the turbulent viscosity using the SA model first requires determining the approximation $\nuTpgd^m$ of the eddy viscosity via a dedicated PGD solver. For this purpose, the integral form of the steady Spalart-Allmaras turbulence model in the high-dimensional space $\Omega \times \bI$ is considered, namely
\begin{equation} \label{eq:weak-SA-PGD}
\begin{aligned}
\int_{\bI}\int_{V_i} \!\! \grad {\cdot}(\bU \nuT) & \, dV  \, d\bI
- \frac{1}{\sigma} \int_{\bI}\int_{V_i}{\!\!  \grad {\cdot} \left( (\nu {+} \nuT) \grad \nuT \right) \, dV  \, d\bI} \\
-& \frac{c_{b2}}{\sigma} \int_{\bI}\int_{V_i}{\!\!  \grad \nuT {\cdot} \grad \nuT \, dV  \, d\bI}  
- c_{b1} \int_{\bI}\int_{V_i}{\!\! \widetilde{S} \nuT \, dV  \, d\bI} \\
&\hspace{90pt} + c_{w1}  \int_{\bI}\int_{V_i}{\!\! f_w \left( \frac{\nuT}{\widetilde{d}} \right)^2 \, dV  \, d\bI} = 0 ,
\end{aligned}
\end{equation}
for each cell $V_i, \, i{=}1,\ldots,N$ in the computational domain.

The PGD solver for the SA equation is obtained by inserting the separated approximation~\eqref{eq:sep-increment-nu} of the eddy viscosity into equation~\eqref{eq:weak-SA-PGD}. Following the rationale utilised for the flow equations, the first term $\nuTpgd^0$ in the PGD expansion is selected to enforce the inhomogeneous Dirichlet boundary conditions on $\GammaI \cup \GammaW$. More precisely, Dirichlet data for $\nuTpgd$ are selected as full order solutions of the SA equation computed using the boundary condition modes of the velocity field. The following modes are determined using a greedy algorithm named \texttt{PGD-SA}: for each new mode $\sigmaNU^m \left[ \fnu^m\psi^m + \de\nuTpgd^m  \right]$, the prediction $\sigmaNU^m \fnu^m\psi^m$ is computed by performing a first iteration of the alternating direction algorithm, whereas the following iterations allow to devise the corresponding correction term $\sigmaNU^m \de\nuTpgd^m$.

It is worth recalling that OpenFOAM employs a segragated approach for the simulation of turbulent flows, by solving independently the flow problem and the SA equation. More precisely, the computation of the eddy viscosity is performed using the velocity field $\bU$ obtained from the RANS equations as input for the solver. The proposed PGD strategy mimics this approach and replaces the velocity field $\bU$ by its rank-$n$ PGD approximation $\upgd^n$ constructed using the procedure described in section~\ref{sc:PGD-RANS}.

Let $\nuBpgd^m {:=} \nuTpgd^{m-1} {+} \sigmaNU^m\fnu^m\psi^m$ denote the estimated rank-$m$ approximation of the eddy viscosity obtained using the $m{-}1$ previously converged modes $\nuTpgd^{m-1}$ and the prediction $\sigmaNU^m\fnu^m\psi^m$ of the $m$-th term. From equation~\eqref{eq:weak-SA-PGD}, it follows that the unknown increment $\sigmaNU^m \de\nuTpgd^m$ can be obtained by solving
\begin{equation}\label{eq:PGD-SA-increment}
\begin{aligned}
\int_{\bI}\int_{V_i}\!\! \grad {\cdot} \Big( \upgd^n \sigmaNU^m\de\nuTpgd^m \Big)  dV  \, d\bI \hspace{12em} & \\
- \frac{1}{\sigma} \int_{\bI}\int_{V_i}{\!\!  \grad {\cdot} \Big( (\nu + \nuBpgd^m + \sigmaNU^m\de\nuTpgd^m) \grad (\sigmaNU^m\de\nuTpgd^m) \! \Big) dV  \, d\bI} & \\
- \frac{1}{\sigma} \int_{\bI}\int_{V_i}{\!\!  \grad {\cdot} \Big( \sigmaNU^m\de\nuTpgd^m \grad \nuBpgd^m \! \Big) dV  \, d\bI} \hspace{7em} & \\
- \frac{c_{b2}}{\sigma} \int_{\bI}\int_{V_i}{\!\!  \grad (\sigmaNU^m\de\nuTpgd^m) {\cdot} \grad (\sigmaNU^m\de\nuTpgd^m) \, dV  \, d\bI} \hspace{3em} & \\
- \frac{2 c_{b2}}{\sigma} \int_{\bI}\int_{V_i}{\!\!  \grad \nuBpgd^m  {\cdot} \grad (\sigmaNU^m\de\nuTpgd^m) \, dV  \, d\bI} \hspace{4em} & \\
- c_{b1} \int_{\bI}\int_{V_i}{\!\! \widetilde{S} \sigmaNU^m\de\nuTpgd^m \, dV  \, d\bI} \hspace{7em} & \\
+ c_{w1}  \int_{\bI}\int_{V_i}{\!\! \frac{f_w}{\widetilde{d}^2} \left(\sigmaNU^m\de\nuTpgd^m \right)^2 \, dV  \, d\bI} \hspace{3em} & \\
+ 2 c_{w1}  \int_{\bI}\int_{V_i}{\!\! \frac{f_w^m}{\widetilde{d}^2} \nuBpgd^m \sigmaNU^m\de\nuTpgd^m \, dV  \, d\bI} &=\mathcal{R}_\nu ,
\end{aligned}
\end{equation}
where the residual $\mathcal{R}_\nu$ is given by
\begin{equation}\label{eq:resSA}
\begin{aligned}
\mathcal{R}_\nu := 
&- \int_{\bI}\int_{V_i}{\grad {\cdot} (\upgd^n \nuBpgd^m) \, dV  \, d\bI} \\
&+ \frac{1}{\sigma} \int_{\bI}\int_{V_i}{\!\!  \grad {\cdot} \Big( (\nu +\nuBpgd^m ) \grad \nuBpgd^m \! \Big) dV  \, d\bI} \\
&+ \frac{c_{b2}}{\sigma} \int_{\bI}\int_{V_i}{\!\!  \grad \nuBpgd^m {\cdot} \grad \nuBpgd^m \, dV  \, d\bI} \\
&+ c_{b1} \int_{\bI}\int_{V_i}{\!\! \widetilde{S} \nuBpgd^m \, dV  \, d\bI} 
 - c_{w1}  \int_{\bI}\int_{V_i}{\!\! \frac{f_w}{\widetilde{d}^2} (\nuBpgd^m)^2 \, dV  \, d\bI} .
\end{aligned}
\end{equation}

Problem~\eqref{eq:PGD-SA-increment} can thus be efficiently solved by alternatively restricting it to the tangent manifold associated with the spatial and the parametric coordinates as described in the following sections.

\begin{remark}\label{rmrk:relaxSA}
Similarly to the observation in remark~\ref{rmrk:relaxNS}, the first, second, fourth, sixth and seventh term in equation~\eqref{eq:PGD-SA-increment} mimic the original SA turbulence model~\eqref{eq:weak-SA-PGD}. Since the remaining integrals are not accounted for in the full order model, and in order to minimise intrusiveness with respect to OpenFOAM core routines, these terms are relaxed during the spatial iteration of the PGD algorithm, by evaluating them using the last computed iteration. This approach is detailed in section~\ref{sc:PGD-SA-spatial}.
\end{remark}

%==========================================================================
\subsubsection{\texttt{PGD-SA}: the spatial iteration}
\label{sc:PGD-SA-spatial}
%==========================================================================

First, the convective velocity field is replaced by its separated approximation \\ $\upgd^n(\bx,\bmu) {=} \sum_{j=1}^n \sigmaU^j \fu^j(\bx) \phi^j(\bmu)$. The SA equation~\eqref{eq:PGD-SA-increment} in the high-dimensional space $\Omega \times \bI$ is then restricted to the spatial direction multiplying it by $\psi^m$. Moreover, the value of the parametric function is set to $\psi^m$ and, from equation~\eqref{eq:PGD-corr-computed}, the increment $\sigmaNU^m\de\nuTpgd^m$ simplifies to $\psi^m \sigmaNU^m \De\fnu$.

The increment $\sigmaNU^m\De\fnu$ thus acts as unknown of the spatial iteration of the PGD procedure for the parametric SA equation. More precisely, a cell-by-cell constant approximation $\sigmaNU^m\De\fnu$ is computed solving 
\begin{equation} \label{eq:PGD-SA-spatial}
\hspace{-10pt}
\begin{aligned}
 \int_{V_i}\!\! \grad {\cdot} &\Big(\!\!\Big(\sum_{j=1}^n \beta_1^j \sigmaU^j \fu^j \Big) \sigmaNU^m \De\fnu \Big)  dV \\
- \frac{1}{\sigma} & \int_{V_i}{\!\!  \grad {\cdot} \Big( \Big[ \Big( \beta_2 D {+} \sum_{j=1}^m \beta_3^j \sigmaNU^j \fnu^j \Big) {+} \beta_4 \sigmaNU^m \De\fnu \Big] \grad (\sigmaNU^m \De\fnu) \! \Big) dV} \\
-& \frac{\beta_4 c_{b2}}{\sigma} \int_{V_i}{\!\!  \grad (\sigmaNU^m \De\fnu) {\cdot} \grad (\sigmaNU^m \De\fnu) \, dV} \\
&- \beta_5 c_{b1} \int_{V_i}{\!\! \Sx{\widetilde{S}^m} \sigmaNU^m \De\fnu \, dV} 
 + \beta_6 c_{w1}  \int_{V_i}{\!\! \frac{\Sx{f_w^m}}{\widetilde{d}^2} \left(\sigmaNU^m \De\fnu \right)^2 \, dV} \\
&\hspace{85pt} = R_{\nu}^m  + \frac{1}{\sigma} \int_{V_i}{\!\!  \grad {\cdot} \Big( \sigmaNU^{k-1} \De\fnu^{k-1} \grad \Big(\sum_{j=1}^m \beta_3^j \sigmaNU^j \fnu^j \Big) \! \Big) dV} \\
&\hspace{95pt} + \frac{2 c_{b2}}{\sigma} \int_{V_i}{\!\!  \grad \Big(\sum_{j=1}^m \beta_3^j \sigmaNU^j \fnu^j \Big)  {\cdot} \grad (\sigmaNU^{k-1} \De\fnu^{k-1}) \, dV} \\
&\hspace{98pt} - 2 c_{w1}  \int_{V_i}{\!\! \frac{\Sx{f_w^m}}{\widetilde{d}^2} \Big( \sum_{j=1}^m \beta_7^j \sigmaNU^j \fnu^j \Big) \sigmaNU^{k-1} \De\fnu^{k-1} \, dV} ,
\end{aligned}
\end{equation}
where $R_{\nu}^m$ denotes the spatial residual of the SA equation and the coefficients $\beta_k, \, k{=}1,\ldots,7$ depend on user-defined data and parametric functions as described in~\ref{sc:appCoeff}. Moreover, the production, $\widetilde{S}^m$, and destruction, $f_w^m$, coefficients are evaluated using the $m$ previously computed modes of $\nuTpgd^m$ and $\Sx{\widetilde{S}^m}$ and $\Sx{f_w^m}$ denote the corresponding spatial modes of these functions.

As commented in remark~\ref{rmrk:relaxSA}, three terms appearing in the high-dimensional problem~\eqref{eq:PGD-SA-increment} do not have a counterpart in the original SA equation available in OpenFOAM. In order for the discussed PGD-ROM implementation to be non-intrusive with respect to the SA solver natively implemented in OpenFOAM, these terms are treated explicitly via a relaxation approach. Hence, the last three terms on the right-hand side of equation~\eqref{eq:PGD-SA-spatial} are evaluated using the last computed increment $\sigmaNU^{k-1} \De\fnu^{k-1}$ in the iterative procedure to solve the SA equation. The left-hand side of the equation~\eqref{eq:PGD-SA-spatial} thus presents the same structure as the original SA equation~\eqref{eq:weak-SA}, where the spatial integrals are now weighted by means of appropriately computed parametric coefficients, and the OpenFOAM strategy for the solution of the turbulence model equation can be exploited.

The spatial residual $R_{\nu}^m$ on the right-hand side of equation~\eqref{eq:PGD-SA-spatial} is obtained starting from the values of the previous terms $\nuTpgd^{m-1}$ in the PGD expansion of the eddy viscosity and the prediction of the $m$-th mode $\sigmaNU^m\fnu^m\psi^m$ currently computed, namely
\begin{equation}\label{eq:SAspatialRes-general}
\begin{aligned}
R_{\nu}^m := 
&- \int_{\bI} \psi^m \int_{V_i}{\grad {\cdot} \left(\upgd^n (\nuTpgd^{m-1} + \sigmaNU^m\fnu^m\psi^m) \right) \, dV  \, d\bI} \\
&+ \frac{1}{\sigma} \int_{\bI} \psi^m \int_{V_i}{\!\!  \grad {\cdot} \Big( (\nu + \nuTpgd^{m-1} + \sigmaNU^m\fnu^m\psi^m ) \grad (\nuTpgd^{m-1} + \sigmaNU^m\fnu^m\psi^m) \! \Big) dV  \, d\bI} \\
&+ \frac{c_{b2}}{\sigma} \int_{\bI} \psi^m \int_{V_i}{\!\!  \grad (\nuTpgd^{m-1} + \sigmaNU^m\fnu^m\psi^m) {\cdot} \grad (\nuTpgd^{m-1} + \sigmaNU^m\fnu^m\psi^m) \, dV  \, d\bI} \\
&+ c_{b1} \int_{\bI} \psi^m \int_{V_i}{\!\! \widetilde{S} (\nuTpgd^{m-1} + \sigmaNU^m\fnu^m\psi^m) \, dV  \, d\bI} \\
&- c_{w1}  \int_{\bI} \psi^m \int_{V_i}{\!\! \frac{f_w}{\widetilde{d}^2} (\nuTpgd^{m-1} + \sigmaNU^m\fnu^m\psi^m)^2 \, dV  \, d\bI} ,
\end{aligned}
\end{equation}
where the parametric function $\psi^m$ in the integrals above stems from the restriction of the high-dimensional residuals $\mathcal{R}_{\nu}$ to the tangent manifold in the spatial direction. By eploiting the separated expression of $\nuTpgd^{m-1}$ in terms of its spatial and parametric modes, the residual can be rewritten as
\begin{equation}\label{eq:SAspatialRes}
\begin{aligned}
R_{\nu}^m
= -& \sum_{j=1}^n \sum_{\ell=1}^m \beta_8^{j\ell} \!\! \int_{V_i}{\grad {\cdot} (\sigmaU^j\fu^j \sigmaNU^\ell\fnu^\ell) \, dV} \\
& + \frac{1}{\sigma} \sum_{\ell=1}^m \int_{V_i}{\!\!  \grad {\cdot} \Big(\!\!\Big( \beta_9^\ell D {+} \sum_{j=1}^m \beta_{10}^{j\ell} \sigmaNU^j \fnu^j \Big) \grad (\sigmaNU^\ell \fnu^\ell) \! \Big) dV} \\
& + \frac{c_{b2}}{\sigma} \sum_{j=1}^m \sum_{\ell=1}^m \beta_{10}^{j\ell} \int_{V_i}{\!\!  \grad (\sigmaNU^j \fnu^j) {\cdot} \grad (\sigmaNU^\ell \fnu^\ell) \, dV} \\
& + c_{b1} \sum_{\ell=1}^m \beta_{11}^\ell \int_{V_i}{\!\! \Sx{\widetilde{S}^m} \sigmaNU^\ell \fnu^\ell \, dV} \\
& - c_{w1}  \sum_{j=1}^m \sum_{\ell=1}^m \beta_{12}^{j\ell} \int_{V_i}{\!\! \frac{\Sx{f_w^m}}{\widetilde{d}^2} \sigmaNU^j \fnu^j \sigmaNU^\ell \fnu^\ell \, dV} .
\end{aligned}
\end{equation}
where the coefficients $\beta_k, \, k{=}8,\ldots,12$, reported in~\ref{sc:appCoeff}, can be precomputed for an efficient implementation of the PGD spatial iteration since they depend soley on user-defined data and parametric functions.

\begin{remark}\label{rmrk:nonSeparable}
It is worth emphasising that neither $\widetilde{S}(\bx,\bmu)$ nor $f_w(\bx,\bmu)$ are separable exactly via an analytical procedure. It follows that equation~\eqref{eq:PGD-SA-spatial} cannot be solved in a complete non-intrusive way using OpenFOAM. The resulting implementation of the \texttt{PGD-SA} algorithm in OpenFOAM is thus minimally intrusive as the structure of equation~\eqref{eq:PGD-SA-spatial} is the same as the original SA equation~\eqref{eq:weak-SA} but a tailored numerical strategy is required to efficiently compute the coefficients and the integral terms involving $\widetilde{S}$ and $f_w$. Suitable procedures include \emph{a posteriori} PGD separation~\cite{Modesto-MZH:15,Diez-DZGH-19} and high-dimensional reconstruction of the functions in the space $\Omega \times \bI$ followed by an interpolation step in $\Omega$. The latter strategy is employed for the simulations in section~\ref{sc:simulations}. 
\end{remark}

%==========================================================================
\subsubsection{\texttt{PGD-SA}: the parametric iteration}
\label{sc:PGD-SA-param}
%==========================================================================

The parametric iteration of the turbulence model is devised by fixing the newly computed spatial mode $\sigmaNU^m \fnu^m {\gets} \sigmaNU^m [\fnu^m {+} \De\fnu]$ of the eddy viscosity. The corresponding parametric increment $\De\psi$ is obtained by restricting the high-dimensional equation~\eqref{eq:PGD-SA-increment} to the parametric direction $\bI$ multiplying it by the spatial function $\sigmaNU^m\fnu^m$.
It follows the algebraic equation
\begin{equation}\label{eq:PGD-SA-param}
\begin{aligned}
 (& -b_4 + b_6 \Smu{f_w^m} )(\De\psi)^2 \\
 &+ \left( - b_2 \zeta - b_5 \Smu{\widetilde{S}^m} + \sum_{j=1}^n b_1^j \phi^j - \sum_{j=1}^m (b_3^j + b_7^j \Smu{f_w^m}) \psi^j \right) \De\psi 
= r_{\nu}^m ,
\end{aligned}
\end{equation}
in which $b_k, \ k=1,\ldots,7$ represent the precomputed coefficients reported in~\ref{sc:appCoeff}, $r_{\nu}^m$ is the residual of the SA equation in the parametric space and $\Smu{\widetilde{S}^m}$ and $\Smu{f_w^m}$ denote the parametric modes of the non-separable functions $\widetilde{S}^m$ and $f_w^m$, respectively. It is worth recalling that $\widetilde{S}^m$ and $f_w^m$ are evaluated using the $m$ previously computed modes of $\nuTpgd^m$ and require appropriate numerical separation strategies to be computed, see remark~\ref{rmrk:nonSeparable}. 
Finally, the right-hand side $r_{\nu}^m$ of equation~\eqref{eq:PGD-SA-param} has the form
\begin{equation}\label{eq:SAparamRes}
r_{\nu}^m :=  \sum_{\ell=1}^m \left( b_9^\ell \zeta +  b_{11}^\ell \Smu{\widetilde{S}^m} + \sum_{j=1}^n b_8^{j\ell} \phi^j + \sum_{j=1}^m (b_{10}^{j\ell} - b_{12}^{j\ell} \Smu{f_w^m} ) \psi^j \right) \psi^\ell  ,
\end{equation}
with the coefficients $b_k, \ k=8,\ldots,12$ depending upon data provided by the user and spatial functions, see~\ref{sc:appCoeff}.

As for the \texttt{PGD-NS} routine, the alternating direction scheme in the \texttt{PGD-SA} algorithm stops when the computed increment is negligible with respect to the amplitude of the mode. Moreover, the PGD enrichment procedure finishes when the relative amplitude of the current mode is negligible with respect to the previously computed ones.

%==========================================================================
\subsection{\texttt{PGD-$\nu_t$}: devising a separated turbulent viscosity}
\label{sc:PGD-nuT}
%==========================================================================

The approximation $\nuTpgd^m$ of the eddy viscosity provided by the \texttt{PGD-SA} routine is employed to construct a separated representation $\turbNUpgd^q$ of the turbulent viscosity, according to the relation~\eqref{eq:nuT}. This procedure is performed by the \texttt{PGD-$\nu_t$} routine, before going back to the \texttt{PGD-NS} algorithm for the RANS equations.

It is worth noticing that the function $f_{v1}$ in equation~\eqref{eq:nuT} is not separable analytically. Hence, as detailed in remark~\ref{rmrk:nonSeparable}, either a numerical PGD separation~\cite{Modesto-MZH:15,Diez-DZGH-19} or a high-dimensional reconstruction of the function in the space $\Omega \times \bI$ to perform interpolation in $\Omega$ and $\bI$ separately, is required. The latter strategy is employed for the simulations in section~\ref{sc:simulations}.

For the sake of readability, consider $f_{v1}^m$ obtained using the $m$ computed modes of $\nuTpgd^m$. The PGD separated expression of this function is given by
\begin{equation}\label{eq:sepFw1}
f_{v1}^m \simeq \Sx{f_{v1}^m} \Smu{f_{v1}^m} ,
\end{equation}
where $\Sx{f_{v1}^m}$ and $\Smu{f_{v1}^m}$ denote the spatial and parametric modes of the function $f_{v1}^m$, respectively, obtained by means of either of the numerical strategies proposed above.

The separated expression $\turbNUpgd^q$ reported in equation~\eqref{eq:sep-turbNu} can thus be devised in terms of elementary arithmetic operations of separated functions~\cite{Diez-DZGH-19}, exploiting the separated nature of all the involved quantities. More precisely, it holds
\begin{equation}\label{eq:sepTurbNU}
 \turbNUpgd^q(\bx,\bmu)  =  \sum_{j=1}^q \sigmaT^j\ft^j(\bx)\xi^j(\bmu) ,
\end{equation}
where the spatial modes $\ft^j$ can be computed as the product of the spatial functions $\Sx{f_{v1}^m}$ and $\fnu^m$, whereas the parametric terms $\xi^j$ can be obtained from the product of the parametric modes $\Smu{f_{v1}^m}$ and $\psi^m$.
It is worth noting that the product of separated functions leads to a number of terms in the PGD expansion larger than the number of modes of its factors. Nonetheless, such operation can be efficiently performed via a separated multiplication~\cite{Diez-DZGH-19} and the result can be compressed to eliminate redundant information~\cite{Modesto-MZH:15,Diez-DZGH-19}.

%==========================================================================
\section{Numerical experiments}
\label{sc:simulations}
%==========================================================================

\begin{figure}[ht]
	\centering
	\subfigure[Experimental set-up]{\includegraphics[width=0.49\textwidth]{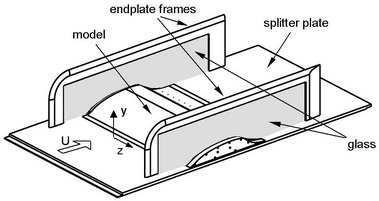}\label{fig:hump_model}}
	\subfigure[Jet location]{\includegraphics[width=0.49\textwidth]{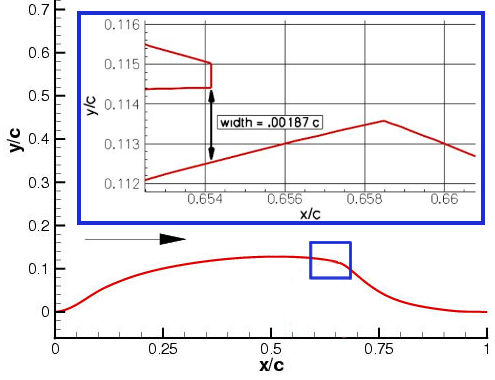}\label{fig:hump_slot}}
	\caption{NASA wall-mounted hump: (a) representation of the experimental set-up and (b) 2D section of the hump, location of the jet (blue rectangle) and detail of the jet opening. Source: {\color{blue}\texttt{https://cfdval2004.larc.nasa.gov/case3.html}}}
	\label{fig:hump_model-hump_slot}
\end{figure}

In this section, numerical simulations of the NASA wall-mounted hump are presented to demonstrate the potential of the proposed methodology. This problem is a quasi-two-dimensional NASA benchmark devised to validate turbulence modelling, starting from an experimental set-up. The domain consists of a Glauert-Goldschmied type body mounted on a splitter plate between two endplates, see figure~\ref{fig:hump_model}. Following~\cite{Seifert-Pack:Hump-FlowControl,Moin-06}, the characteristic length of the problem is set equal to the chord length of the hump $c {=} \SI{0.42}{m}$, whereas its maximum height is $\SI{0.0537}{m}$ and its span is $\SI{0.5842}{m}$. Flow separation along the hump is controlled via a suction jet acting through an opening located at $65\%$ of the chord $c$, as detailed in figure~\ref{fig:hump_slot}. In the experimental set-up the opening is connected to a plenum, on the bottom of which suction pumps are installed; for the numerical simulations in the following sections, the plenum is removed and the suction effect is imposed as a boundary condition on the opening via a mass flow rate of $1.518\times 10^{-2}\SI{}{kg/s}$ for the jet. 

In the analysis of this problem, the quantity of interest is represented by the effect of the suction jet on the flow separation and on the position of the reattachment point. Experimental and numerical studies~\cite{Greenblatt-PYHSW:Hump,Rumsey:Hump-RANS} verified the quasi-two-dimensional nature of the phenomena identifying minor three-dimensional effects located near the endplates. 
%Concerning CFD simulations, in~\cite{Rumsey:Hump-RANS} the authors highlighted that, regardless of the turbulence model employed, RANS solutions of the NASA hump problem tend to overestimate the recirculation bubble and to locate the reattachment point downstream with respect to the position identified by the experiments, with an error between $10\%$ and $20\%$. 
Henceforth, the PGD results will be compared to the full order OpenFOAM approximation, considered as \emph{reference solution}. 

The NASA wall-mounted hump problem being quasi-two-dimensional, in the following sections both the 2D and the 3D cases are studied. A parametric flow control problem is devised by varying the maximum amplitude of the suction jet between $10\%$ and $100\%$ of the module of a peak velocity $\hat{U}$. In two dimensions, a sinusoidal velocity profile is defined as
\begin{equation}\label{eq:Hump-jetProfile}
\bU^{\text{jet}}_{\!\! \hat{y}} = \mu \frac{\hat{U}}{2}(1-\cos(2\pi \hat{x})) ,
\end{equation}
where $\hat{x}$ is the normalised curvilinear abscissa of the jet patch, that is $\hat{x} \in \left[0 , 1\right]$, and the resulting jet is pointing in the direction $\hat{y}$ orthogonal to the boundary.
Similarly, in the 3D case the jet defined on the plane $(\hat{x},\hat{z})$ and pointing in the orthogonal direction $\hat{y}$ is
\begin{equation}\label{eq:Hump-jetProfile-3D}
\bU^{\text{jet}}_{\!\! \hat{y}} = \mu \frac{\hat{U}}{4}(1-\cos(2\pi \hat{x}))(1-\cos(2\pi \hat{z})),
\end{equation}
where the normalised coordinate $\hat{z}$ is
\begin{equation}\label{eq:Hump-hatZ}
\hat{z}= 
\begin{cases}
0 													& \text{ for } z < 0.4 c \\
\displaystyle\frac{5z-2c}{c} 								& \text{ for } 0.4 c \leq z \leq 0.6 c \\
0 													& \text{ for } z > 0.6 c .
\end{cases}
\end{equation}

It is worth noting that the module of the peak velocity $\hat{U}$ is selected such that the ratio between the mass flow rate of the jet and of the inlet is $1.5 \times 10^{-3}$, reproducing the value in the experimental set-up of the NASA wall-mounted hump with a plenum~\cite{Seifert-Pack:Hump-FlowControl}. In addition, both in~\eqref{eq:Hump-jetProfile} and~\eqref{eq:Hump-jetProfile-3D}, the interval of the parametric variable is defined as $\I {=} [0.1,\,1]$.

%________________________________________________________________________
\subsection{Two-dimensional NASA wall-mounted hump with parametrised jet}
\label{sc:simulation-2D}
%________________________________________________________________________

\begin{figure}%[h]
	\centering
	\includegraphics[width=0.8\textwidth]{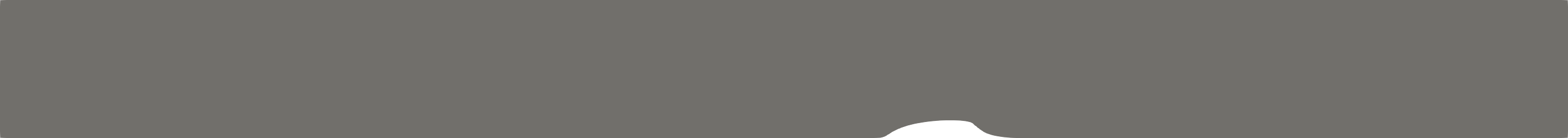}
	\caption{Computational domain for the two-dimensional NASA wall-mounted hump.}
	\label{fig:hump2D}
\end{figure}

The computational domain for the two-dimensional NASA wall-mounted hump is a channel of height $c$, extending $6.39 c$ upstream and $5 c$ downstream as displayed in figure~\ref{fig:hump2D}. The resulting mesh consists of $114,000$ cells. 
Homogeneous Dirichlet boundary conditions are imposed on both the velocity and the eddy viscosity on the bottom wall and on the hump. A symmetry condition is imposed on the top wall, whereas on the outlet free traction is enforced. At the inlet, a parabolic profile is imposed for both velocity and eddy viscosity. More precisely, the variables range between a null value on the bottom wall and a maximum value at $y {=} \SI{0.02}{m}$. For the velocity, the peak value is $\SI{34.6}{m/s}$, whereas the free-stream eddy viscosity $\nuT {=} 3\nu$ is selected. The kinematic viscosity is $\nu {=} 1.55274\times 10^{-5}\SI{}{m^2/s}$, thus the resulting Reynolds number is approximately $\text{Re} {=} 936,000$. On the jet patch, the velocity profile~\eqref{eq:Hump-jetProfile} with $\hat{U} {=} \SI{23.4}{m/s}$ is imposed and a homogeneous Neumann condition is considered for the eddy viscosity. 
It is worth noting that on the hump the mesh is such that $y^+<1$, whence no wall treatment is required for the turbulent viscosity.

In order to impose the inhomogeneous Dirichlet boundary data, two modes are introduced. In particular, recall that the boundary condition on the jet patch depends upon the parameter $\mu$. For this purpose, two spatial modes, associated with the extreme values of the parametric interval $\I {=} [0.1,\,1]$, are computed using \texttt{simpleFoam} with SA turbulence model. The corresponding parametric functions are selected such that
\begin{equation}\label{eq:parModesDef}
\phi^1(\mu)= \begin{cases}
1 , & \text{ if } \mu = 1\\
0 , & \text{ if } \mu = 0.1
\end{cases}
\quad
\text{ and }
\quad
\phi^2(\mu)= \begin{cases}
0 , & \text{ if } \mu = 1\\
1 , & \text{ if } \mu = 0.1
\end{cases} 
\end{equation}
to define a linear variation of the Dirichlet data between the extreme values associated with the computed spatial modes.
More precisely, the first spatial mode is a full order solution with the jet acting at $100\%$ of the mass flow rate ($\mu {=} 1$) and the second one is associated with the jet acting at $10\%$, that is $\mu {=} 0.1$. The corresponding parametric modes are given by
\begin{equation}\label{eq:parModesHump}
\phi^1(\mu)=\frac{10\mu  -1}{9} 
\quad 
\text{ and } 
\quad
 \phi^2(\mu)= 1 - \phi^1(\mu) .
\end{equation}

\begin{figure}
	\centering
	\includegraphics[width=0.6\textwidth]{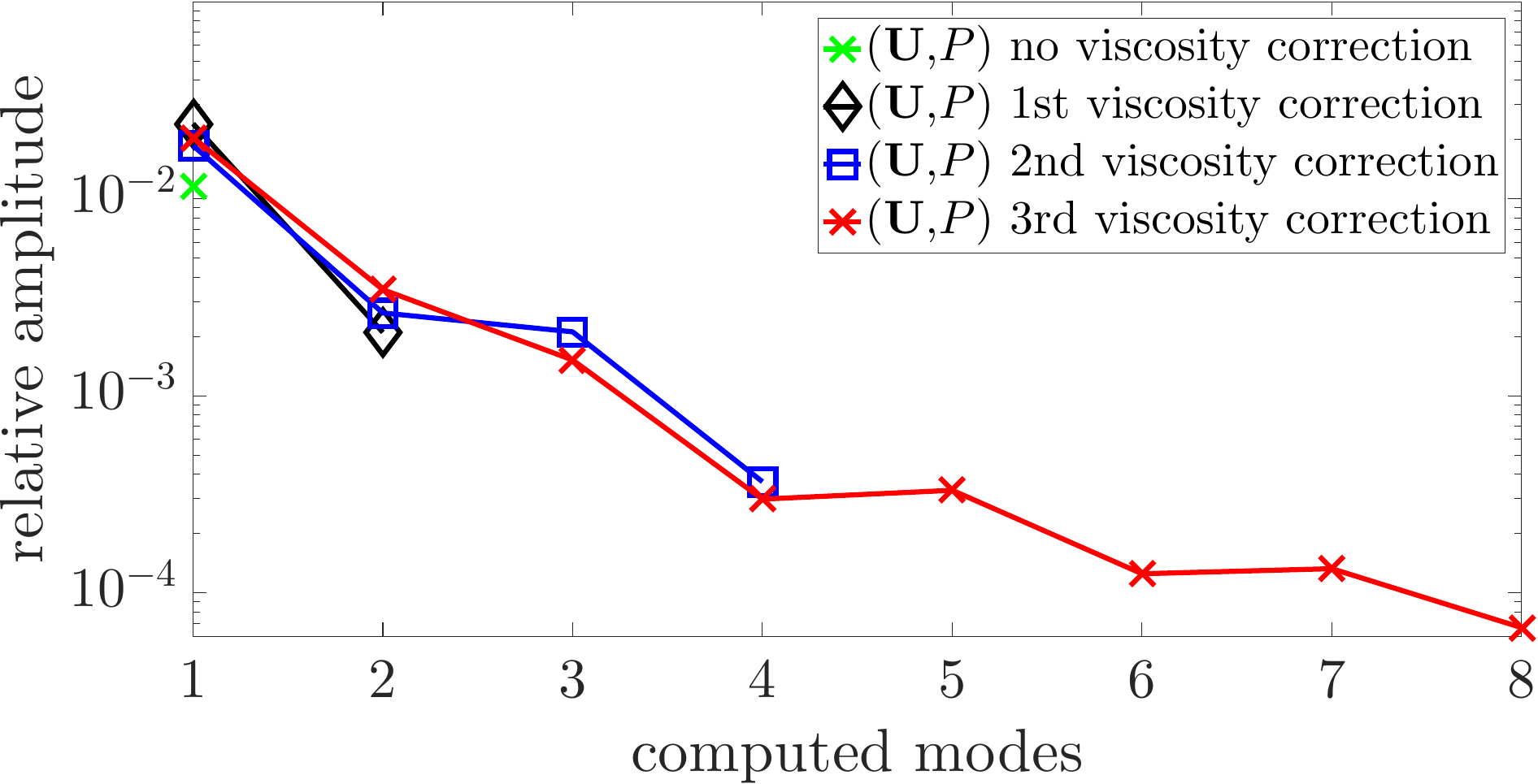}
	\caption{Relative amplitude of the computed velocity-pressure modes as the turbulent viscosity is updated.}
	\label{fig:hump-ampl}
\end{figure}

The tolerance for the enrichment of the flow variables is set to $\eta_u^\star {=} \eta_p^\star {=} 10^{-4}$, whereas the tolerance for the turbulence model is selected as $\eta_{\nu}^\star {=} 10^{-2}$. The criterion to update the turbulent viscosity is detailed in remark~\ref{rmrk:turbTolerance}, with $\gamma {=} 1$.
The turbulent \texttt{pgdFoam} algorithm achieves convergence with eight velocity-pressure modes computed using \texttt{PGS-NS} and three corrections by means of \texttt{PGD-SA} and \texttt{PGD-$\nu_t$}. Each \texttt{PGD-SA} loop reached the prescribed tolerance within three computed modes. The relative amplitude of the computed modes, as the turbulent viscosity is updated, is reported in figure~\ref{fig:hump-ampl}. Following~\cite{Tsiolakis-TGSOH-20}, a measure of the relative amplitude accounting for both velocity and pressure modes is employed via the definition
\begin{equation}\label{eq:ampCombination}
\sigma_{(U,P)}^n := \left[
\left( \frac{\sigmaU^n}{\sum_{j=1}^{n} \sigmaU^j} \right)^2
+
\left( \frac{\sigmaP^n}{\sum_{j=1}^{n} \sigmaP^j} \right)^2
\right]^{\tfrac{1}{2}} .
\end{equation}
It is worth recalling that each time the relative amplitude of the modes drops by one order of magnitude, the separated representation of the turbulent viscosity is updated via the \texttt{PGD-SA} and \texttt{PGD-$\nu_t$} routines (see remark~\ref{rmrk:turbTolerance}) and the PGD approximation for velocity and pressure is recomputed using the updated turbulent viscosity. 

\begin{figure}[ht]
	\centering
	\subfigure[Influence of turbulent viscosity update]{\includegraphics[width=0.49\textwidth]{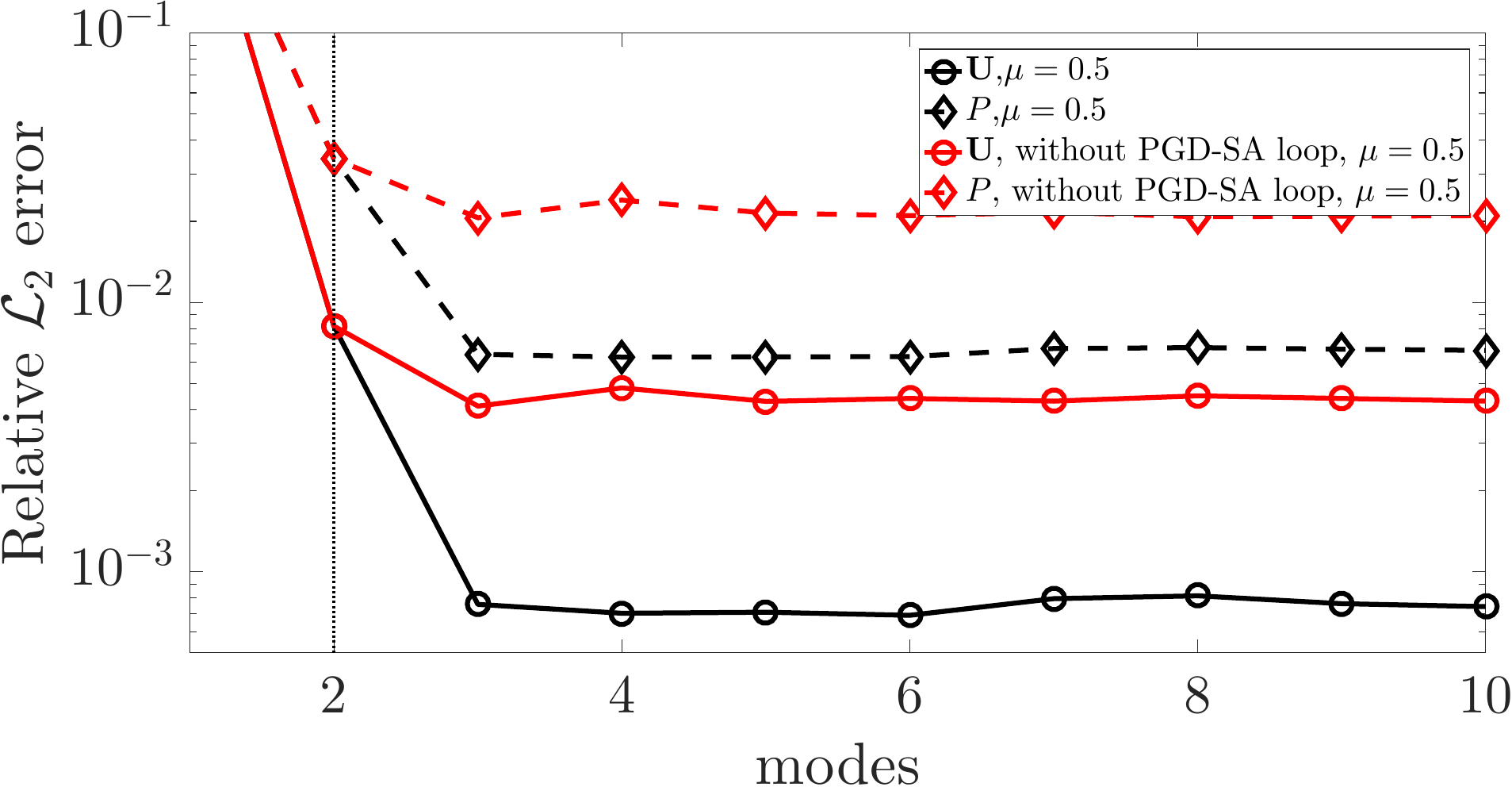}\label{fig:hump-L2-2D-PGD-SA-comparison}}
	\subfigure[Relative $\eltwo(\Omega)$ error]{\includegraphics[width=0.49\textwidth]{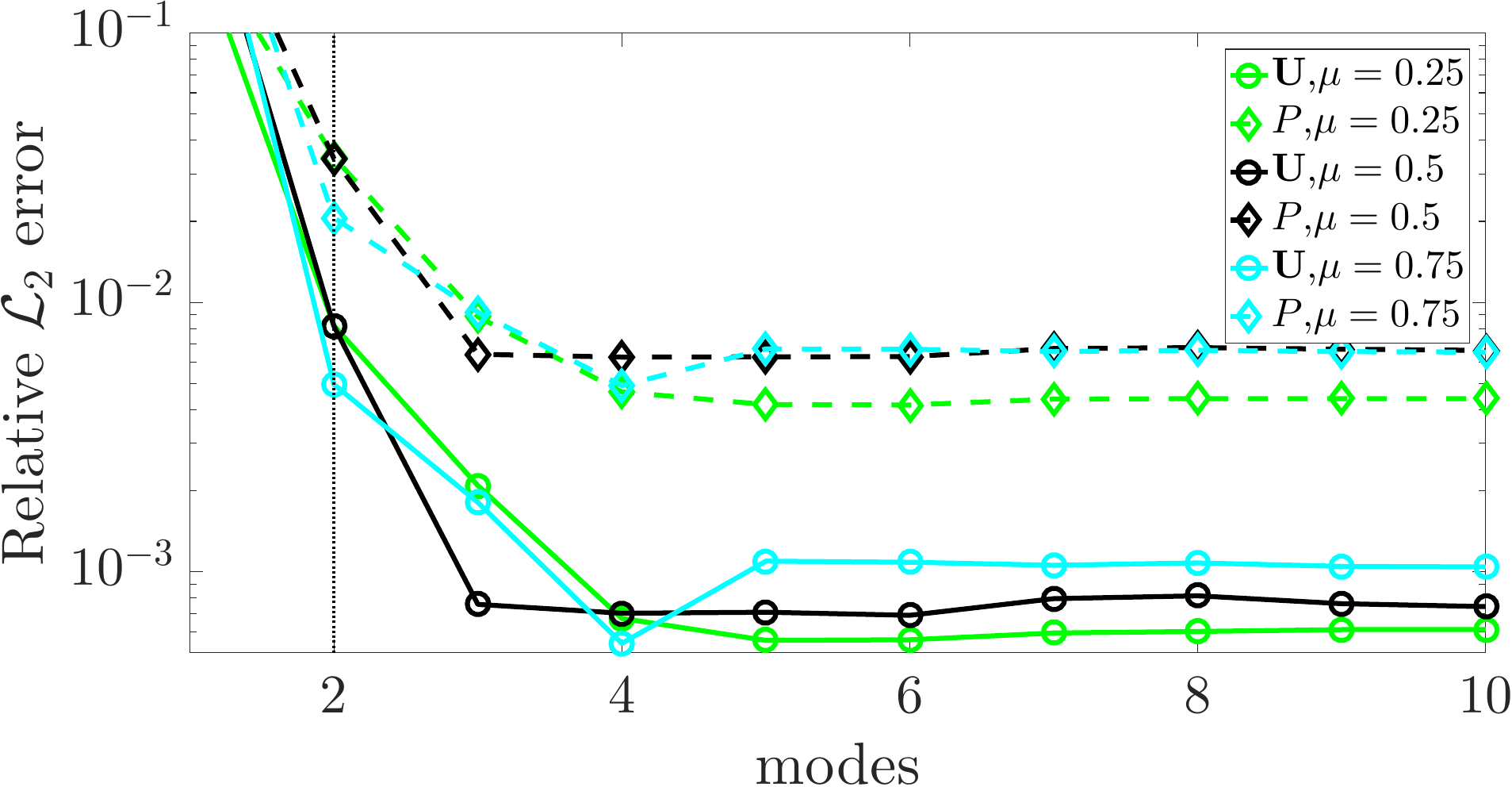}\label{fig:hump-L2-2D}}
	\caption{Accuracy of the PGD approximations of velocity and pressure with respect to the full order solutions as a function of the number of modes. (a) Influence of turbulent viscosity update for the case of $\mu {=} 0.5$. (b) Relative $\eltwo(\Omega)$ error for different values of $\mu$. The vertical lines separates the two boundary condition modes and the computed modes.}
	\label{fig:hump-L2}
\end{figure}

The importance of updating the PGD approximation of the turbulent viscosity to correctly compute the turbulent velocity and pressure fields is displayed in figure~\ref{fig:hump-L2-2D-PGD-SA-comparison}. Considering the result of the full order \texttt{simpleFoam} with SA turbulence model for $\mu {=} 0.5$ as a reference solution, figure~\ref{fig:hump-L2-2D-PGD-SA-comparison} compares the relative $\eltwo(\Omega)$ error of the PGD approximation computed via the \texttt{PGD-NS}, \texttt{PGD-SA} and \texttt{PGD-$\nu_t$} strategy described in algorithm~\ref{alg:PGD-RANS-SA-OF} with the one obtained by omitting the turbulent viscosity update. Without the \texttt{PGD-SA} and \texttt{PGD-$\nu_t$} routines, the error of both velocity and pressure approximations stagnates from the first computed mode and the overall value is one order of magnitude larger than the one achieved by the methodology in algorithm~\ref{alg:PGD-RANS-SA-OF}.
Figure~\ref{fig:hump-L2-2D} reports the relative $\eltwo(\Omega)$ error of the PGD approximation with turbulent viscosity update for three configurations, that is $\mu {=} 0.25$, $\mu {=} 0.5$ and $\mu {=} 0.75$. The results clearly display that the PGD approximation achieves comparable accuracy throughout the parametric interval $\I$ using two boundary condition modes and three computed modes. The following modes only introduce minor corrections to the solution as identified by their corresponding amplitudes, see figure~\ref{fig:hump-ampl}.

\begin{figure}[ht]
	\centering
	\begin{tabular}[c]{@{}c@{}c@{ }c@{ }c@{ }}
		$\upgd$ & 
		\parbox[c]{0.3\textwidth}{\includegraphics[width=0.3\textwidth]{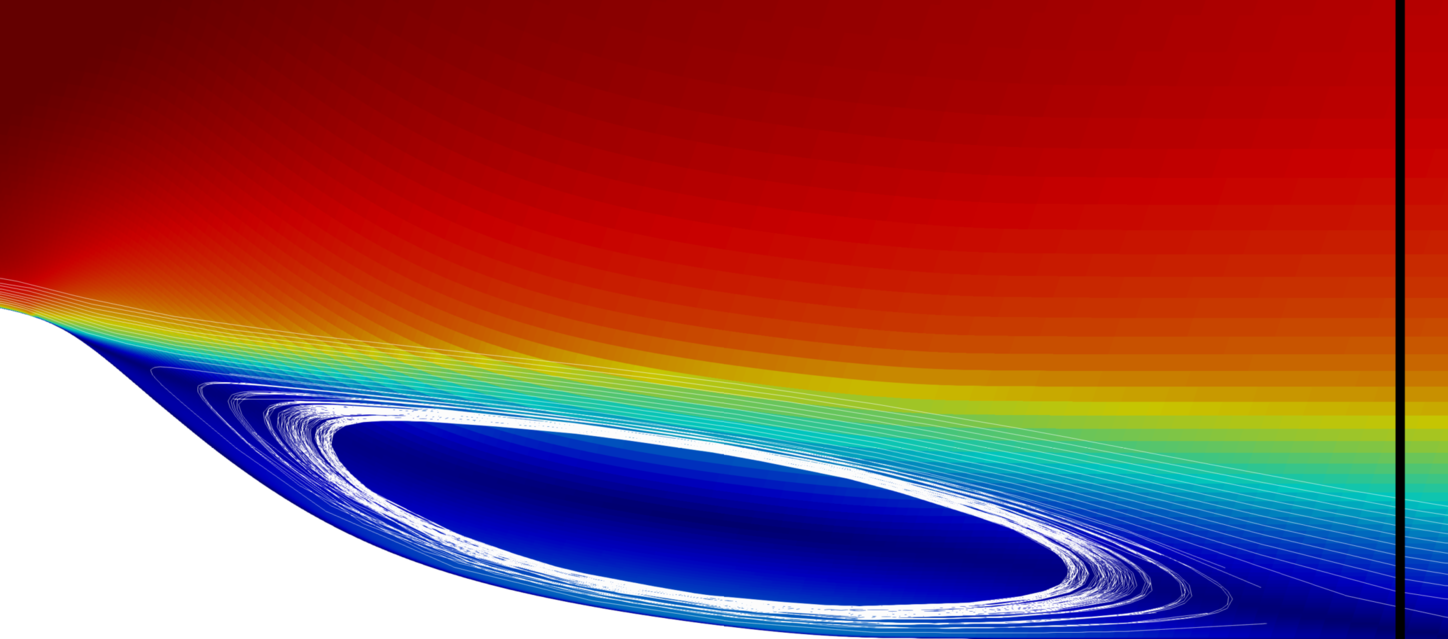}} &
		\parbox[c]{0.3\textwidth}{\includegraphics[width=0.3\textwidth]{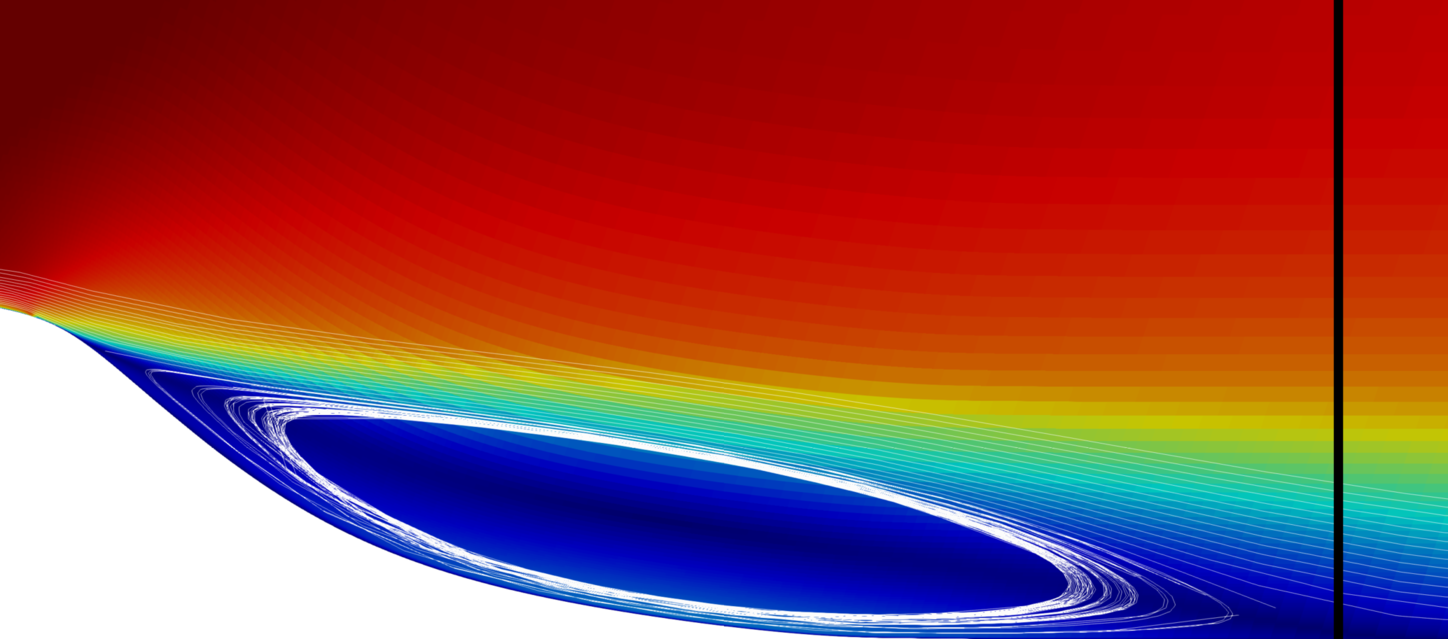}} &
		\parbox[c]{0.3\textwidth}{\includegraphics[width=0.3\textwidth]{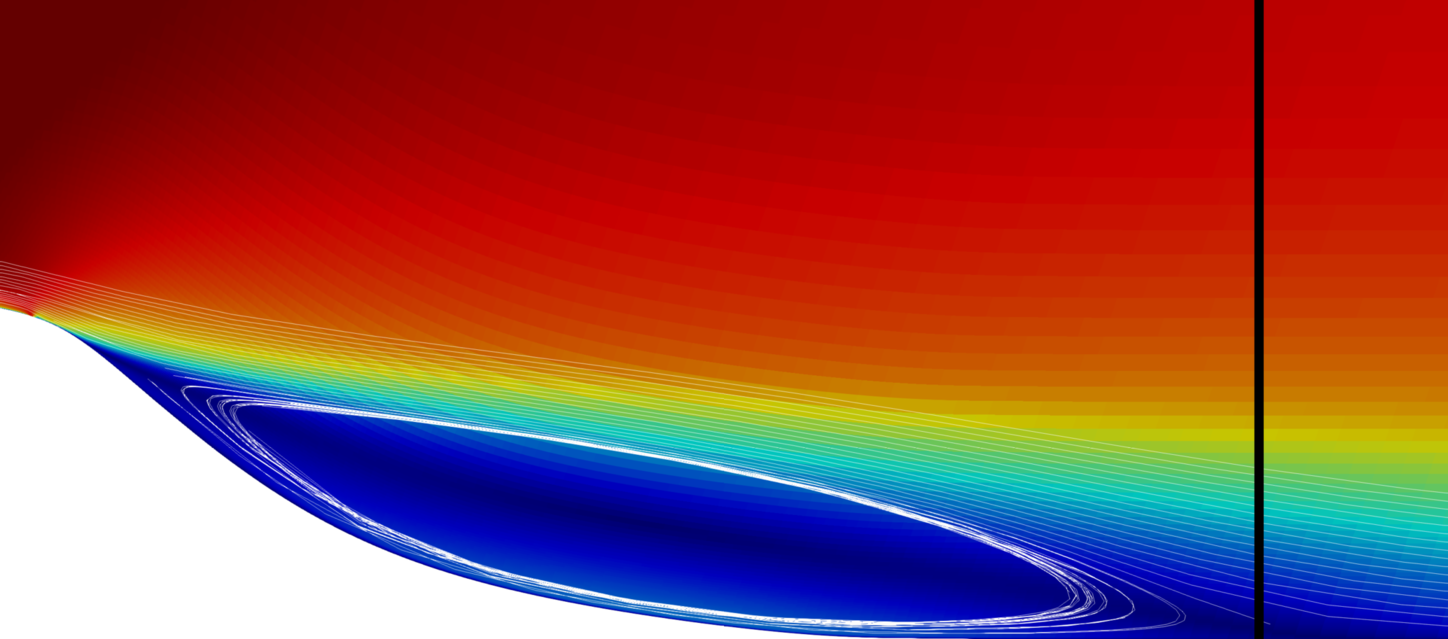}} \\[2em]
		$\uref$ &
		\parbox[c]{0.3\textwidth}{\includegraphics[width=0.3\textwidth]{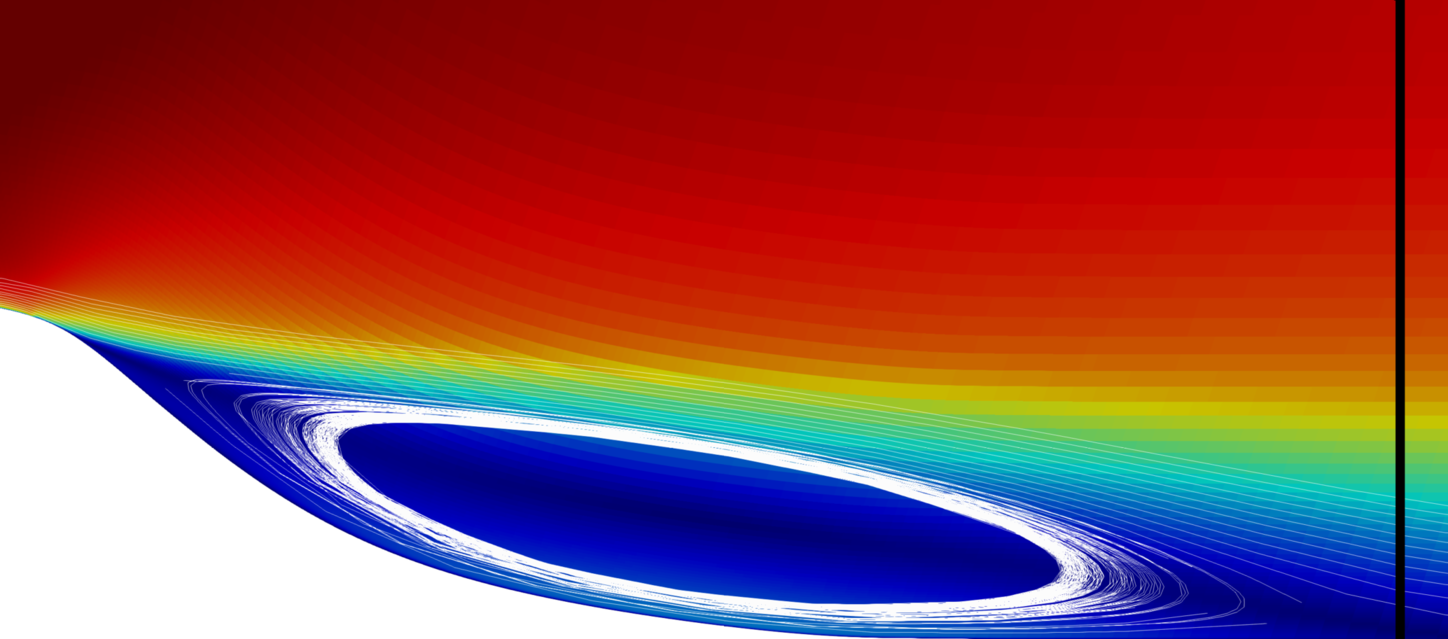}} &
		\parbox[c]{0.3\textwidth}{\includegraphics[width=0.3\textwidth]{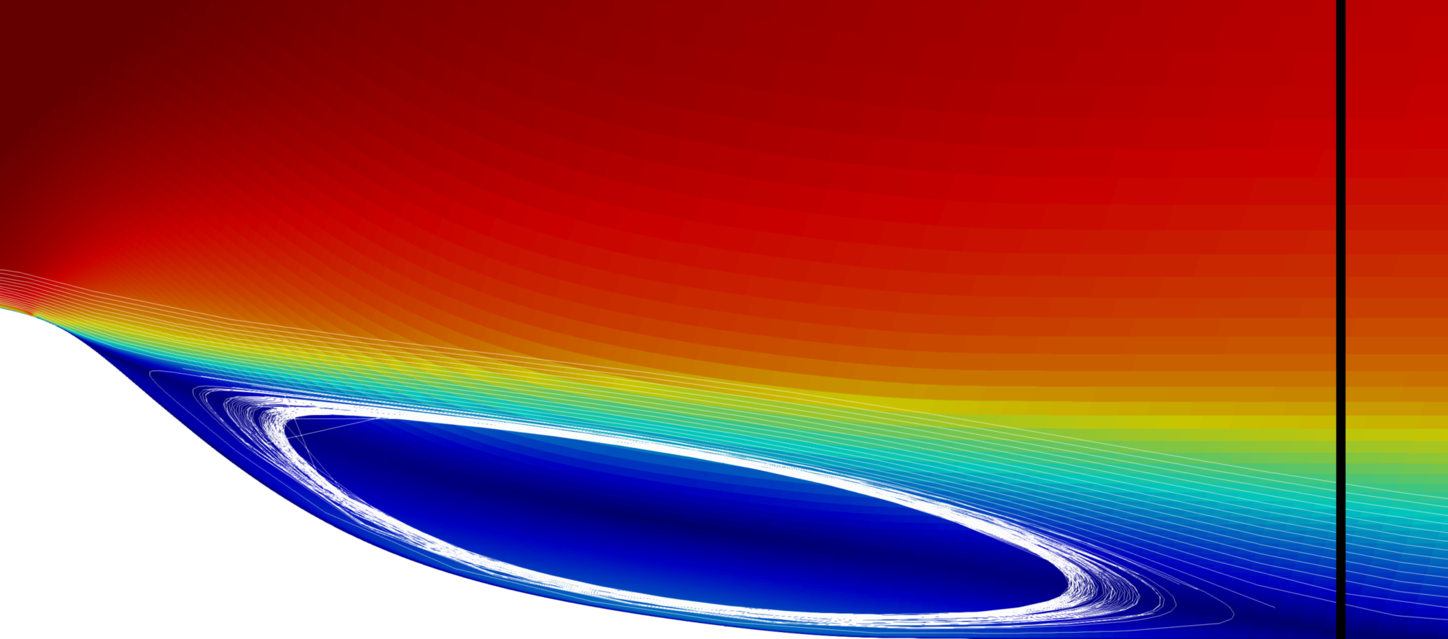}} &
		\parbox[c]{0.3\textwidth}{\includegraphics[width=0.3\textwidth]{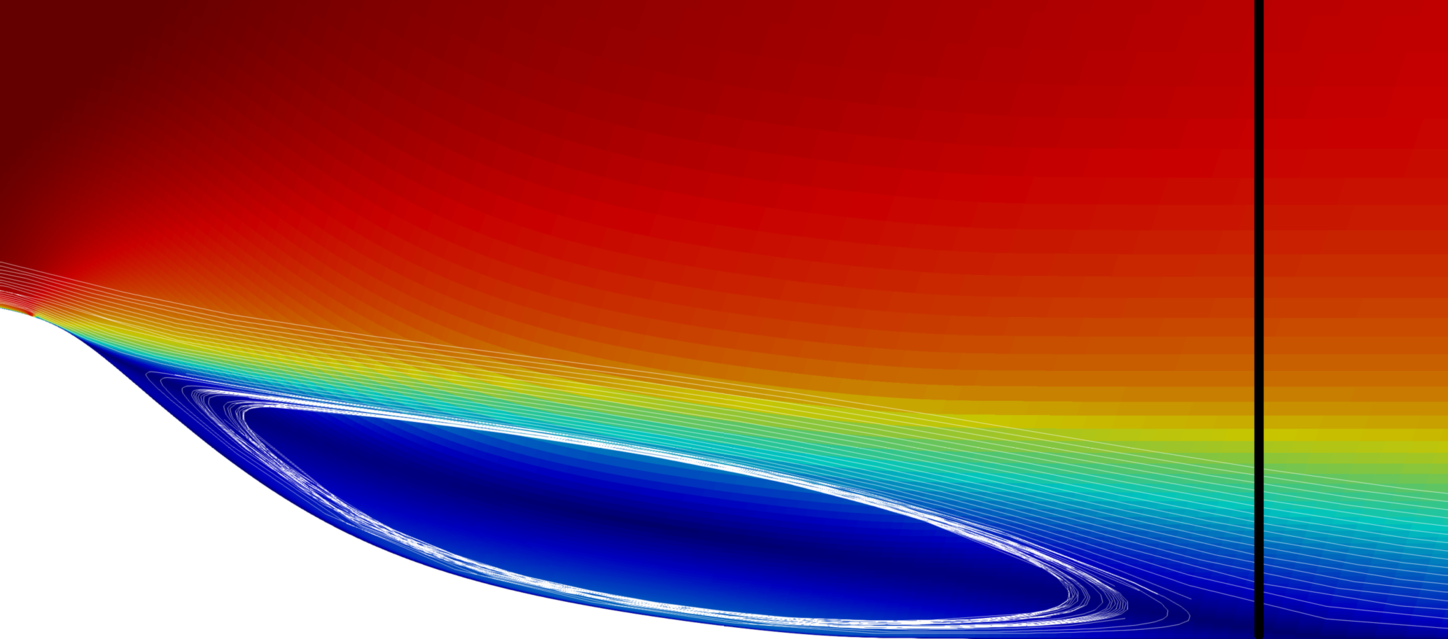}} \\
		  & $\mu{=}0.25$ & $\mu{=}0.5$ & $\mu{=}0.75$ \\
	\end{tabular}
	\includegraphics[width=0.6\textwidth]{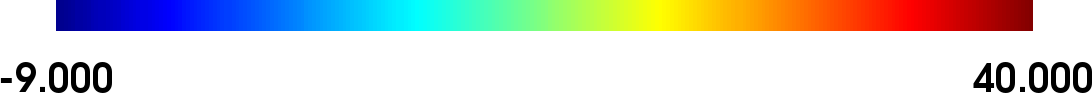}
\caption{Comparison of the PGD approximation (top) and the full order solution (bottom) of the recirculation bubble for $\mu{=}0.25$, $\mu{=}0.5$ and $\mu{=}0.75$. The vertical line denotes the position of the reattachment point.}
\label{fig:hump-velocity-2D}
\end{figure}

As mentioned in the problem statement, the quantities of interest in this study are the position of the reattachment point and the effect of the suction jet on the recirculation bubble.
Figure~\ref{fig:hump-velocity-2D} displays the velocity field after the hump and the recirculation bubble for three values of the parameter $\mu$. The influence of the jet in reducing the flow separation and moving the reattachment point towards the hump is well captured by the PGD approximation which is in excellent agreement with the full order solution.
\begin{figure}[ht]
	\centering
	\begin{tabular}[c]{@{}c@{}c@{ }c@{ }c@{ }}
		$\ppgd$ & 
		\parbox[c]{0.3\textwidth}{\includegraphics[width=0.3\textwidth]{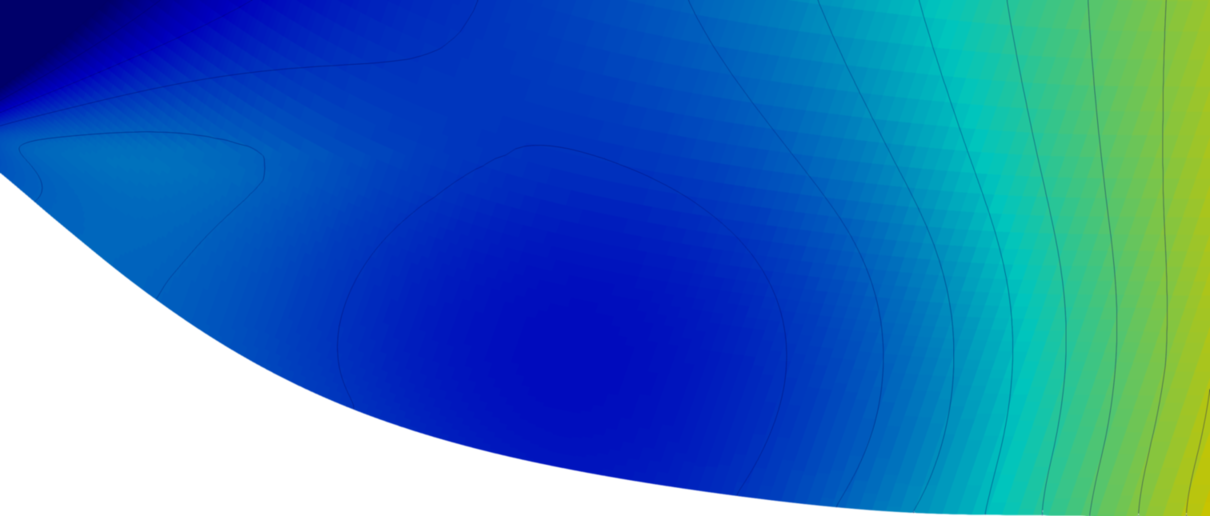}} &
		\parbox[c]{0.3\textwidth}{\includegraphics[width=0.3\textwidth]{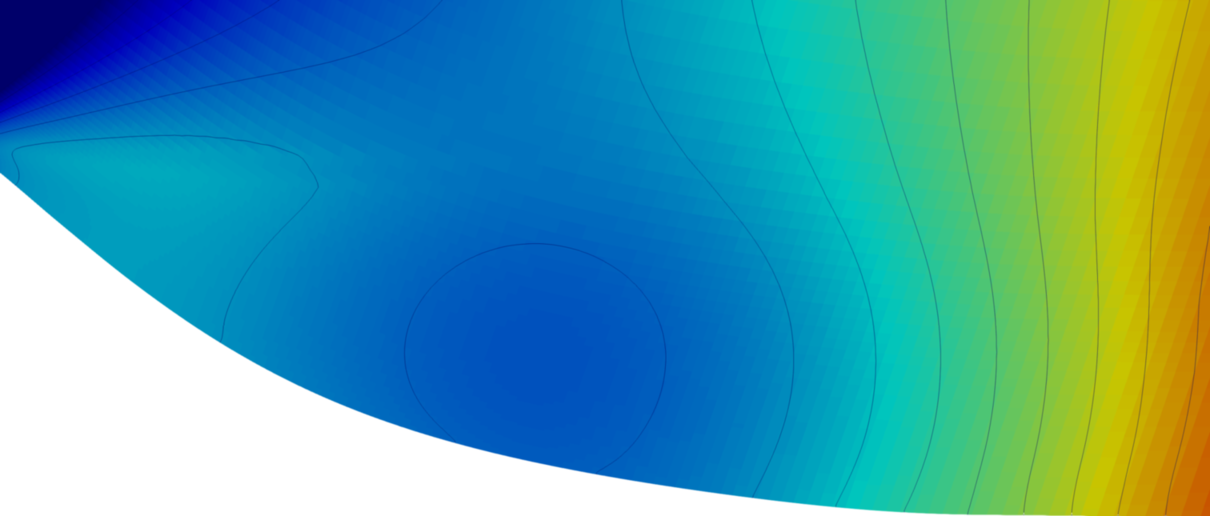}} &
		\parbox[c]{0.3\textwidth}{\includegraphics[width=0.3\textwidth]{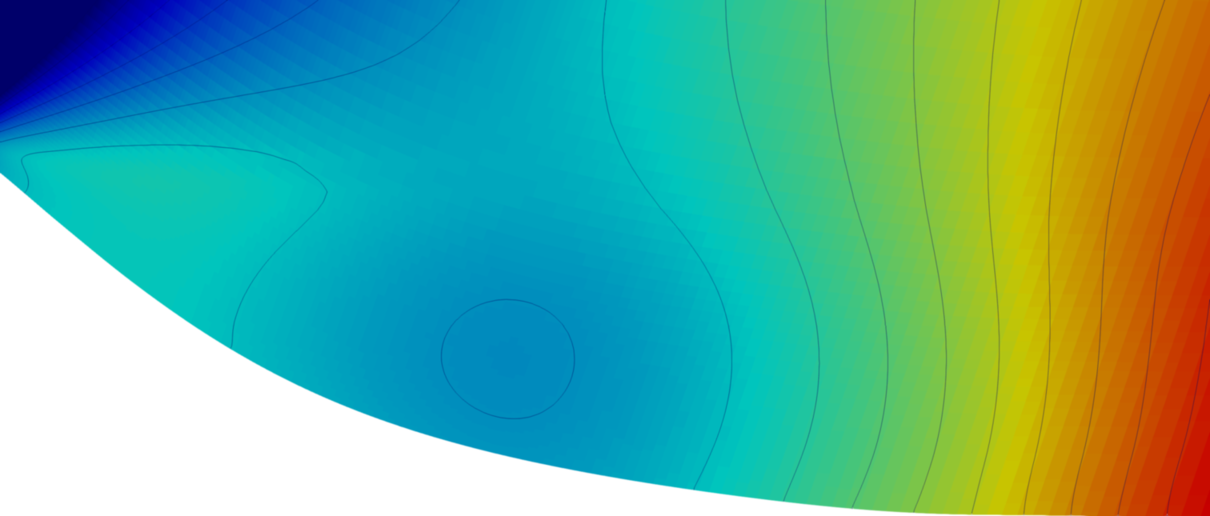}} \\[2em]
		$\pref$ &
		\parbox[c]{0.3\textwidth}{\includegraphics[width=0.3\textwidth]{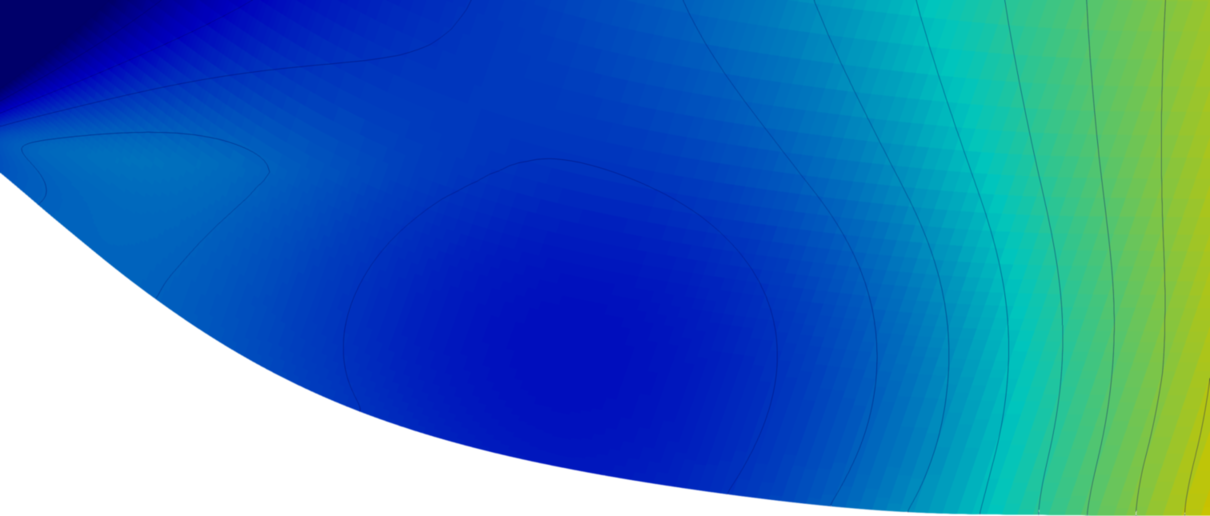}} &
		\parbox[c]{0.3\textwidth}{\includegraphics[width=0.3\textwidth]{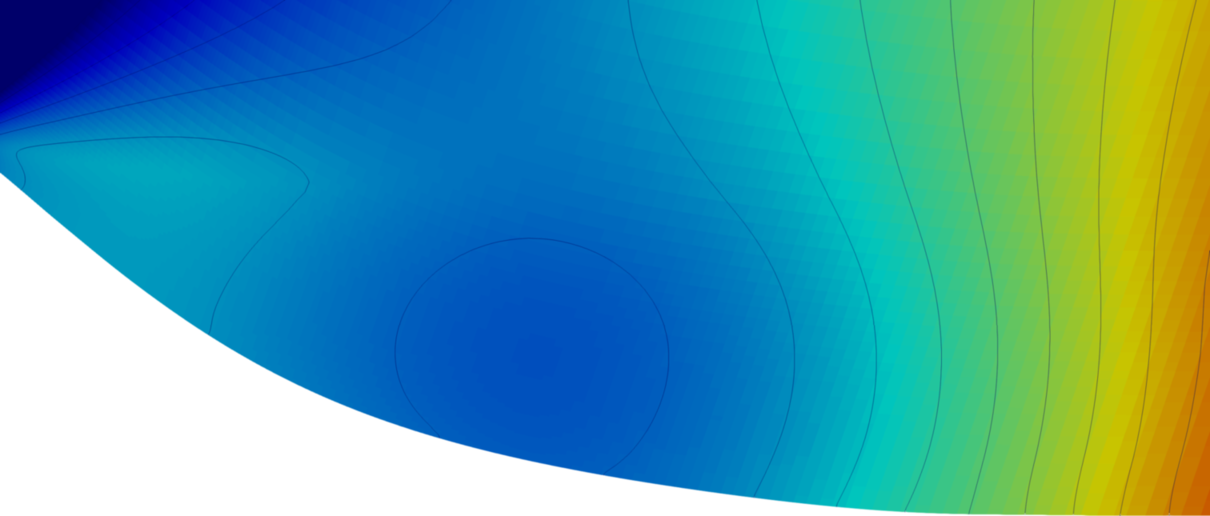}} &
		\parbox[c]{0.3\textwidth}{\includegraphics[width=0.3\textwidth]{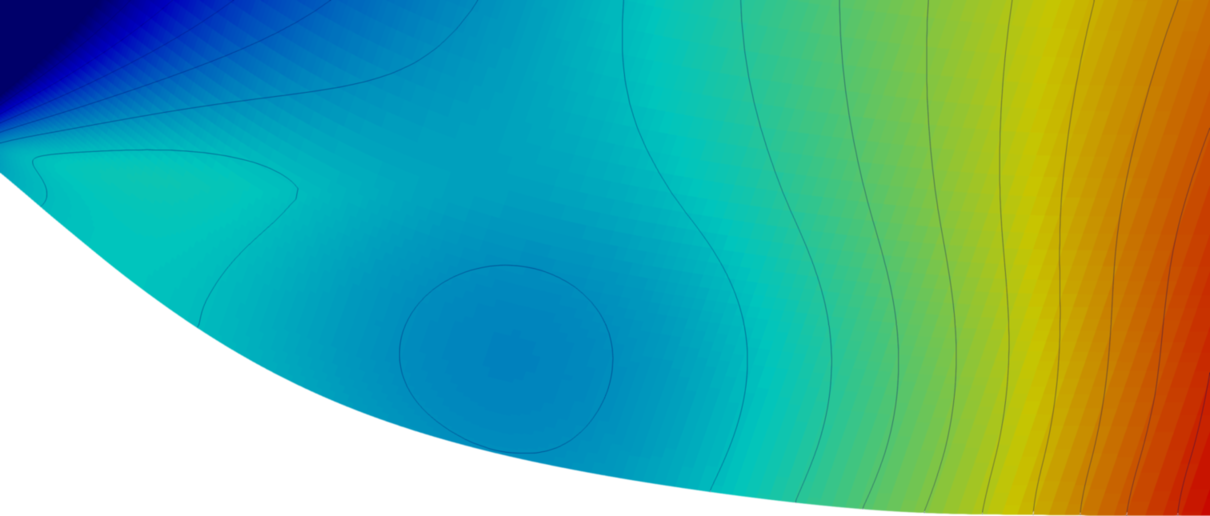}} \\
		  & $\mu{=}0.25$ & $\mu{=}0.5$ & $\mu{=}0.75$ \\
	\end{tabular}
	\includegraphics[width=0.6\textwidth]{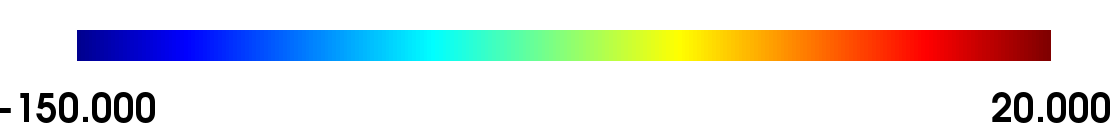}
\caption{Comparison of the PGD approximation (top) and the full order solution (bottom) of the pressure field after the hump for $\mu{=}0.25$, $\mu{=}0.5$ and $\mu{=}0.75$.}
\label{fig:hump-pressure-2D}
\end{figure}
\begin{figure}[ht]
	\centering
	\begin{tabular}[c]{@{}c@{}c@{ }c@{ }c@{ }}
		$\nu_{t_{\texttt{PGD}}}$ & 
		\parbox[c]{0.3\textwidth}{\includegraphics[width=0.3\textwidth]{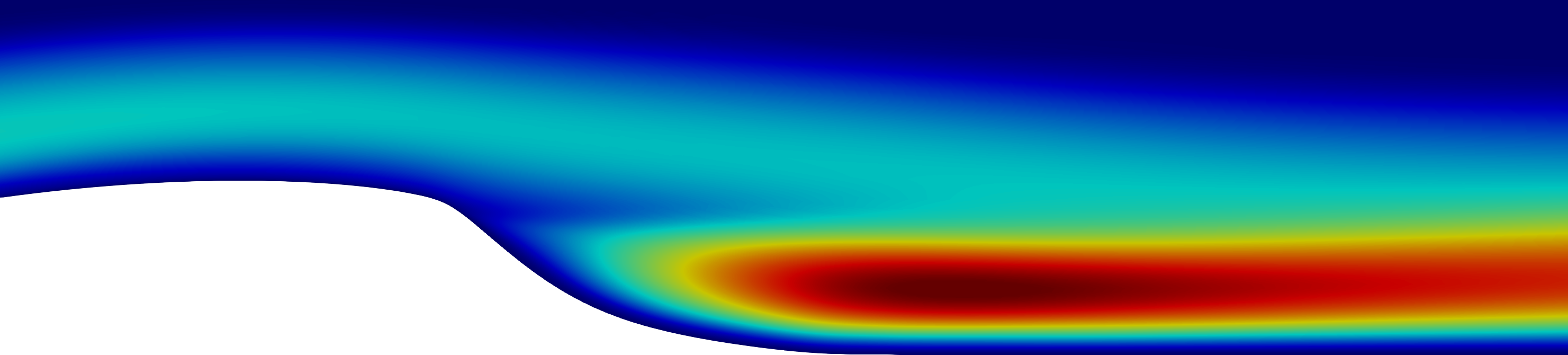}} &
		\parbox[c]{0.3\textwidth}{\includegraphics[width=0.3\textwidth]{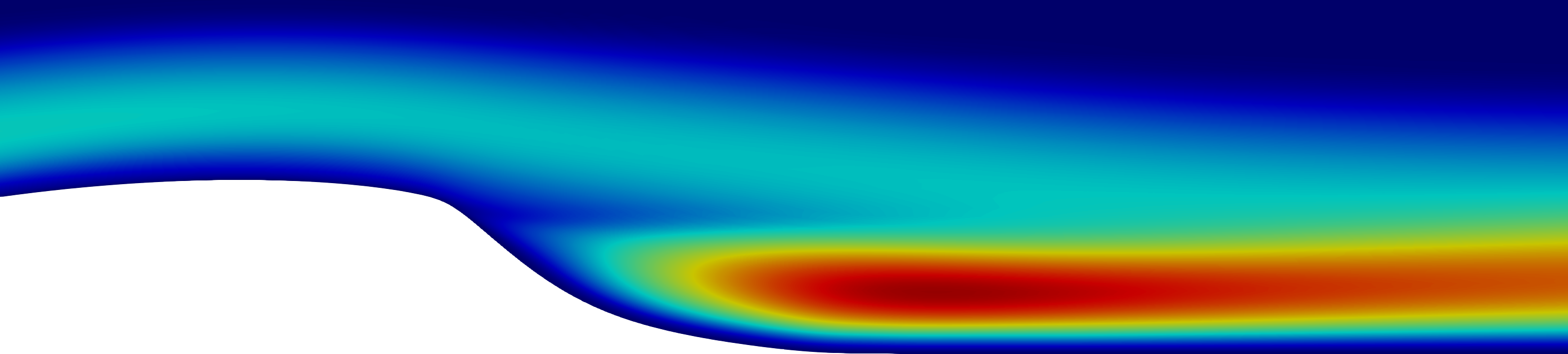}} &
		\parbox[c]{0.3\textwidth}{\includegraphics[width=0.3\textwidth]{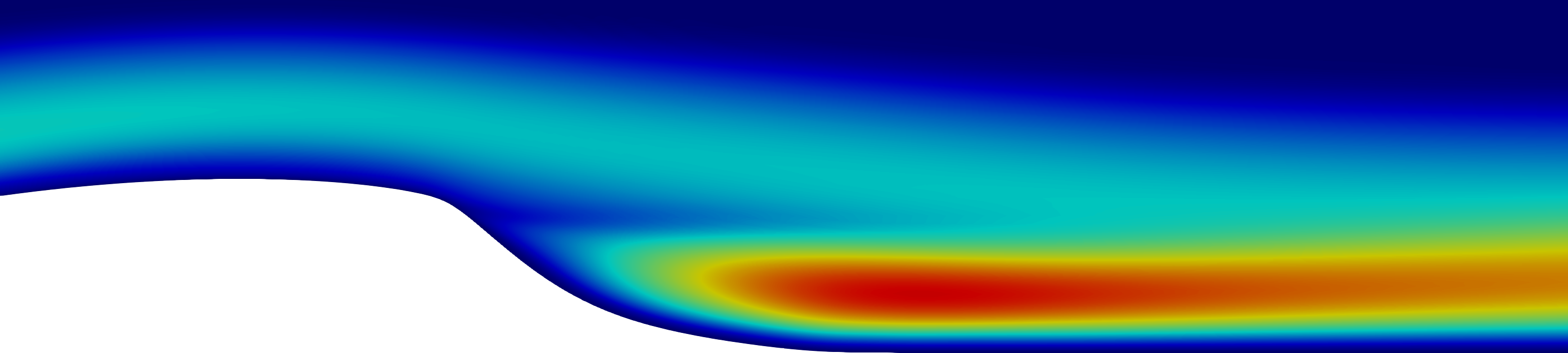}} \\[2em]
		$\nu_{t_{\texttt{REF}}}$ &
		\parbox[c]{0.3\textwidth}{\includegraphics[width=0.3\textwidth]{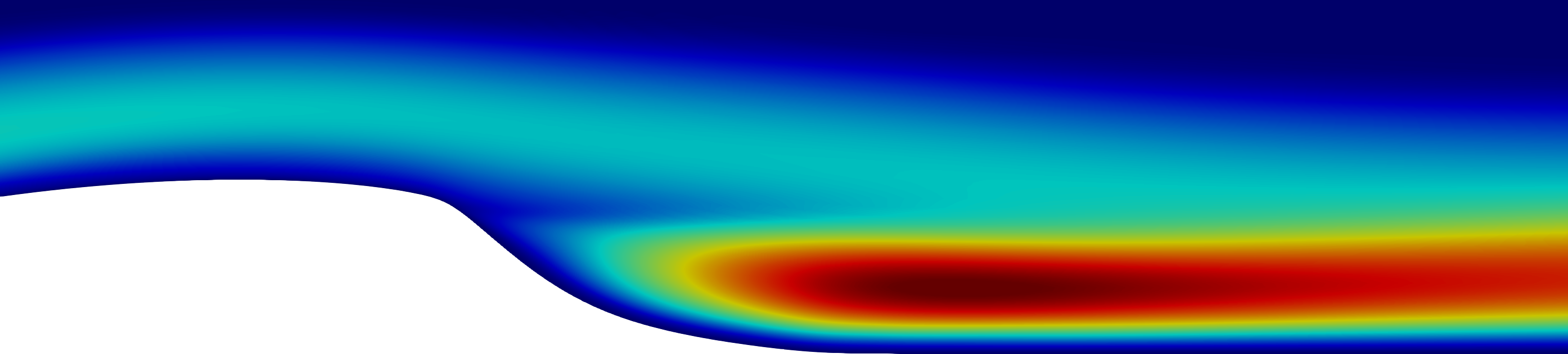}} &
		\parbox[c]{0.3\textwidth}{\includegraphics[width=0.3\textwidth]{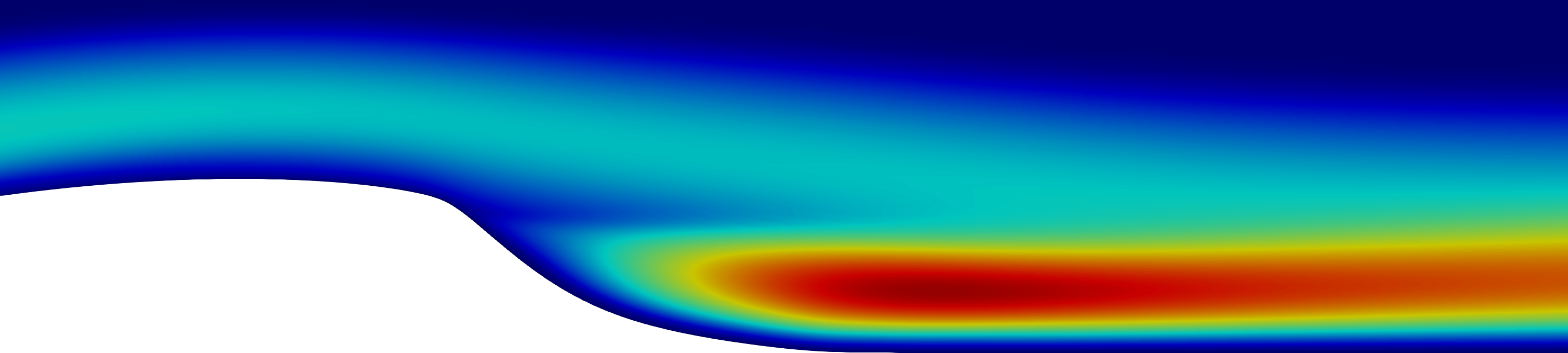}} &
		\parbox[c]{0.3\textwidth}{\includegraphics[width=0.3\textwidth]{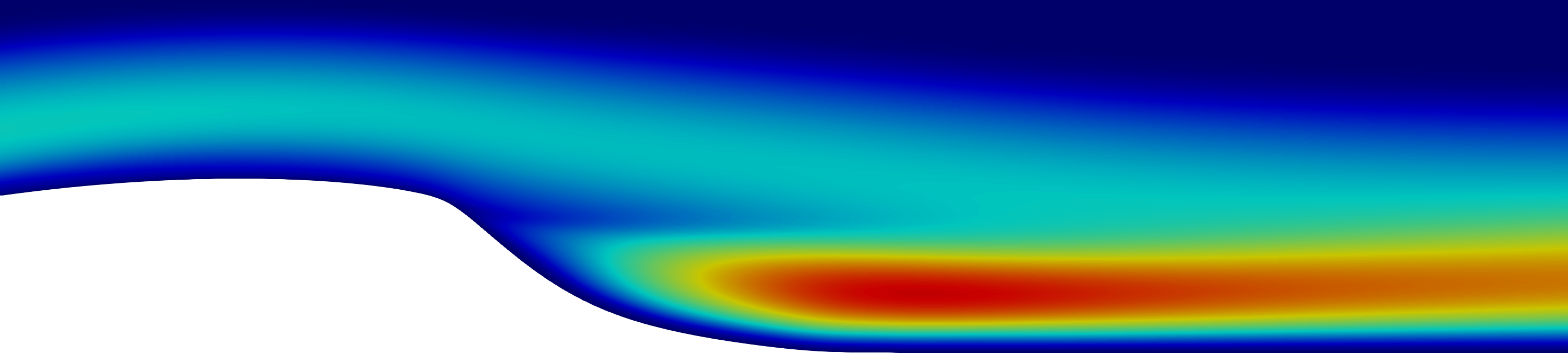}} \\
		  & $\mu{=}0.25$ & $\mu{=}0.5$ & $\mu{=}0.75$ \\
	\end{tabular}
	\includegraphics[width=0.6\textwidth]{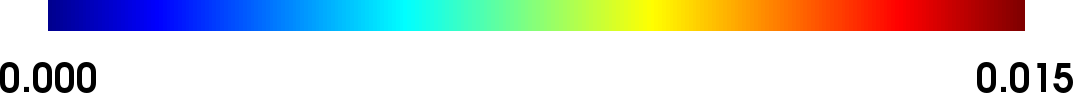}
\caption{Comparison of the PGD approximation (top) and the full order solution (bottom) of the turbulent viscosity after the hump for $\mu{=}0.25$, $\mu{=}0.5$ and $\mu{=}0.75$.}
\label{fig:hump-viscosity-2D}
\end{figure}

Qualitative comparisons of the pressure field and the turbulent viscosity for different values of the parameter $\mu$ are presented in figure~\ref{fig:hump-pressure-2D} and~\ref{fig:hump-viscosity-2D}, respectively. Using eight computed modes, the PGD approximation is able to accurately approximate localised variations in the flow pattern, throughout the interval $\I$.

In addition, the accuracy of the PGD-ROM is evaluated by quantitatively comparing quantities of engineering interest with the reference values provided by the OpenFOAM full order solver.
Table~\ref{tab:reattach-2D} reports the estimated position of the reattachment point using the reduced model and the corresponding value obtained using the full order \texttt{simpleFoam} solver with SA turbulence model. For the three values of the parameter $\mu$ reported, the online evaluations of the PGD-ROM with eight computed modes provide errors in the quantity of interest below $0.2\%$. The results show excellent agreement with the reference solution confirming the accuracy of the PGD approximation, even for the computation of surrogate models of relevant physical quantities.
\begin{table}[ht]
\begin{center}
	\begin{tabular}{|c||c|c|c|}
		\hline 
		$ \mu $	   &	$0.25$       &  $0.50$       &  $0.75$      \\
		\hline 
		  PGD          &	$1.183 c$  &  $1.156 c$   &  $1.129 c$ \\
		  full order &	$1.184 c$  &  $1.154 c$   &  $1.131 c$ \\
		  \hline 
		  Relative error         &  $0.84 \times 10^{-3}$ & $0.17 \times 10^{-2}$ & $0.17 \times 10^{-2}$ \\
		\hline 
	\end{tabular}
	\caption{2D NASA wall-mounted hump: position of the reattachment point computed using the PGD approximation and the full order solver for different values of $\mu$.}
	\label{tab:reattach-2D}
\end{center}
\end{table}

\begin{figure}[ht]
	\centering
	\subfigure[Skin friction coefficient]{\includegraphics[width=0.49\textwidth]{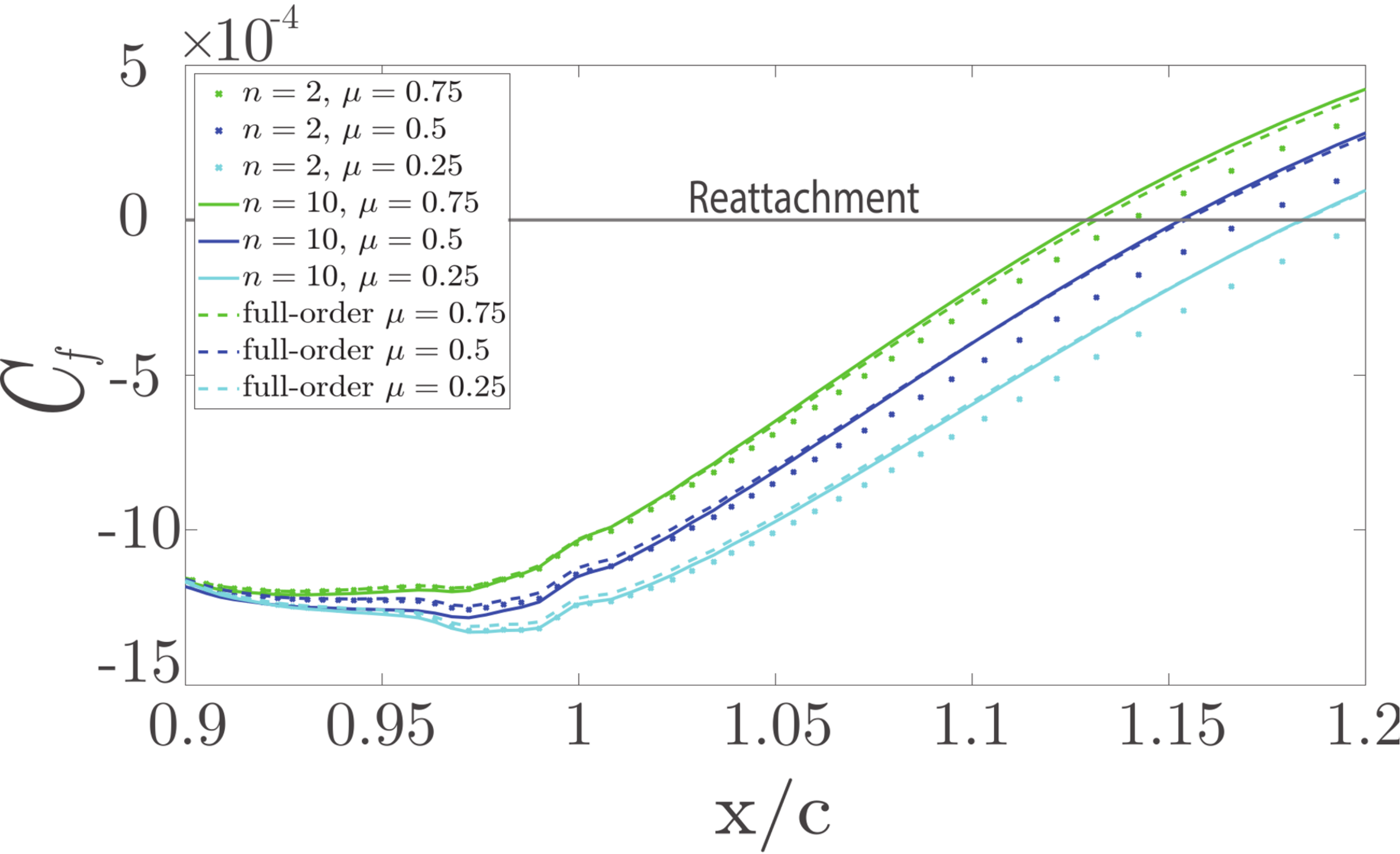}\label{fig:hump-Cf-2D}}
	\subfigure[Pressure coefficient]{\includegraphics[width=0.49\textwidth]{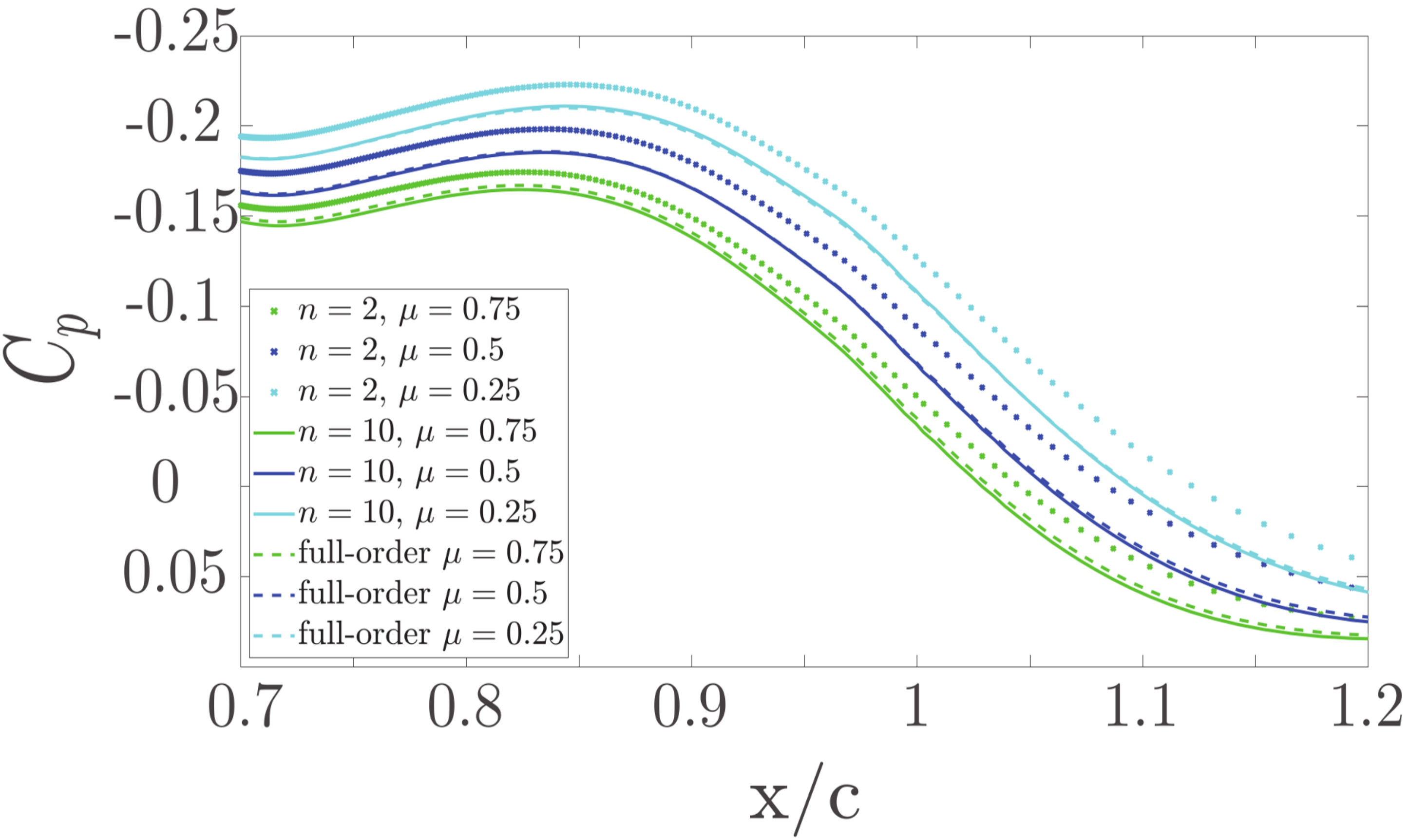}\label{fig:hump-Cp-2D}}
	\caption{Comparison of (a) the skin friction coefficient and (b) the pressure coefficient of the full order solution and the PGD approximation, for different number of PGD modes and for the three values of the parameter $\mu$.}
	\label{fig:hump-Cf-Cp}
\end{figure}

Finally, figure~\ref{fig:hump-Cf-2D} and~\ref{fig:hump-Cp-2D} report the skin friction coefficient and the pressure coefficient, respectively. Both figures focus on the area after the jet and compare the PGD approximation using different number of modes with the full order solution. 
First, it is worth noting that the PGD results based on the boundary condition modes, i.e. $n {=} 2$, provide approximations for the two quantities of interest which are qualitatively comparable with the full order solution.
Introducing eight computed modes ($n {=} 10$), the resulting PGD-based surrogate models for the skin friction coefficient and the pressure coefficient show perfect agreement with the full order solutions, confirming the capability of the proposed approach to construct accurate parametric representations of quantities of interest.

%________________________________________________________________________
\subsection{Three-dimensional NASA wall-mounted hump with parametrised jet}
\label{sc:simulation-3D}
%________________________________________________________________________

\begin{figure}
	\centering
	\includegraphics[width=0.8\textwidth]{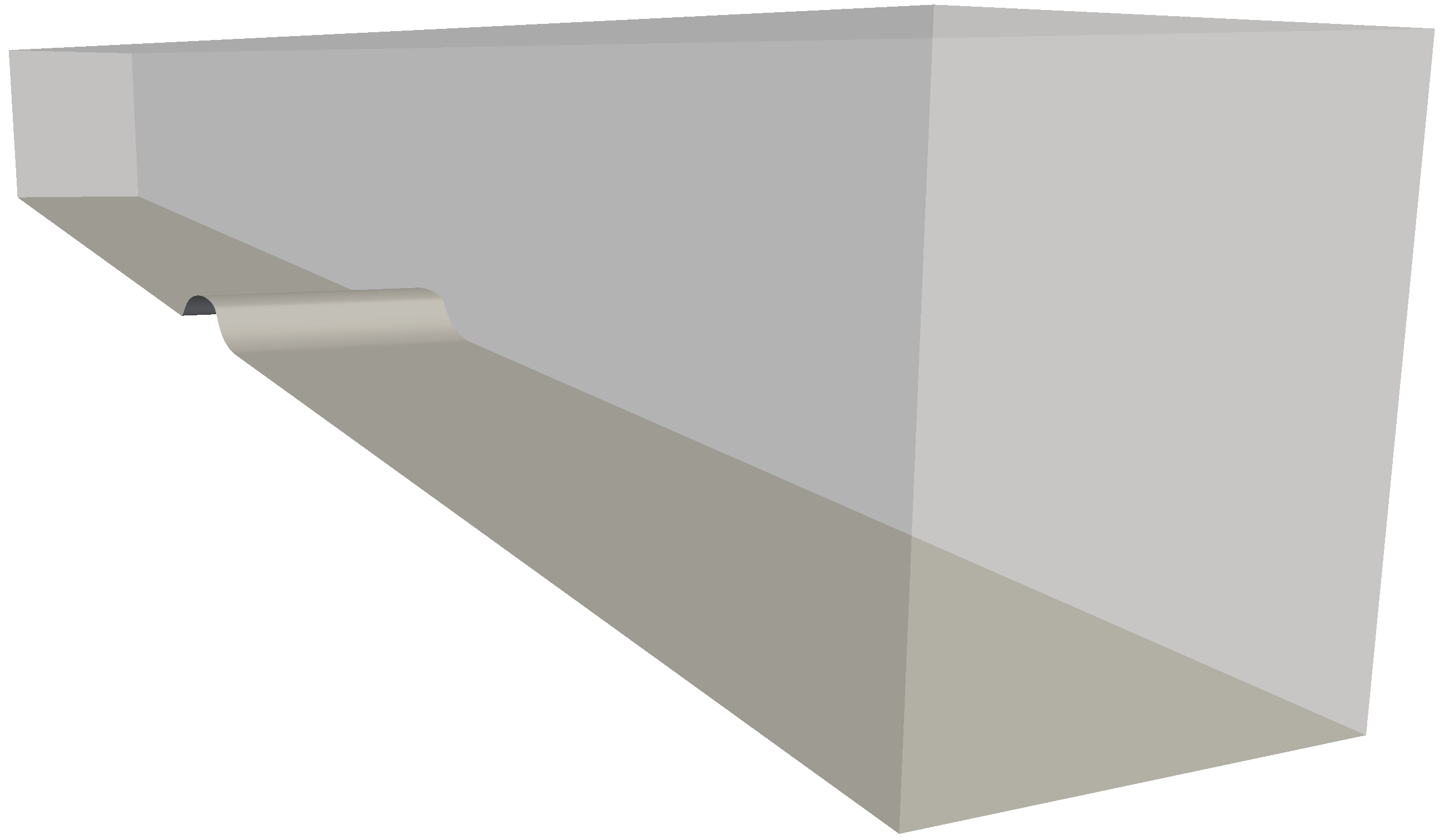}
	\caption{Computational domain for the three-dimensional NASA wall-mounted hump.}
	\label{fig:hump-domain-3D}
\end{figure}

The computational domain for the three-dimensional problem, see figure~\ref{fig:hump-domain-3D}, is obtained by extruding the 2D domain described in the previous section in the $z$ direction by $0.8$ chord lengths.
The problem inherits the set of boundary conditions utilised in the 2D case. On the additional external surfaces, slip boundary conditions are imposed. The peak value of the inlet velocity is set to $\SI{3.46}{m/s}$ and the profile of the jet suction is defined as in~\eqref{eq:Hump-jetProfile-3D} with $\hat{U} {=} \SI{2.34}{m/s}$. The kinematic viscosity being $\nu {=} 1.55274\times 10^{-5}\SI{}{m^2/s}$, the Reynolds number for the 3D case is $\text{Re} {=} 93,600$ and the computational mesh consists of $2.34$ million cells.

Similarly to the two-dimensional case, the boundary conditions are enforced using the two parametric modes in~\eqref{eq:parModesHump} and two spatial modes corresponding to the \texttt{simpleFoam} solutions with SA turbulence model for $\mu {=} 0.1$ and $\mu {=} 1$.

The values $\eta_u^\star {=} \eta_p^\star {=} 0.5 \times 10^{-3}$ and $\eta_{\nu}^\star {=} 10^{-2}$ are considered for the tolerance of the enrichment loops of the flow variables and the turbulent viscosity, respectively. To reduce the overall cost of the \texttt{PGD-NS}, \texttt{PGD-SA} and \texttt{PGD-$\nu_t$} procedure, the number of turbulent viscosity updates is reduced by considering a lower initial tolerance in criterion~\eqref{eq:criterionTurbUpdate}, namely $\gamma {=} 2$.

\begin{figure}[ht]
	\centering
	\includegraphics[width=0.6\textwidth]{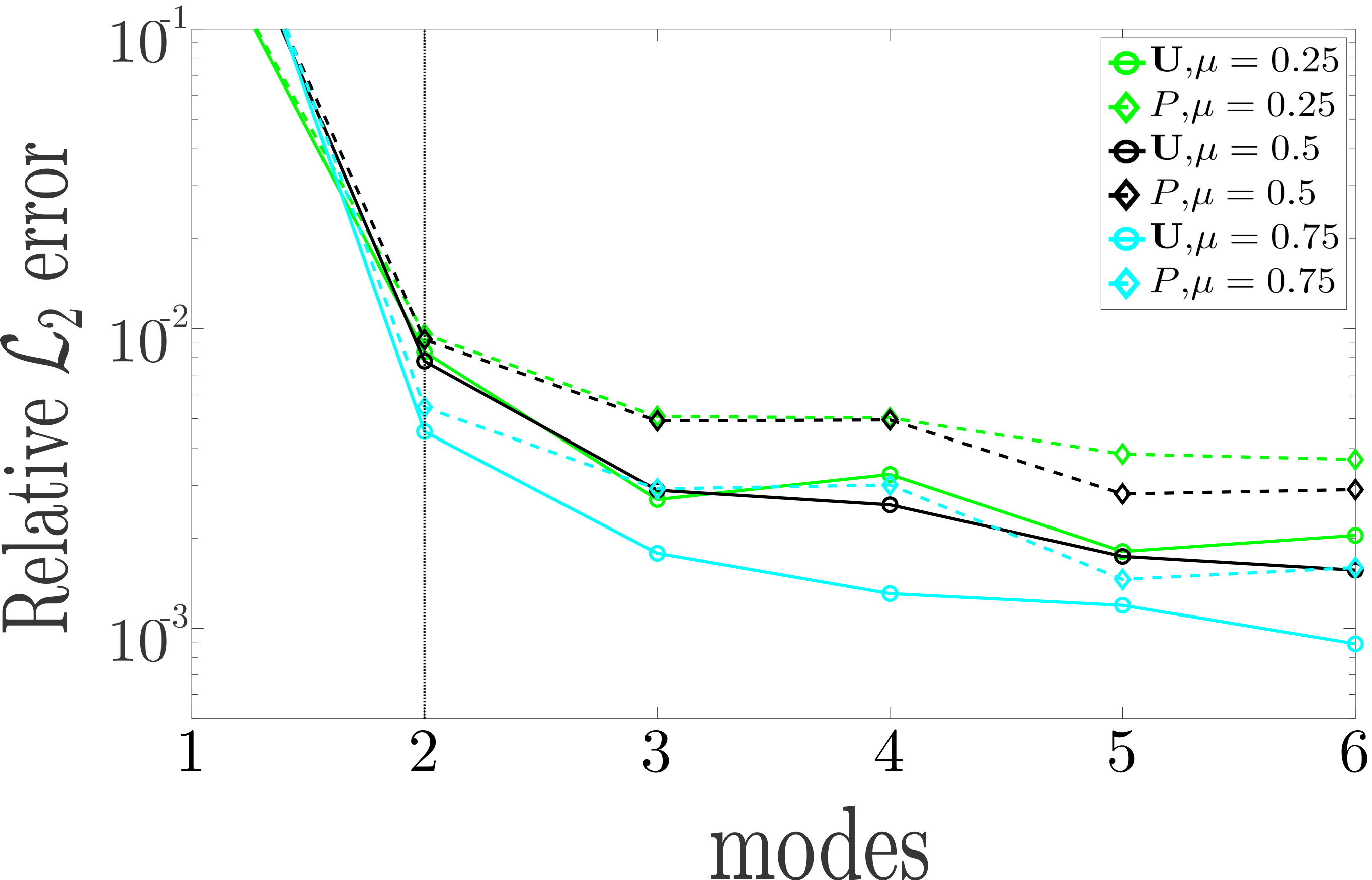}
	\caption{Relative $\eltwo(\Omega)$ error of the PGD approximations of velocity and pressure with respect to the full order solutions as a function of the number of modes for different values of $\mu$. The vertical line separates the two boundary condition modes and the computed modes.}
	\label{fig:hump-L2-3D}
\end{figure}

\begin{figure}[ht]
	\centering
	\subfigure[$\mu{=}0.25$]{\includegraphics[width=0.49\textwidth]{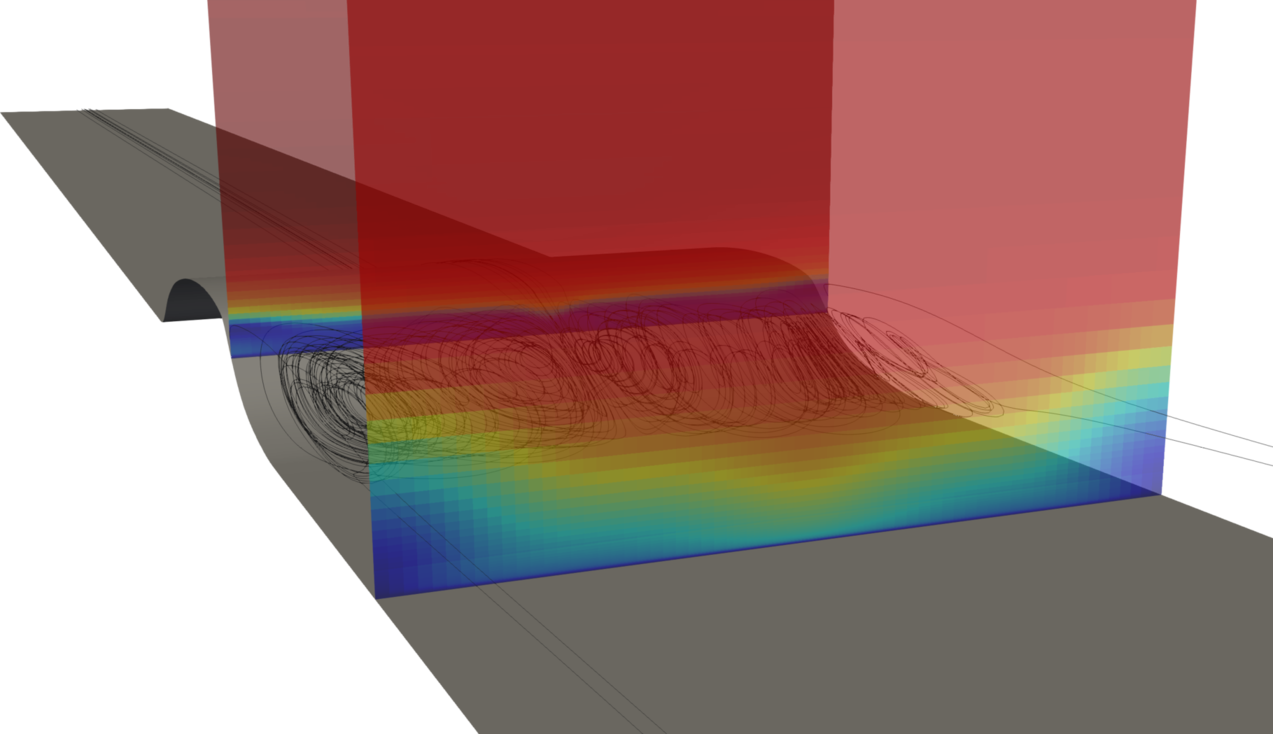}}
	\subfigure[$\mu{=}0.75$]{\includegraphics[width=0.49\textwidth]{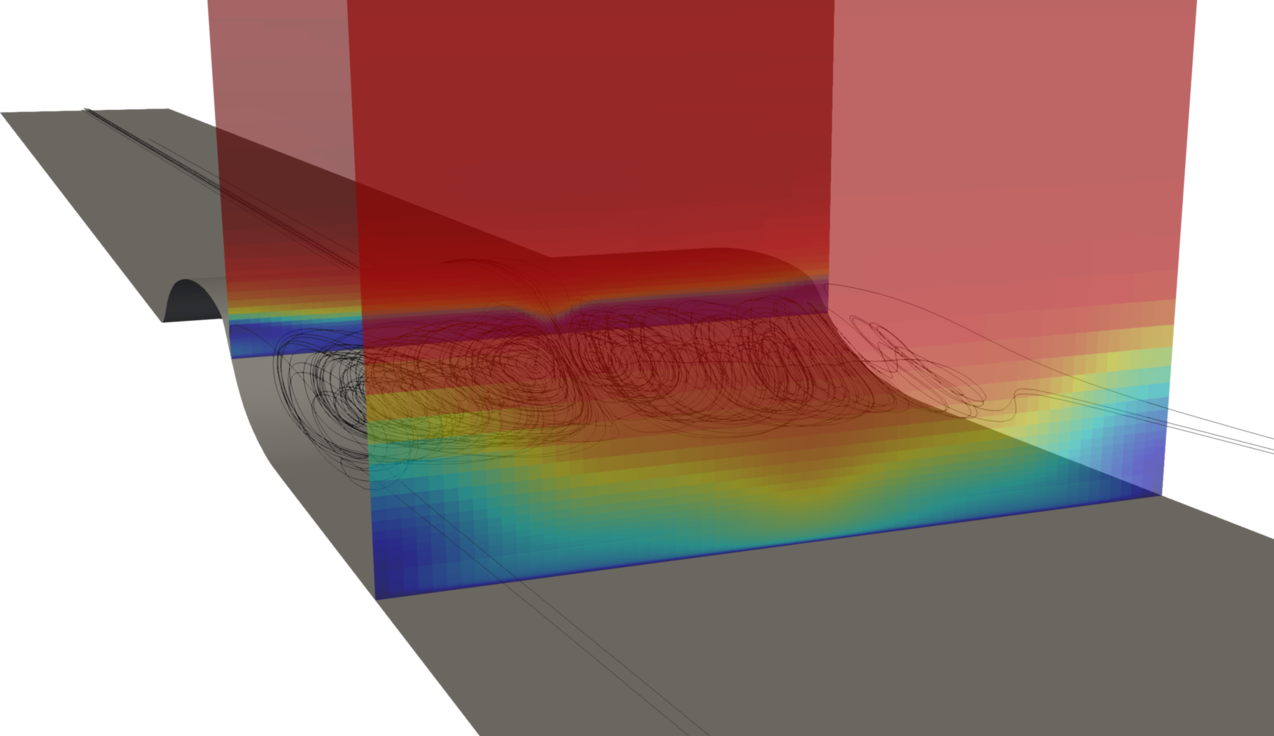}}
	\caption{Comparison of the PGD approximation of the velocity profile on the hump for $\mu{=}0.25$ and $\mu{=}0.75$.}
	\label{fig:hump-velocity-3D}
\end{figure}

\begin{figure}[ht]
	\centering
	\begin{tabular}[c]{@{}c@{}c@{ }c@{ }}
		$\upgd$ & 
		\parbox[c]{0.45\textwidth}{\includegraphics[width=0.45\textwidth]{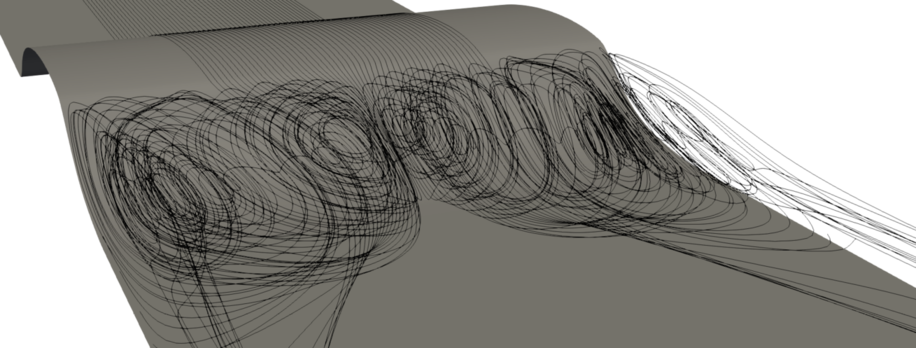}} &
		\parbox[c]{0.45\textwidth}{\includegraphics[width=0.45\textwidth]{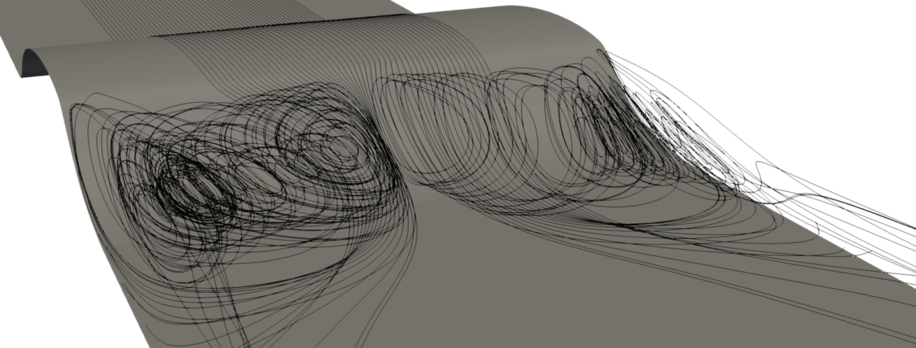}} \\[2.5em]
		$\uref$ &
		\parbox[c]{0.45\textwidth}{\includegraphics[width=0.45\textwidth]{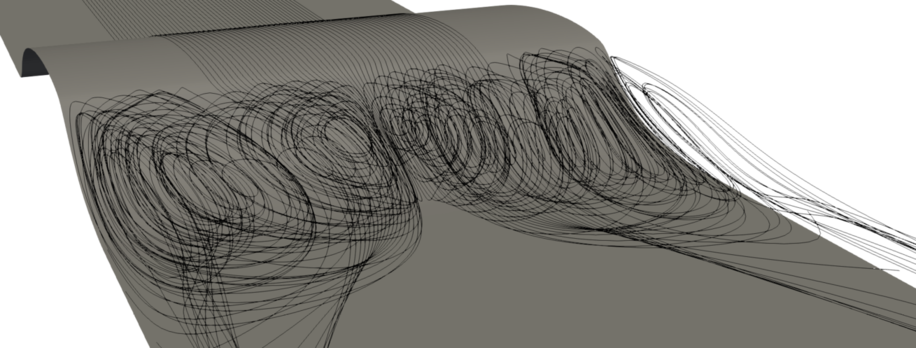}} &
		\parbox[c]{0.45\textwidth}{\includegraphics[width=0.45\textwidth]{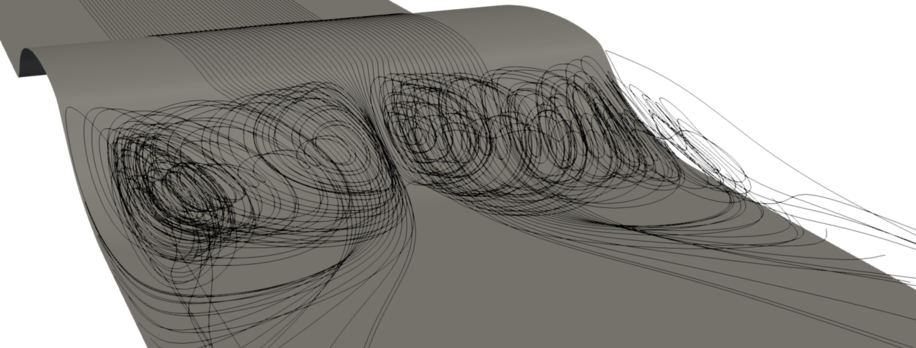}} \\
		  & $\mu{=}0.25$ & $\mu{=}0.75$ \\
	\end{tabular}
\caption{Detail of the vortex structure in the recirculation region computed using the PGD approximation (top) and the full order solution (bottom) for $\mu{=}0.25$ and $\mu{=}0.75$.}
	\label{fig:hump-streamlines-3D}
\end{figure}

Algorithm~\ref{alg:PGD-RANS-SA-OF} achieves convergence with four modes computed by the \texttt{PGS-NS} routine and two \texttt{PGD-SA} and \texttt{PGD-$\nu_t$} corrections. Each \texttt{PGD-SA} loop reached the prescribed tolerance within two computed modes. The PGD approximation is then compared with the corresponding full order solution provided by \texttt{simpleFoam} with the SA turbulence model: the relative $\eltwo(\Omega)$ error for $\mu {=} 0.25$, $\mu {=} 0.5$ and $\mu {=} 0.75$ is displayed in figure~\ref{fig:hump-L2-3D}, reporting that the reduced order model is able to provide errors in velocity and pressure below $0.1 \%$ and $0.5 \%$, respectively.
The resulting PGD-ROM is thus employed to analyse the physical phenomena involved in the turbulent flow over the hump. Figure~\ref{fig:hump-velocity-3D} displays the velocity profile on the hump, computed using the PGD, for different values of the parameter $\mu$. In addition, a qualitative comparison of the streamlines and the recirculation effects computed using the reduced model and the full order OpenFOAM solution are reported in figure~\ref{fig:hump-streamlines-3D}. The results display that the recirculation effects are reduced when increasing the suction jet and the PGD is able to capture the vortex structure with comparable accuracy with respect to the full order solution.

\begin{figure}[ht]
	\centering
	\begin{tabular}[c]{@{}c@{}c@{ }c@{ }c@{ }}
		$\btau_{\!\! w_{\texttt{PGD}}}$ & 
		\parbox[c]{0.3\textwidth}{\includegraphics[width=0.3\textwidth]{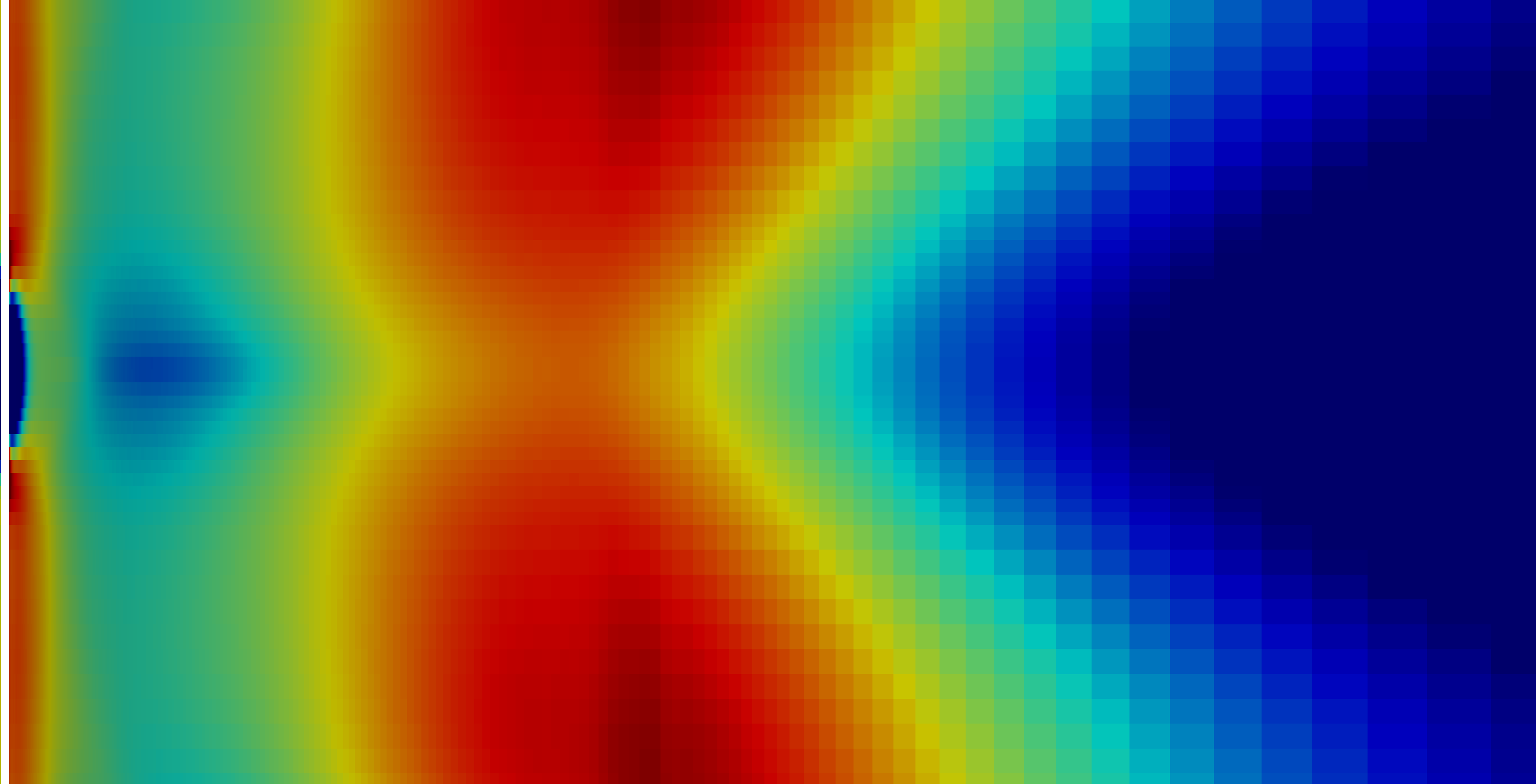}} &
		\parbox[c]{0.3\textwidth}{\includegraphics[width=0.3\textwidth]{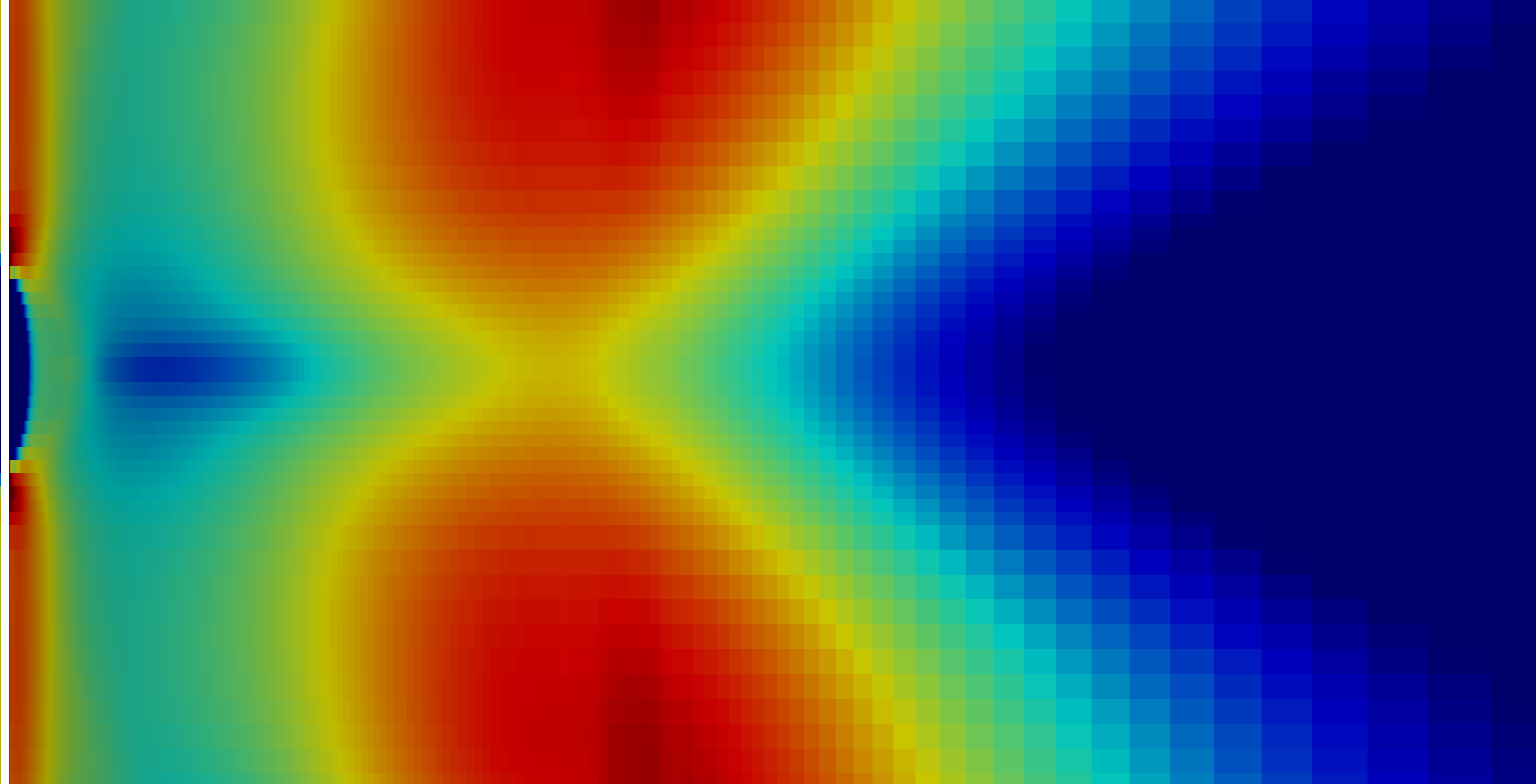}} &
		\parbox[c]{0.3\textwidth}{\includegraphics[width=0.3\textwidth]{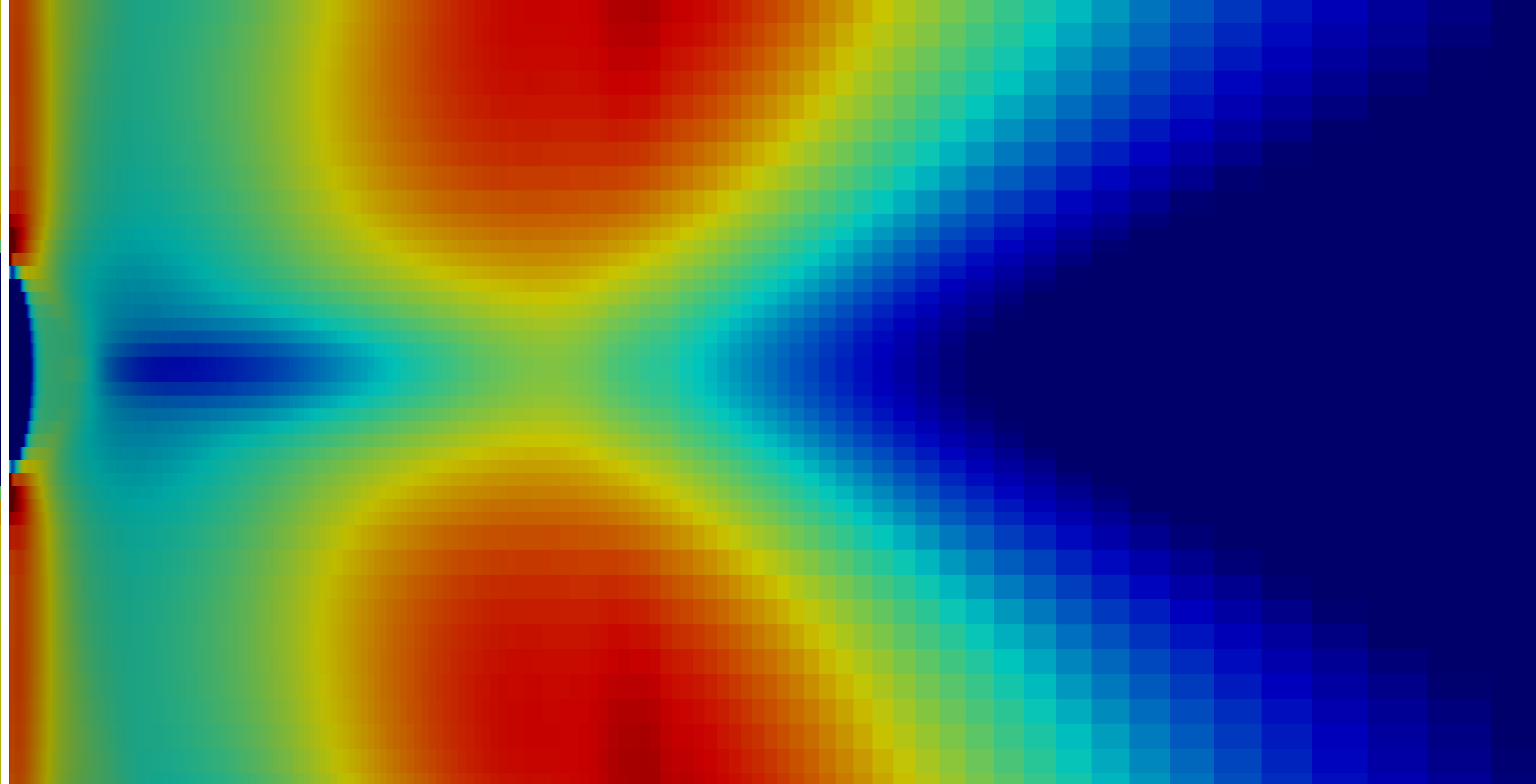}} \\[2em]
		$\btau_{\!\! w_{\texttt{REF}}}$ &
		\parbox[c]{0.3\textwidth}{\includegraphics[width=0.3\textwidth]{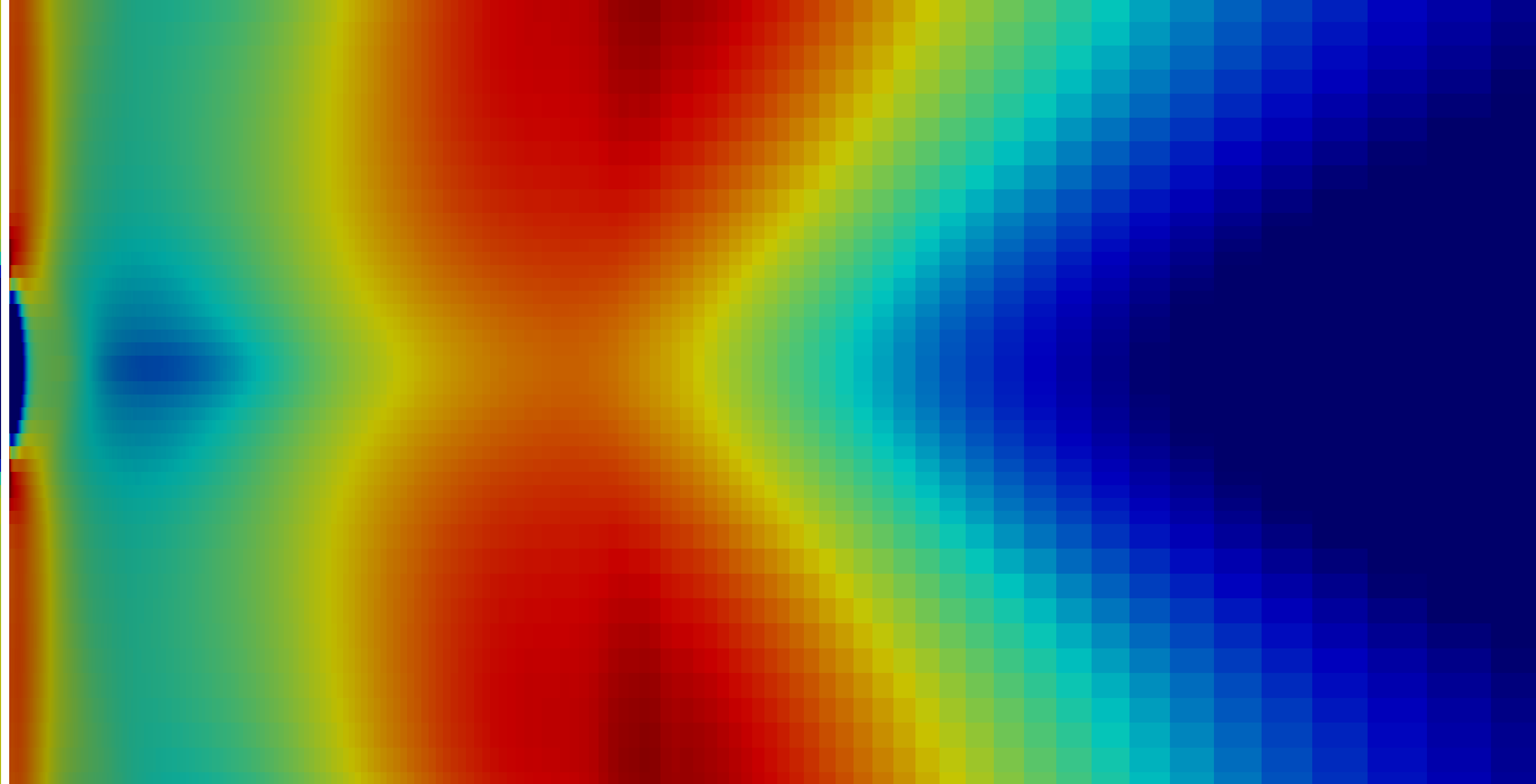}} &
		\parbox[c]{0.3\textwidth}{\includegraphics[width=0.3\textwidth]{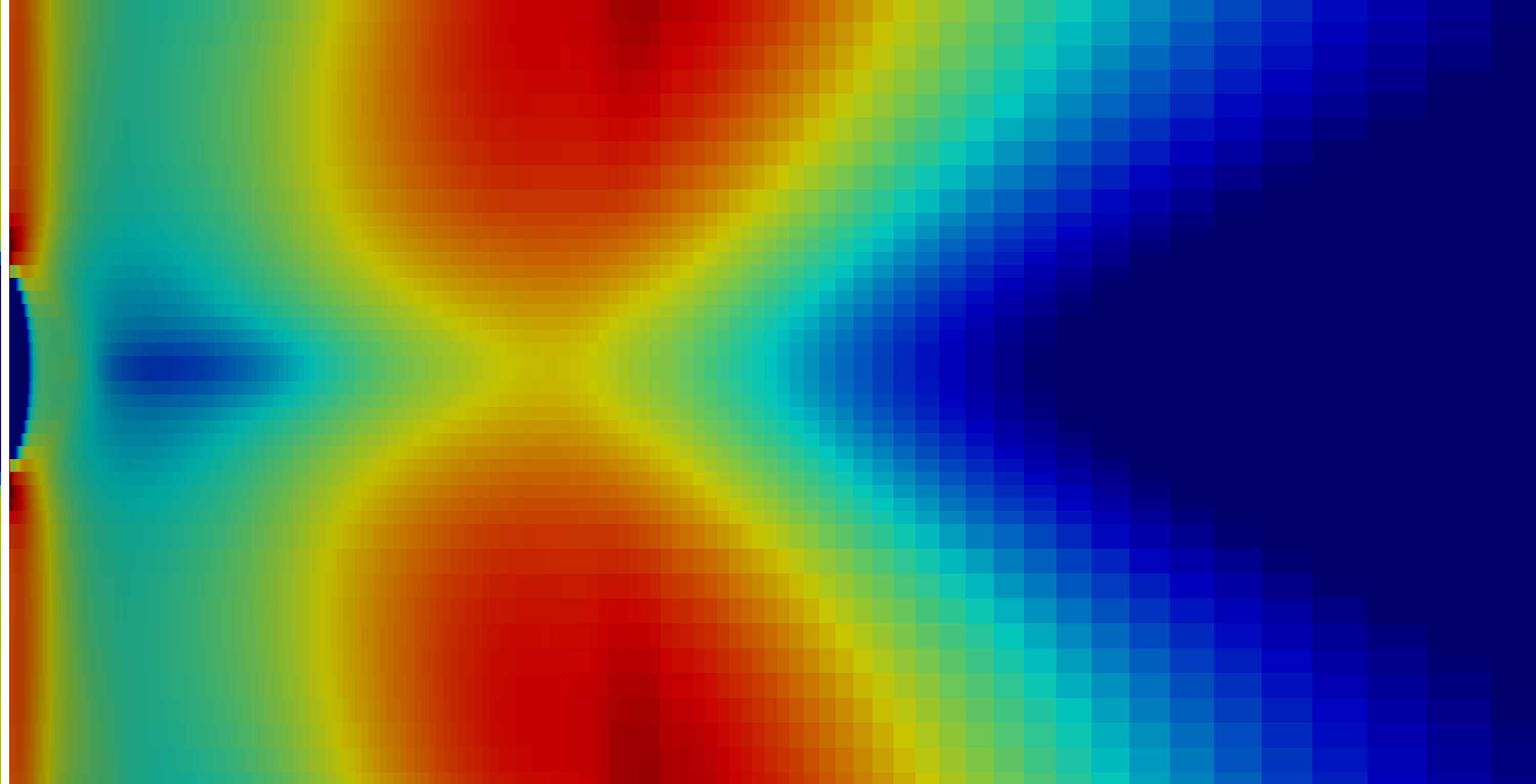}} &
		\parbox[c]{0.3\textwidth}{\includegraphics[width=0.3\textwidth]{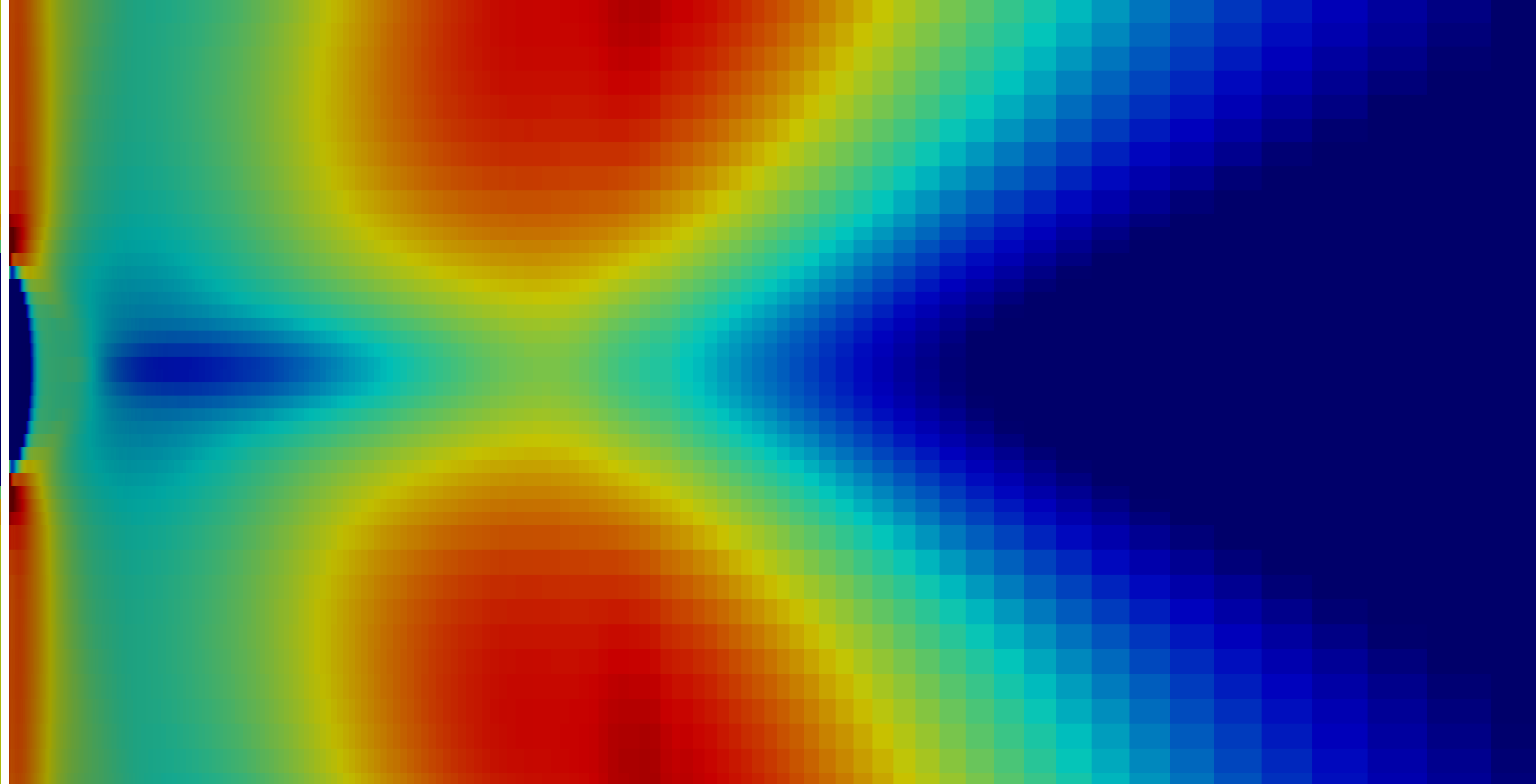}} \\
		  & $\mu{=}0.25$ & $\mu{=}0.5$ & $\mu{=}0.75$ \\
	\end{tabular}
	\includegraphics[width=0.6\textwidth]{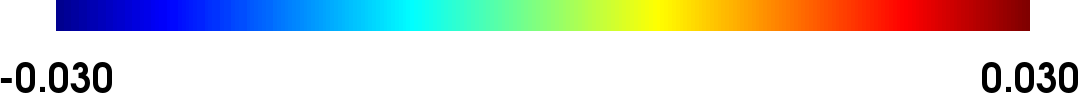}
\caption{Comparison of the PGD approximation (top) and the full order solution (bottom) of the wall shear stress on the bottom wall for $\mu{=}0.25$, $\mu{=}0.5$ and $\mu{=}0.75$. Detail of the region starting from the jet patch up to $1.6 c$ downstream.}
	\label{fig:hump-Cf-3D}
\end{figure}

The capability of the proposed PGD-ROM strategy to treat complex problems of engineering interest is thus confirmed by the following analysis focusing on relevant physical quantities. In figure~\ref{fig:hump-Cf-3D}, the top view of the wall shear stress on the bottom wall is reported in the region from the jet patch up to $1.6 c$ downstream, highlighting the effect of the suction jet on flow recirculation.
A qualitative comparison between the reduced order and the full order solution confirms the ability of the PGD to accurately reproduce the turbulent flow in the entire range of values $\I$ of the parameter. In particular, these conclusions hold true both for the flow variables and for given physical quantities computed starting from them.

Finally, the position of the reattachment point in correspondance of the location of the peak of the jet profile is reported in table~\ref{tab:reattach-3D} for different values of the parameter $\mu$. For each tested configuration, the PGD approximation with four computed modes shows excellent agreement with the full order solver, with relative errors below $0.5\%$. The results thus display the capability of the PGD-ROM strategy to devise a surrogate model for a quantity of physical interest, robust throughout the parametric domain $\I$, and with an accuracy acceptable for industrial applications.
\begin{table}[ht]
\begin{center}
	\begin{tabular}{|c||c|c|c|}
		\hline 
		$ \mu $	   &	$0.25$       &  $0.50$       &  $0.75$      \\
		\hline 
		  PGD          &	$1.102 c$  &  $1.062 c$   &  $1.024 c$ \\
		  full order  &	$1.103 c$  &  $1.059 c$   &  $1.019 c$ \\
		  \hline 
		  Relative error         &  $0.91 \times 10^{-3}$ & $0.28 \times 10^{-2}$ & $0.49 \times 10^{-2}$ \\
		\hline 
	\end{tabular}
	\caption{3D NASA wall-mounted hump: position of the reattachment point computed using the PGD approximation and the full order solver for different values of $\mu$.}
	\label{tab:reattach-3D}
\end{center}
\end{table}

%==========================================================================
\section{Conclusion}
\label{sc:Conclusion}
%==========================================================================

A PGD strategy to compute parametric solutions of turbulent incompressible flow problems in OpenFOAM has been proposed. The methodology is based on the incompressible Reynolds-averaged Navier-Stokes equations with Spalart-Allmaras turbulence model and mimics the segragated approach implemented in the industrially-validated solver OpenFOAM to devise a minimally intrusive PGD-ROM for convection-dominated flow problems of industrial interest.
First, the velocity and pressure modes are computed using the non-intrusive PGD strategy \texttt{PGD-NS} developed in~\cite{Tsiolakis-TGSOH-20} using a seed value for the turbulent viscosity. The PGD approximation of the velocity is then used to improve the turbulent viscosity representation via the minimally intrusive \texttt{PGD-SA} and \texttt{PGD-$\nu_t$} routines. Finally, the resulting separated turbulent viscosity is utilised to recompute the PGD expansions of velocity and pressure. 
The importance of an accurate approximation of the turbulent viscosity has been verified by comparing the solution of the above algorithm with the one computed without solving the SA equation: the latter solution quickly stagnates providing errors of one order of magnitude larger than the proposed methodology.

The developed strategy has been validated in two and three spatial dimensions using a benchmark problem of turbulent external flow, the NASA wall-mounted hump, with $\text{Re} {=} 93,600$ and $\text{Re} {=} 936,000$. A flow control problem of industrial interest has been devised by introducing a suction jet on the hump to reduce the recirculation effects. 
The proposed PGD-based reduced order model has proved to be able to compute a reduced basis with no \emph{a priori} knowledge of the solution, for convection-dominated viscous incompressible flows achieving both qualitative and quantitative agreement with the full order solution computed via \texttt{simpleFoam} with SA turbulence model, throughout the interval of the parametric variable.
More precisely, the reduced model provided accurate approximations of the velocity and pressure fields, with relative $\eltwo$ errors below $0.1 \%$ and $1 \%$, respectively. In addition, it proved to be able to capture localised flow features and estimate quantities of engineering interest with errors below $0.5 \%$.
The reported results thus highlight the robustness of the proposed PGD methodology in presence of turbulent phenomena and its capability to devise accurate approximations of the physical variables involved in the parametric problem (i.e. velocity and pressure), the variables modelling turbulent effects (i.e. eddy and turbulent viscosity), as well as relevant physical quantities, including the position of the reattachment point, the skin friction coefficient and the pressure coefficient.

Finally, it is worth noticing that the minimally intrusive nature of the proposed method represents a promising starting point for the construction of PGD-ROM strategies based on CFD software validated by the industry, beyond incompressible flows. This may be of interest e.g. for parametric compressible flow problems with applications to the aerospace industry. In this context, future investigations will have to deal with the additional difficulty of the treatment of shock waves whose position and intensity may depend upon the value of the parameters of the problem under analysis.

%==========================================================================
\section*{Acknowledgements}
%==========================================================================
This work was partially supported by the European Union's Horizon 2020 research and innovation programme under the Marie Sk\l odowska-Curie Actions (Grant agreement No. 675919) that financed the PhD fellowship of V.T..
M.G. acknowledges the support of the Serra H\'unter Programme of the Generalitat de Catalunya.
M.G., R.S. and A.H. were supported by the Spanish Ministry of Economy and Competitiveness (Grant agreement No. DPI2017-85139-C2-2-R). 
M.G. and A.H. are also grateful for the financial support provided by the Spanish Ministry of Economy and Competitiveness through the Severo Ochoa programme for centres of excellence in RTD (Grant agreement No. CEX2018-000797-S) and the Generalitat de Catalunya (Grant agreement No. 2017-SGR-1278).
R.S. also acknowledges the support of the Engineering and Physical Sciences Research Council (Grant number: EP/P033997/1).

%==========================================================================
%\bibliographystyle{plain}
%\bibliographystyle{elsarticle-num}
\bibliographystyle{unsrt}
\bibliography{Ref-PGD}
%==========================================================================

%==========================================================================
\appendix

%==========================================================================
\section{Classical and \emph{predictor-corrector} PGD algorithms}
\label{sc:appPredCorr}
%==========================================================================

In this appendix, a comparison of two PGD strategies, the classical one~\cite{Chinesta-Keunings-Leygue} and the one based on a \emph{predictor-corrector} approach~\cite{Tsiolakis-TGSOH-20}, is presented using an abstract variational framework with the goal of highlighting main differences, advantages and disadvantages of each solution.

Consider a, possibly nonlinear, partial differential equation (PDE) whose variational form is: seek $v(\bx) \in \V$ such that
\begin{equation}\label{eq:varPb}
\A(w,v(\bx)) = \ell(w) , \quad \forall w \in \V ,
\end{equation}
where $v(\bx)$ is the unknwon solution belonging to an appropriately defined functional space $\V$, $w$ is a test function and $\A$ and $\ell$ account for the differential operator and the independent term of the problem, respectively.

Given the set of parameters $\bmu \in \bI \subset \RR^M$, the solution of the corresponding parametric PDE is given by the function $v(\bx,\bmu) \in \Vmu := \V \otimes \eltwo(\I_1) \otimes \cdots \otimes \eltwo(\I_M)$ satifying
\begin{equation}\label{eq:varPbPar}
\Amu(w,v(\bx,\bmu)) = \Lmu(w)  , \quad \forall w \in \Vmu ,
\end{equation}
where
\begin{equation}\label{eq:paramVarForm}
\Amu(w,v) := \int_{\bI} \A(w,v) d\bmu
\quad \text{ and } \quad
\Lmu(w) := \int_{\bI} \ell(w) d\bmu .
\end{equation}

First, the classical PGD algorithm for the computation of the separated solution of equation~\eqref{eq:varPbPar} is recalled. The PGD approximation of the high-dimensional unknown function $v(\bx,\bmu)$ is thus defined as
\begin{equation}\label{eq:pgdApproxV}
v(\bx,\bmu) \simeq \vpgd^n(\bx,\bmu) = \vpgd^{n-1}(\bx,\bmu) + \sigmaV^n \fv^n(\bx)\varphi^n(\bmu) ,
\end{equation}
where $\fv^n(\bx)$ and $\varphi^n(\bmu)$ represent the $n$-th spatial and parametric modes, respectively and $ \sigmaV^n$ is its corresponding amplitude. It  is worth noticing that the previously introduced modes  are such that $\| \fv^n \| {=} \| \varphi^n \| {=} 1$, whereas the computed modes before the normalisation procedure are denoted by $\tilde{\fv}^{\! n}(\bx)$ and $\tilde{\varphi}^n(\bmu)$.

Algorithm~\ref{alg:classicPGD} reports the flowchart of the classical PGD algorithm described in~\cite{Chinesta-Keunings-Leygue}. A greedy strategy is employed to compute the terms in the rank-$n$ PGD approximation. More precisely, the algorithm assumes that the approximation $\vpgd^{n-1}$ is known to determine the $n$-th term in the PGD expansion using an alternating direction scheme. For this purpose, equation~\eqref{eq:varPbPar} is alternately projected upon the tangent manifold along the parametric (Algorithm 2 - Line 6) and spatial  (Algorithm 2 - Line 9) directions to determine $\tilde{\varphi}^n$ and $\tilde{\fv}^{\! n}$, respectively.
\begin{algorithm}
\caption{Classical PGD algorithm}\label{alg:classicPGD}
\begin{algorithmic}[1]
\REQUIRE{Stopping criterion $\eta_v^\star$ for the PGD enrichment. Stopping criterion \texttt{AD\_stopCrit} for the alternating direction scheme.}
\STATE{Compute boundary condition modes.}
\STATE{Set $n \gets 1$ and initialise the amplitude $\sigma_v^1 \gets 1$.}
%___Enrichment loop
\WHILE{$\sigma_v^n > \eta_v^\star\,\sigma_v^1$}
%___AD loop
\WHILE{\texttt{AD\_stopCrit} not fulfilled}
\STATE{Set the value of the spatial mode $\tilde{\fv}^{\! n}$.}
\STATE{Compute the parametric mode $\tilde{\varphi}^n$ by solving \\ $\Amu(w,\tilde{\fv}^{\! n}\tilde{\varphi}^n) = \Lmu(w) - \Amu(w,\vpgd^{n-1})$, projected along the parametric direction.}
\STATE{Normalise the parametric mode $\varphi^n = \dfrac{\tilde{\varphi}^n}{\|\tilde{\varphi}^n\|}$.}
\STATE{Set the value of the normalised parametric mode $\varphi^n$.}
\STATE{Compute the spatial mode $\tilde{\fv}^{\! n}$ by solving \\ $\Amu(w,\tilde{\fv}^{\! n}\varphi^n) = \Lmu(w) - \Amu(w,\vpgd^{n-1})$, projected along the spatial direction.}
\STATE{Compute the  amplitude $\sigma_v^n = \| \tilde{\fv}^{\! n} \|$.}
\STATE{Normalise the spatial mode $\fv^n = \dfrac{\tilde{\fv}^{\! n}}{\sigmaV^n}$.}
\ENDWHILE
\STATE{Update the mode counter $n \gets n+1$.}
\ENDWHILE
\end{algorithmic}
\end{algorithm}
Several variations of algorithm~\ref{alg:classicPGD} have been proposed in the literature and great attention has been devoted to the control of the alternating direction loop (Algorithm 2 - Line 4). The most common approach is to perform a fixed number of iterations in the alternating direction loop, see e.g.~\cite{Chinesta-Keunings-Leygue}. Arnoldi-type methods~\cite{Nouy:08} only perform one iteration of the alternating direction loop and then eliminate the redundant information by means of a realignment procedure. Recently, stopping criteria for the alternating direction loop requiring a target accuracy on the relative increments between the last two computed modes and the relative variation in the amplitude of the modes were also explored~\cite{Diez-DZGH-19}.

Stemming from the above  framework, the \emph{predictor-corrector} PGD algorithm, see~\cite{Tsiolakis-TGSOH-20}, replaces the separated approximation in equation~\eqref{eq:pgdApproxV} by
\begin{equation}\label{eq:pgdApproxVpredCorr}
\vpgd^n(\bx,\bmu) = \vpgd^{n-1}(\bx,\bmu) + \sigmaV^n \left[ \fv^n(\bx)\varphi^n(\bmu) + \De( \fv^n(\bx)\varphi^n(\bmu) ) \right] ,
\end{equation}
where $\fv^n(\bx)$  and $\varphi^n(\bmu)$ represent the prediction of the $n$-th modes,  whereas the correction $\De( \fv^n(\bx)\varphi^n(\bmu))$ is defined as
\begin{equation}\label{eq:vCorr}
\De( \fv^n(\bx)\varphi^n(\bmu) ) := \De\fv(\bx)\varphi^n(\bmu)+\fv^n(\bx)\De\varphi(\bmu)  .
\end{equation}
In addition, the computed predictions before the normalisation procedure are denoted by $\tilde{\fv}^{\! n}(\bx)$ and $\tilde{\varphi}^n(\bmu)$ and the corresponding corrections are represented by $\De\tilde{\fv}(\bx)$ and $\De\tilde{\varphi}(\bmu)$.

The \emph{predictor-corrector} PGD strategy (see algorithm~\ref{alg:predCorrPGD}) thus splits the computation of each new mode in two stages: first, the \emph{prediction} step solves one parametric problem (Algorithm 3  - Line 5) and one spatial problem (Algorithm 3  - Line 8)  to compute $\tilde{\varphi}^n(\bmu)$ and $\tilde{\fv}^{\! n}(\bx)$, respectively; second, the \emph{correction} stage performs multiple iterations of the alternating direction method to determine the \emph{best} corrections $\De\tilde{\varphi}(\bmu)$ (Algorithm 3  - Line 11)  and $\De\tilde{\fv}(\bx)$ (Algorithm 3  - Line 14),  given the base solution represented by the predicted modes $\tilde{\varphi}^n(\bmu)$ and $\tilde{\fv}^{\! n}(\bx)$.
\begin{algorithm}
\caption{\emph{Predictor-corrector} PGD algorithm}\label{alg:predCorrPGD}
\begin{algorithmic}[1]
\REQUIRE{Stopping criterion $\eta_v^\star$ for the PGD enrichment. Stopping criterion \texttt{AD\_stopCrit} for the alternating direction scheme.}
\STATE{Compute boundary condition modes.}
\STATE{Set $n \gets 1$ and initialise the amplitude $\sigma_v^1 \gets 1$.}
%___Enrichment loop
\WHILE{$\sigma_v^n > \eta_v^\star\,\sigma_v^1$}
%___AD loop
\STATE{Set the value of the spatial mode $\tilde{\fv}^{\! n}$.}
\STATE{Compute the parametric prediction $\tilde{\varphi}^n$ by solving \\ $\Amu(w,\tilde{\fv}^{\! n}\tilde{\varphi}^n) = \Lmu(w) - \Amu(w,\vpgd^{n-1})$, projected along the parametric direction.}
\STATE{Normalise the parametric prediction $\varphi^n = \dfrac{\tilde{\varphi}^n}{\|\tilde{\varphi}^n\|}$.}
\STATE{Set the value of the normalised parametric prediction $\varphi^n$.}
\STATE{Compute the spatial prediction $\tilde{\fv}^{\! n}$ by solving \\ $\Amu(w,\tilde{\fv}^{\! n}\varphi^n) = \Lmu(w) - \Amu(w,\vpgd^{n-1})$, projected along the spatial direction.}
\WHILE{\texttt{AD\_stopCrit} not fulfilled}
\STATE{Set the value of the spatial mode $\tilde{\fv}^{\! n}$.}
\STATE{Compute the parametric correction $\De\tilde{\varphi}$ by solving \\ $\Amu(w,\tilde{\fv}^{\! n}\De\tilde{\varphi}) = \Lmu(w) - \Amu(w,\vpgd^{n-1}) - \Amu(w,\tilde{\fv}^{\! n}\varphi^n)$, projected along the parametric direction.}
\STATE{Update the normalised parametric mode $\varphi^n \gets \dfrac{\varphi^n+\De\tilde{\varphi}}{\|\varphi^n+\De\tilde{\varphi}\|}$.}
\STATE{Set the value of the normalised parametric mode $\varphi^n$.}
\STATE{Compute the spatial correction $\De\tilde{\fv}$ by solving \\ $\Amu(w,\De\tilde{\fv}\varphi^n) = \Lmu(w) - \Amu(w,\vpgd^{n-1}) - \Amu(w,\tilde{\fv}^{\! n}\varphi^n)$, projected along the spatial direction.}
\STATE{Compute the  amplitude $\sigma_v^n = \| \tilde{\fv}^{\! n}+\De\tilde{\fv} \|$.}
\STATE{Normalise the spatial mode $\fv^n = \dfrac{\tilde{\fv}^{\! n}+\De\tilde{\fv}}{\sigmaV^n}$.}
\ENDWHILE
\STATE{Update the mode counter $n \gets n+1$.}
\ENDWHILE
\end{algorithmic}
\end{algorithm}

It is worth noticing that the \emph{predictor-corrector} strategy in algorithm~\ref{alg:predCorrPGD} requires the solution of one extra set of parametric and spatial problems per mode with respect to the classical PGD procedure in algorithm~\ref{alg:classicPGD}. Nonetheless, in presence  of nonlinear problems like the RANS and the SA equations studied in this work, the \emph{predictor-corrector} strategy provides increased accuracy in the approximation of the high-dimensional nonlinear differential operator $\Amu$, by successively  correcting the base solution identified  by the predictions. This allows to reduce the number of iterations required by the alternating direction method to converge and the overall cost of the PGD algorithm. On the contrary, numerical experiments have shown that the computational gain of the \emph{predictor-corrector} PGD is negligible when linear problems are treated. Finally, the \emph{predictor-corrector} PGD algorithm also provides a natural stopping criterion for the alternating direction scheme, when the norm of the computed corrections is small with respect to the amplitude of the mode under analysis.

%==========================================================================
\section{PGD spatial and parametric coefficients}
\label{sc:appCoeff}
%==========================================================================

%===============================
\subsection*{PGD for the Navier-Stokes equations}
%===============================
The coefficients for the spatial iteration~\eqref{eq:PGD-NS-spatial} are
\begin{equation}\label{eq:NScoef-spatial}
\begin{alignedat}{5}
  \alpha_0 & := \int_{\bI}{ \left[\phi^n\right]^3 \, d\bI} , \quad
  && \alpha_1 && := \int_{\bI}{ \left[\phi^n\right]^2 \zeta \, d\bI} , \quad
  && \alpha_2 && := \int_{\bI}{ \left[\phi^n\right]^2 \, d\bI} , \\
  \alpha_3^j & := \int_{\bI}{ [\phi^n]^2 \phi^j \, d\bI}, \quad
  && \alpha_4^{j\ell} && := \int_{\bI}{ \phi^n \phi^j \phi^\ell \, d\bI} , \quad
  && \alpha_5^\ell && := \int_{\bI}{ \phi^n \phi^\ell \zeta \, d\bI} , \\
  \alpha_6^\ell & := \int_{\bI}{ \phi^n \phi^\ell \, d\bI} , \quad 
  && \alpha_7^j && := \int_{\bI}{ \left[\phi^n\right]^2 \xi^j \, d\bI} , \quad
  && \alpha_8^{j\ell} && := \int_{\bI}{ \phi^n \phi^\ell \xi^j \, d\bI} .
\end{alignedat}
\end{equation}

The coefficients for the parametric iteration~\eqref{eq:PGD-NS-param} are
\begin{equation}\label{eq:NScoef-param}
\begin{aligned}
  a_0 &:= \int_{V_i}{\!\!\! \sigmaU^n\fu^n {\cdot} \bigl[\grad {\cdot} (\sigmaU^n\fu^n {\otimes} \sigmaU^n\fu^n)\bigr] \, dV} , \\
  a_1 &:=  \int_{V_i}{\!\!\! \sigmaU^n\fu^n {\cdot} \bigl[\grad {\cdot} (D \grad (\sigmaU^n\fu^n))\bigr] \, dV} , \\
  a_2 &:= \int_{V_i}{\!\!\! \sigmaU^n\fu^n {\cdot} \grad (\sigmaP^n\fp^n)  \, dV} 
			 + \!\! \int_{V_i}{\!\!\! \sigmaP^n\fp^n \grad {\cdot} (\sigmaU^n\fu^n) \, dV} , \\
  a_3^j &:= \int_{V_i}{\!\!\! \sigmaU^n\fu^n {\cdot} \bigl[\grad {\cdot} (\sigmaU^n\fu^n {\otimes} \sigmaU^j\fu^j) + \grad {\cdot} (\sigmaU^j\fu^j {\otimes} \sigmaU^n\fu^n)\bigr] \, dV} , \\
  a_4^{j\ell} &:= \int_{V_i}{\!\!\! \sigmaU^n\fu^n {\cdot} \bigl[\grad {\cdot} (\sigmaU^j\fu^j {\otimes} \sigmaU^\ell\fu^\ell)\bigr] \, dV} , \\
  a_5^\ell &:= \int_{V_i}{\!\!\! \sigmaU^n\fu^n {\cdot} \bigl[\grad {\cdot} (D \grad (\sigmaU^\ell\fu^\ell))\bigr] \, dV} , \\
  a_6^\ell &:= \int_{V_i}{\!\!\! \sigmaU^n\fu^n {\cdot} \grad (\sigmaP^\ell\fp^\ell)  \, dV} , \\  
  a_7^\ell &:=  \int_{V_i}{\!\!\! \sigmaP^n\fp^n \grad {\cdot} (\sigmaU^\ell\fu^\ell) \, dV}  , \\
  a_8^j &:=  \int_{V_i}{\!\!\! \sigmaU^n\fu^n {\cdot} \bigl[\grad {\cdot} (\sigmaT^j\ft^j \grad (\sigmaU^n\fu^n))\bigr] \, dV} , \\
  a_9^{j\ell} &:= \int_{V_i}{\!\!\! \sigmaU^n\fu^n {\cdot} \bigl[\grad {\cdot} (\sigmaT^j\ft^j \grad (\sigmaU^\ell\fu^\ell))\bigr] \, dV} .
  \end{aligned}
\end{equation}

%=================================
\subsection*{PGD for the Spalart-Allmaras equations}
%=================================
The coefficients for the spatial iteration~\eqref{eq:PGD-SA-spatial} are
\begin{equation}\label{eq:SAcoef-spatial}
\begin{alignedat}{3}
  \beta_1^j & := \int_{\bI}{ \left[\psi^m\right]^2 \phi^j \, d\bI} , \quad
  && \beta_2 && := \int_{\bI}{ \left[\psi^m\right]^2 \zeta \, d\bI} , \\
  \beta_3^j & := \int_{\bI}{ \left[\psi^m\right]^2 \psi^j\, d\bI} , \quad
  && \beta_4 && := \int_{\bI}{ \left[\psi^m\right]^3 \, d\bI} , \\
  \beta_5 & := \int_{\bI}{ \left[\psi^m\right]^2 \Smu{\widetilde{S}^m} \, d\bI} , \quad
  && \beta_6 && := \int_{\bI}{ \left[\psi^m\right]^3 \Smu{f_w^m} \, d\bI} , \\
  \beta_7^j & := \int_{\bI}{ [\psi^m]^2 \psi^j \Smu{f_w^m} \, d\bI} , \quad
  && \beta_8^{j\ell} && := \int_{\bI}{ \psi^m \psi^\ell \phi^j \, d\bI} , \\
  \beta_9^\ell & := \int_{\bI}{ \psi^m \psi^\ell \zeta \, d\bI} , \quad
  && \beta_{10}^{j\ell} && := \int_{\bI}{ \psi^m \psi^\ell \psi^j \, d\bI} , \\
 \beta_{11}^\ell & := \int_{\bI}{ \psi^m \psi^\ell \Smu{\widetilde{S}^m} \, d\bI} , \quad
  && \beta_{12}^{j\ell} && := \int_{\bI}{ \psi^m \psi^\ell \psi^j \Smu{f_w^m} \, d\bI} .
\end{alignedat}
\end{equation}

The coefficients for the parametric iteration~\eqref{eq:PGD-SA-param} are
\begin{equation}\label{eq:SAcoef-param}
\begin{aligned}
  b_1^j :=& \, \int_{V_i}{\!\!\! \sigmaNU^m\fnu^m \bigl[\grad {\cdot} (\sigmaU^j\fu^j \sigmaNU^m\fnu^m) \bigr] \, dV} , \\
  b_2 :=& \, \frac{1}{\sigma} \int_{V_i}{\!\!\! \sigmaNU^m\fnu^m \bigl[\grad {\cdot} (D \grad (\sigmaNU^m \fnu^m)) \bigr]  \, dV} , \\
  b_3^j :=& \, \frac{1}{\sigma} \int_{V_i}{\!\!\! \sigmaNU^m\fnu^m \bigl[\grad {\cdot} (\sigmaNU^j \fnu^j \grad (\sigmaNU^m \fnu^m)) 
  + \grad {\cdot} (\sigmaNU^m \fnu^m \grad (\sigmaNU^j \fnu^j)) \bigr]  \, dV} \\
  &+ \frac{2 c_{b2}}{\sigma} \int_{V_i}{\!\!\! \sigmaNU^m\fnu^m \bigl[\grad (\sigmaNU^j \fnu^j ) {\cdot} \grad (\sigmaNU^m \fnu^m) \bigr]  \, dV} , \\
  b_4 :=& \, \frac{1}{\sigma} \int_{V_i}{\!\!\! \sigmaNU^m\fnu^m \bigl[\grad {\cdot} (\sigmaNU^m \fnu^m \grad (\sigmaNU^m \fnu^m)) \bigr]  \, dV} \\
  &+ \frac{c_{b2}}{\sigma} \int_{V_i}{\!\!\! \sigmaNU^m\fnu^m \bigl[\grad (\sigmaNU^m \fnu^m ) {\cdot} \grad (\sigmaNU^m \fnu^m) \bigr]  \, dV} , \\  
  b_5 :=& \, c_{b1} \int_{V_i}{\!\!\! \sigmaNU^m\fnu^m \bigl[\Sx{\widetilde{S}^m} \sigmaNU^m\fnu^m\bigr] \, dV} , \\
  b_6 :=& \, c_{w1} \int_{V_i}{\!\!\! \sigmaNU^m\fnu^m \Bigl[\frac{\Sx{f_w^m}}{\widetilde{d}^2} (\sigmaNU^m\fnu^m)^2 \Bigr] \, dV} , \\
  b_7^j :=& \, 2 c_{w1} \int_{V_i}{\!\!\! \sigmaNU^m\fnu^m \Bigl[\frac{\Sx{f_w^m}}{\widetilde{d}^2} \sigmaNU^j\fnu^j\sigmaNU^m\fnu^m \Bigr] \, dV} , \\
  b_8^{j\ell} :=& \, \int_{V_i}{\!\!\! \sigmaNU^m\fnu^m \bigl[\grad {\cdot} (\sigmaU^j\fu^j \sigmaNU^\ell\fnu^\ell) \bigr] \, dV} , \\
  b_9^\ell :=& \, \frac{1}{\sigma} \int_{V_i}{\!\!\! \sigmaNU^m\fnu^m \bigl[\grad {\cdot} (D \grad (\sigmaNU^\ell \fnu^\ell)) \bigr]  \, dV} , \\
  b_{10}^{j\ell} :=& \, \frac{1}{\sigma} \int_{V_i}{\!\!\! \sigmaNU^m\fnu^m \bigl[\grad {\cdot} (\sigmaNU^j \fnu^j \grad (\sigmaNU^\ell \fnu^\ell)) \bigr]  \, dV} \\
  &+ \frac{c_{b2}}{\sigma} \int_{V_i}{\!\!\! \sigmaNU^m\fnu^m \bigl[\grad (\sigmaNU^j \fnu^j ) {\cdot} \grad (\sigmaNU^\ell \fnu^\ell) \bigr]  \, dV} , \\
  b_{11}^\ell :=& \, c_{b1} \int_{V_i}{\!\!\! \sigmaNU^m\fnu^m \bigl[\Sx{\widetilde{S}^m} \sigmaNU^\ell\fnu^\ell\bigr] \, dV} , \\
  b_{12}^{j\ell} :=& \, c_{w1} \int_{V_i}{\!\!\! \sigmaNU^m\fnu^m \Bigl[\frac{\Sx{f_w^m}}{\widetilde{d}^2} \sigmaNU^j\fnu^j\sigmaNU^\ell\fnu^\ell \Bigr] \, dV} .
  \end{aligned}
\end{equation}

\end{document}